\documentclass[a4paper,11pt]{article}
\usepackage{jheppub} % for details on the use of the package, please see the JINST-author-manual
\usepackage{lineno}
  
\usepackage{amsmath}
\usepackage{xr-hyper} 
\usepackage{hyperref}
\usepackage{bookmark}
\usepackage{mathtools}
\usepackage{bbm}
\usepackage{cleveref}
\usepackage{esint}
\crefformat{equation}{(#2#1#3)}
\crefrangeformat{equation}{(#3#1#4--#5#2#6)}
\crefmultiformat{equation}{(#2#1#3}{, #2#1#3)}{, #2#1#3}{, #2#1#3)}
\crefname{section}{Section}{Sections}
\Crefname{section}{Section}{Sections}
\crefname{table}{Table}{Tables}
\Crefname{table}{Table}{Tables}
\crefname{figure}{Figure}{Figures}
\Crefname{figure}{Figure}{Figures}
\crefname{appendix}{Appendix}{Appendices}
\Crefname{appendix}{Appendix}{Appendices}

\usepackage{textcomp}
\usepackage[T1, T2A]{fontenc}% T2A for Cyrillic font encoding
\usepackage[russian, english]{babel}
\usepackage{amsmath, amssymb}
\usepackage{physics}
\usepackage{tensor}
\usepackage{color}
\usepackage{enumerate}
\usepackage{subcaption}
\usepackage{graphicx}
\usepackage{etoolbox}
\usepackage{ifthen}
\usepackage{pifont}
\usepackage[nodayofweek,level]{datetime}
\usepackage{fontawesome}
\usepackage{varwidth}
\usepackage{cancel}
\usepackage[table]{xcolor} % colours + \cellcolor
\usepackage{amsmath,amssymb} % \mathfrak, etc.
\usepackage{graphicx} % \resizebox
\usepackage{tikz}
\definecolor{solA}{rgb}{0.368,0.507,0.710}
\definecolor{solB}{rgb}{0.881,0.611,0.142}
\definecolor{solC}{rgb}{0.560,0.692,0.195}
\definecolor{solD}{rgb}{0.923,0.386,0.209}
\definecolor{solE}{rgb}{0.528,0.471,0.701}
\definecolor{solF}{rgb}{0.772,0.432,0.102}

\DeclareMathAlphabet{\mathpzc}{OT1}{pzc}{m}{it}

\newcommand{\iu}{\ensuremath{\mathrm{i}}}
\newcommand{\eu}{\ensuremath{\mathrm{e}}}
\newcommand{\curly}[1]{\ensuremath{\mathpzc{#1}}}

\usepackage{bbm}

\title{\boldmath Macroscopic origin of the topologically twisted index on $T^2 \times \Sigma_{\mathfrak{g}}$}

\author[a]{Nikolay Bobev,}
\author[b]{Vasil Dimitrov,}
\author[c,d]{and Dario Martelli}
\affiliation[a]{Institute for Theoretical Physics and Leuven Gravity Institute, KU Leuven, Celestijnenlaan 200D, B-3001 Leuven, Belgium}
\affiliation[b]{INRNE, Bulgarian Academy of Sciences, Tsarigradsko Chaussee 72, 1784 Sofia, Bulgaria}
\affiliation[c]{Dipartimento di Matematica, Universit\`a di Torino, Via Carlo Alberto 10, 10123 Torino, Italy}
\affiliation[d]{INFN, Sezione di Torino, Via Pietro Giuria 1, 10125 Torino, Italy}
\emailAdd{nikolay.bobev@kuleuven.be, vasild@inrne.bas.bg, dario.martelli@unito.it}

\abstract{
  We construct a novel family of non-supersymmetric, asymptotically locally AdS$_5$ solutions of five-dimensional minimal gauged supergravity. In Lorentzian signature the solutions describe finite temperature rotating electro-magnetically charged black strings with $S^1\times \Sigma_{\mathfrak{g}}$ horizon topology. In Euclidean signature the solutions are completely smooth and are asymptotic to $T^2\times \Sigma_{\mathfrak{g}}$. We present a detailed analysis of the thermodynamic properties of these gravitational backgrounds, compute their regularized on-shell action and analyze the extremal and supersymmetric limits. The supersymmetric, but non-extremal, limit of the Euclidean backgrounds is a family of complex black saddles and it provides the holographic dual description of the topologically twisted index of 4d $\mathcal{N}=1$ SCFTs placed on $T^2\times \Sigma_{\mathfrak{g}}$. We show that the supergravity regularized on-shell action in this limit is in agreement with the large $N$ limit of the topologically twisted index in the dual SCFT. 
}

\begin{document}
\maketitle
\flushbottom
 
%%%%%%%
\section{Introduction}\label{sec:intro}
%%%%%%%
Euclidean solutions of supergravity provide a pathway to a better understanding of the gravitational path integral with important applications to AdS/CFT and black hole
 physics~\cite{Cabo-Bizet:2018ehj,Bobev:2020pjk}. In the context of holography regular supersymmetric Euclidean supergravity backgrounds, which are locally asymptotic to AdS, offer a concrete arena to test and explore holography by making direct contact with the partition functions of the dual SCFT placed on compact Euclidean manifolds. These SCFT partition functions can often be efficiently computed via supersymmetric localization by reducing them to a matrix model which is then amenable to an explicit analysis via standard large $N$ methods. From the perspective of supersymmetric black holes, Euclidean supergravity backgrounds provide a precise way to understand the subtleties associated with the zero temperature thermodynamics, opening the door towards a detailed understanding of black hole entropy and microstates. The goal of this work is to construct and study such ``Euclidean black saddle'' solutions in 5d minimal $\mathcal{N}=2$ gauged supergravity which describe black strings in AdS$_5$ and provide the gravitational description of the topologically twisted index (TTI) in the holographically dual 4d $\mathcal{N}=1$ SCFT.

Part of the motivation for this work is to find a supergravity description of the TTI. This is a supersymmetric partition function that can be defined for 4d $\mathcal{N}=1$ QFTs 
placed on the manifold $T^2\times \Sigma_{\mathfrak{g}}$, with a background ${\rm U}(1)$ R-symmetry gauge field and non-zero flux threading $\Sigma_{\mathfrak{g}}$. Here $T^2$ is a 2d torus and $\Sigma_{\mathfrak{g}}$ is a Riemann surface of genus $\mathfrak{g}$ which for simplicity we take to be compact and smooth. Supersymmetry is preserved by employing the ${\rm U}(1)$ R-symmetry to perform a partial topological twist on $\Sigma_{\mathfrak{g}}$ \`a la Witten. For 4d SCFTs with a Lagrangian formulation the TTI can be computed by supersymmetric localization after reducing the path integral to a matrix model~\cite{Benini:2016hjo}. In a holographic context one needs to study the large $N$ limit of this matrix model which appears to be quite involved. Fortunately, the TTI simplifies significantly if one first takes a Cardy-like limit of a small thermal circle on $T^2$ in which case the leading term takes the following compact form~\cite{Hosseini:2016cyf,Hosseini:2019lkt}
\begin{equation}\label{eq:TTItauintro}
\log Z_{T^2\times \Sigma_{\mathfrak{g}}}(\Delta_I,\mathfrak{n}_I) = \frac{\iu \pi}{12\tau} c_{l}(\Delta_I,\mathfrak{n}_I)\,.
\end{equation}
Here $c_{l}$ is a trial central charge for an effective 2d $\mathcal{N}=(0,2)$ SCFT on $T^2$ obtained after a partial topological twist and dimensional reduction on $\Sigma_{\mathfrak{g}}$. The superconformal value of $c_{l}$ is determined by $c$-extremization~\cite{Benini:2012cz,Benini:2013cda}. Here we will be interested in the simplest possible limit of the TTI in which all flavor symmetry fugacities $\Delta_{I}$ are switched off and the background magnetic fluxes $\mathfrak{n}_I$ are aligned with the 4d superconformal ${\rm U}(1)$ R-symmetry. This limit of the TTI is universal since it exists for any 4d $\mathcal{N}=1$ SCFT that admits a topological twist on $\Sigma_{\mathfrak{g}}$ independent of the detailed structure of its flavor symmetry. For the universal limit of the TTI the right hand side of~\eqref{eq:TTItauintro} simplifies to $c_{l}=\frac{32}{3}(\mathfrak{g}-1)a$ where $a$ is the conformal anomaly of the 4d SCFT. This is a simple and universal result for the SCFT partition function and our goal will be to reproduce it from a dual supergravity calculation.

The universality of the TTI described above should be mirrored in the bulk supergravity description. This suggests to look for the holographic dual of the TTI as a solution of 5d $\mathcal{N}=2$ minimal gauged supergravity which contains precisely the bosonic fields, i.e. the metric and the ${\rm U}(1)$ graviphoton, that couple to the 4d $\mathcal{N}=1$ energy momentum multiplet used in the topological twist. This is a familiar predicament in the holographic studies of supersymmetric partition functions of SCFTs placed on compact Euclidean manifolds. Indeed, to construct the 5d supergravity solutions dual to the TTI we draw inspiration from the studies of the holographic dual to the 4d $\mathcal{N}=1$ superconformal index (SCI) on 
$S^1\times_{\omega_{1,2}}S^3$ and the 3d $\mathcal{N}=2$ TTI on $S^1\times \Sigma_{\mathfrak{g}}$. As explained in \cite{Cabo-Bizet:2018ehj,Cassani:2019mms} and \cite{Bobev:2020pjk}, see also~\cite{Cassani:2025sim} for a recent review, in both of these setups one finds the relevant Euclidean supersymmetric solution by taking a supersymmetric but not extremal limit of the non-supersymmetric and non-extremal black hole backgrounds presented in~\cite{Chong:2005hr} and~\cite{Romans:1991nq}, respectively. The resulting supersymmetric Euclidean black saddle solutions are in general complex but in the extremal limits reduce to the supersymmetric AdS black holes of \cite{Gutowski:2004ez} and \cite{Cacciatori:2009iz}.

Guided by these SCFT and supergravity results we are led to consider Euclidean solutions of 5d minimal $\mathcal{N}=2$ gauged supergravity with a $T^2\times \Sigma_{\mathfrak{g}}$ conformal boundary. To incorporate the topological twist we need to have magnetic flux on $\Sigma_{\mathfrak{g}}$ and for general values of $\tau$ in~\eqref{eq:TTItauintro} we need to add an appropriate rotation parameter. These considerations lead to an ansatz for the 5d supergravity fields that has both electric and magnetic components of the gauge field and off-diagonal components in the metric. The resulting equations of motion reduce to a complicated non-linear system of ODEs which we are not able to solve in closed form. Nevertheless we can make substantial progress. We show that the equations of motion (EOMs) admit a consistent IR series expansion in the bulk of the 5d space where the metric and gauge field are regular and the ``thermal circle'' inside the $T^2$ part of the geometry smoothly shrinks to zero size while the $\Sigma_{\mathfrak{g}}$ Riemann surface remains of finite volume. A similar asymptotic analysis near the UV AdS$_5$ boundary also reveals a consistent and smooth perturbative expansion of the EOMs compatible with the Fefferman-Graham form. Equipped with these series expansions we then employ numerics to show that there are smooth solutions of the EOMs that interpolate between these two asymptotic regions. This analysis results in a large family of novel Euclidean non-supersymmetric solutions of 5d $\mathcal{N}=2$ gauged supergravity. These backgrounds can be interpreted as the Euclidean manifestation of non-supersymmetric, rotating black string solutions that carry both magnetic and electric dipole charge and have $S^1\times \Sigma_{\mathfrak{g}}$ horizon topology. We note that while the topology of the horizon is that of a black ring, the asymptotic region of the solution is that of an asymptotically \textit{locally} AdS$_5$ background, whose topology is different from $\mathbb{R}\times S^3$. This is why we refer to these backgrounds as black strings rather than black rings.

To analyze these new solutions in detail we make use of both the UV and IR series expansions, together with a number of integrals of motion (IOMs).
  While we do not have explicit analytic solutions of the EOMs we manage to use these ingredients to find analytic expressions for the mass, entropy, angular momentum, electric and magnetic charges of the solutions, together with their thermodynamically dual fugacities, in terms of the independent UV and/or IR expansion parameters. Using these expressions we explicitly check the validity of the, so-called, quantum statistical relation. We encounter several subtleties in this analysis that stem from the presence of a Chern-Simons term for the ${\rm U}(1)$ gauge field in the 5d $\mathcal{N}=2$ supergravity action. This, combined with the fact that we study a compact Riemann surface $\Sigma_{\mathfrak{g}}$ with magnetic flux, leads to subtleties in the definition of electric charges, see~\cite{Marolf:2000cb,Copsey:2005se}, and the calculation of black hole entropy, see~\cite{Tachikawa:2006sz,Hanaki:2007mb}. The intricacies in these calculations essentially boil down to the fact that the Chern-Simons term is not gauge invariant, which in turn also leads to an on-shell action that is not gauge invariant, together with the non-trivial topology of $\Sigma_{\mathfrak{g}}$ that necessitates the introduction of local patches in explicit calculations.
  See \cite{Colombo:2025yqy} for recent discussions of this subtlety in the context of 5d minimal gauged supergravity.

To study the supersymmetric limit of these thermal Euclidean backgrounds we employ the general classification of supersymmetric solutions of 5d $\mathcal{N}=2$ minimal gauged supergravity in~\cite{Gauntlett:2003fk}. We then find the restrictions imposed by supersymmetry on the parameters that control the UV and IR expansion of the general black string backgrounds. Using these constraints we study the thermodynamics of the black string solutions in the supersymmetric limit. We find that there is a family of supersymmetric non-extremal backgrounds that are smooth and have an on-shell action that reproduces the SCFT result for the TTI in~\eqref{eq:TTItauintro}. These solutions in general do not have real entropy and thus do not admit a clear Lorentzian interpretation. As is by now familiar from similar holographic setups, one can impose an additional relation on the gravitational parameters of the supersymmetric solution that makes it extremal and with real entropy. In this limit, the supergravity backgrounds can be interpreted as supersymmetric Lorentzian rotating black strings with $S^1\times \Sigma_{\mathfrak{g}}$ horizon topology and finite entropy.

We note that black string solutions in 5d $\mathcal{N}=2$ gauged supergravity have been discussed previously in the literature. Static magnetically charged black strings were presented in~\cite{Chamseddine:1999xk,Klemm:2000nj} and further generalized in~\cite{Bernamonti:2007bu}. In~\cite{Hong:2021bzg} an attempt to construct the non-extremal supersymmetric background dual to the TTI was discussed and a near-horizon limit of the solution was presented. We discuss how to obtain the supergravity backgrounds in~\cite{Bernamonti:2007bu} and~\cite{Hong:2021bzg} as particular limits of our solutions. Further holographic studies of the TTI can be found in~\cite{Hosseini:2016cyf,Hosseini:2020vgl,Hosseini:2019lkt,Hong:2018viz}. The novelty in our construction is that we have non-extremal black strings with angular momentum, together with electric dipole and magnetic monopole charges. We note in passing that a particular limit of our supersymmetric solutions leads to a supergravity background that is not asymptotically locally AdS$_5$ but rather takes the form of a product between a 3d non-extremal supersymmetric version of the BTZ background and the Riemann surface $\Sigma_{\mathfrak{g}}$.

The 5d supergravity backgrounds we construct can be uplifted to solutions of type IIB or 11d supergravity by using the consistent truncation results in~\cite{Gauntlett:2007ma,Gauntlett:2007sm}. This provides an embedding of our construction in string/M-theory and a direct relation to various classes of holographic SCFTs realized on the worldvolume of D3- or M5-branes. From the perspective of the branes the construction involves wrapping two of the worldvolume directions on $\Sigma_{\mathfrak{g}}$ and the topological twist is realized by an appropriate fibration of the normal directions to the brane over the Riemann surface. Indeed, we discuss these top-down constructions in \cref{sec:TTI_5d} where we summarize three classes of such holographic SCFTs and their conformal anomalies which ultimately determine the leading term in the TTI in~\eqref{eq:TTItauintro}. 

We continue our exposition in the next section where we discuss Euclidean AdS$_4$ solutions of 4d supergravity dual to 3d SCFTs on $S^1\times \Sigma_{\mathfrak{g}}$. This serves as a blueprint for the much more involved construction and analysis of the 5d minimal $\mathcal{N}=2$ gauged supergravity solutions that are of main interest in this work. We present a brief summary of the QFT results for the TTI of 4d $\mathcal{N}=1$ SCFTs focusing on the large $N$ and Cardy limit relevant for our holographic discussion in \cref{sec:TTI_5d}. In \cref{sec:thermo_bs} we present the construction of the novel family of non-supersymmetric Euclidean solutions of 5d minimal $\mathcal{N}=2$ gauged supergravity and study their on-shell action and thermodynamics in detail. The supersymmetric limit of these new solutions is discussed in \cref{sec:susy_solutions} where we also show that the supersymmetric on-shell action agrees with the results in the dual SCFT. We conclude our discussion and outline some problems for future work in \cref{sec:conclusion}. The appendices contain further details on some of the technical aspects of the 5d supergravity analysis.

%%%%%%%%%%%%%%%%%%%%%%%%%%%%%%%%%%

%
\section{The gravity dual of the TTI on $S^1 \times \Sigma_{\mathfrak{g}}$}\label{sec:4d}

The TTI is a supersymmetric partition function that can be defined for 3d $\mathcal{N}=2$ SCFTs placed on $S^1_{\beta}\times\Sigma_{\mathfrak{g}}$ with a partial topological twist on $\Sigma_{\mathfrak{g}}$. Similarly to the Witten index it is independent of the size $\beta$ of the $S^1$ and can be computed efficiently by supersymmetric localization~\cite{Benini:2015noa,Benini:2016hjo,Closset:2016arn}. Our goal in this section is to summarize how the TTI for 3d $\mathcal{N}=2$ SCFTs with a holographic dual is realized in terms of Euclidean supersymmetric solutions in 4d $\mathcal{N}=2$ minimal gauged supergravity. The relevant supersymmetric backgrounds are obtained from Euclidean non-extremal and non-supersymmetric dyonic solutions of the 4d supergravity theory which upon specific limits of the mass, electric and magnetic charge parameters reduce to finite $\beta$ supersymmetric solutions. Taking the extremal limit $\beta \to \infty$ of these backgrounds then leads to BPS solutions that admit an interpretation as supersymmetric Lorentzian black holes. The results presented in this section are mostly well-known but we summarize them here since they serve as a blueprint for the more complicated subsequent discussion on the TTI for 4d $\mathcal{N}=1$ SCFTs and its dual gravitational description.

\subsection{Dyonic black holes}\label{sec:dyonic_4d}

The bosonic action of 4d $\mathcal{N}=2$ minimal gauged supergravity in Lorentzian signature is
\begin{align}\label{eq:action_4d}
  S ={}& \frac{1}{16 \pi G_4} \int_{M_4} \qty[\qty(R_4 + \frac{6}{\ell_4^2}) \star_4 - \, 2 F \wedge \star_4 F] + \frac{\vartheta}{8\pi^2} \int_{M_4} F \wedge F \,,
\end{align}
where $\ell_4$ is the scale of $\text{AdS}_4$ and we have included the topological $\vartheta$ term. The $\vartheta$ term is locally a total derivative, so it does not influence the equations of motion or the supersymmetry variations, but it nevertheless plays a role in the thermodynamics of the magnetically charged solutions we study. The equations of motion derived from \eqref{eq:action_4d} read
\begin{align}\label{eq:eom_4d}
  \begin{aligned}
    0 ={}& (R_4)_{\mu\nu} + \frac{3}{\ell_4^2} g_{\mu\nu} + \frac{1}{2} g_{\mu\nu} F_{\rho\sigma} F^{\rho\sigma} - 2 F\indices{_\mu^\rho} F_{\nu\rho} \,, \\
    0 ={}& \dd[]{(\star_4 F)} \,. 
  \end{aligned}
\end{align}
We begin with static dyonic non-supersymmetric black hole solutions of the form
\begin{align}\label{eq:static_dyonic_non-susy}
  \begin{aligned}
    \dd[]{s}^2_4 ={}& - U(r) \dd[]{t}^2 + \frac{\dd[]{r}^2}{U(r)} + r^2 \dd[]{s}^2_\Sigma \,, \\
    U(r) ={}& \kappa + \frac{r^2}{\ell_4^2} + \frac{\mathfrak{p}^2 + q^2}{r^2} + \frac{\mu}{r} \,, \\
    F ={}& - \frac{q}{r^2} \dd[]{t} \wedge \dd[]{r} - \mathfrak{p} \qty(\star_\Sigma) \,,
  \end{aligned}
\end{align}
where $\qty{\mu, q, \mathfrak{p}}$ are integration constants associated with mass, electric charge and magnetic charge, respectively. We will simultaneously treat the cases
\begin{align}\label{eq:genus_table}
  \begin{aligned}
    \mathfrak{g} ={}& 0: & \Sigma \cong{}& S^2 \,, & \kappa &= 1 \,, & \eta_\Sigma ={}& 2 \,, \\
    \mathfrak{g} ={}& 1: & \Sigma \cong{}& T^2 \,, & \kappa &= 0 \,, & \eta_\Sigma ={}& 1 \,, \\
    \mathfrak{g} >{}& 1: & \Sigma \cong{}& \Sigma_{\mathfrak{g} > 1} \,, & \kappa &= -1 \,, & \eta_\Sigma ={}& 2(\mathfrak{g} - 1) \,,
  \end{aligned}
\end{align}
where $\mathfrak{g}$ is the genus of the Riemann surface $\Sigma$, $\kappa$ is its constant Gaussian curvature, and $\eta_\Sigma$ is its normalized volume
\begin{align}
  \eta_\Sigma = \frac{1}{2\pi}\int_\Sigma \qty(\star_\Sigma) \,. 
\end{align}
The black hole horizon is located at the largest positive root $r_+$ of $U(r)$
\begin{align}
  U(r_+) ={}& 0 \quad \implies \quad \mu = - r_+ \qty(\kappa + \frac{r_+^2}{\ell_4^2} + \frac{\mathfrak{p}^2 + q^2}{r_+^2}) \,. 
\end{align}
Indeed, in the vicinity of $r_+$, expressed in terms of a new radial coordinate $r = r_+ + R^2$, the metric takes the form
\begin{align}
  \dd[]{s}^2_4 ={}& \frac{\beta}{\pi} \qty(\dd[]{R}^2 - R^2 \qty(\frac{2\pi}{\beta})^2 \dd[]{t}^2) + r_+^2 \dd[]{s}^2_\Sigma + \order{R^2} \,,
\end{align}
where
\begin{align}
  \beta ={}& \frac{4\pi r_+}{\kappa + \frac{3 r_+^2}{\ell_4^2} - \frac{\mathfrak{p}^2 + q^2}{r_+^2}} \,. 
\end{align}
In Euclidean signature, with $t = -\iu t_E$, this is the geometry of a smoothly capped cigar times $\Sigma$, provided that we identify
\begin{align}
  t_E \sim t_E + \beta \,. 
\end{align}
The parameter $\beta$ therefore plays the role of inverse temperature. The gauge field takes the form
\begin{align}
  A ={}& - \frac{q}{r} \dd[]{t} - \mathfrak{p} \, \omega_\Sigma - \Phi \dd[]{t} \,, 
\end{align}
where $\omega_\Sigma$ is the potential one-form for $\qty(\star_\Sigma)$ and the integration constant $\Phi$ plays the role of electrostatic potential. 
To see this, we use the standard procedure of evaluating the difference in the electromagnetic potential between the horizon and asymptotic infinity to find
\begin{align}
  \Phi_{\text{phys}} ={}& \eval{\iota_V A}_{H}^{} - \eval{\iota_V A}_{\infty}^{} = - \qty(\frac{q}{r_+} + \Phi) - (-\Phi) = - \frac{q}{r_+} \,, \qquad V = \partial_t \,,
\end{align}
together with the regular-gauge condition $\iota_V A|_H = 0$, which fixes
\begin{align}
  \Phi ={}& - \frac{q}{r_+} \,, 
\end{align}
so that $\Phi_{\text{phys}} = \Phi$.

%%%%%%%%
\subsection{Thermodynamic quantities}\label{sec:thermo_quantities_4d}
%%%%%%%%

It is instructive to derive the thermodynamic quantities of the black holes presented above in some detail.

First, we bring the field configuration \eqref{eq:static_dyonic_non-susy} into Fefferman-Graham form
\begin{align}\label{eq:metric_fg_form}
  \begin{aligned}
  \dd[]{s}^2_4 ={}& \frac{\ell_4^2}{z^2} \dd[]{z}^2 + \dd[]{s}^2_h \,, \quad \dd[]{s}^2_h = \frac{\ell_4^2}{z^2} \qty[ - u(z) \dd[]{t}^2 + v(z) \dd[]{s}^2_\Sigma] \,, \\
  A ={}& \widetilde{a}_t(z) \dd[]{t} - \mathfrak{p} \, \omega_\Sigma \,,
  \end{aligned}
\end{align}
where $h_{ij}$ is the induced metric on surfaces of constant radial coordinate $z$, and the conformal boundary is located at $z = 0$. The change of radial coordinate is determined by
\begin{align}\label{eq:change_to_FG_diff_equ}
  \frac{\dd[]{(r(z))}^2}{U(r(z))} ={}& \frac{\ell_4^2}{z^2} \dd[]{z}^2 \quad \implies \quad (r'(z))^2 = \frac{\ell_4^2}{z^2} U(r(z)) \,. 
\end{align}
Since the above differential equation does not appear to admit a closed form solution, we proceed perturbatively around $z = 0$
\begin{align}\label{eq:radial_to_fg_radial}
  r(z) ={}& \frac{\ell_4}{z} - \frac{\ell_4 \kappa}{4} z - \frac{\mu}{6} z^2 - \frac{\mathfrak{p}^2 + q^2}{8 \ell_4} z^3 - \frac{\mu\kappa}{60} z^4 - \frac{9(\mathfrak{p}^2 + q^2)\kappa + 2 \mu^2}{288 \ell_4} z^5 + \dots \,, 
\end{align}
where we have fixed the coefficient of the leading term to a convenient value and the remaining coefficients are then determined by~\eqref{eq:change_to_FG_diff_equ}. Comparing the field configurations in \eqref{eq:metric_fg_form} and \eqref{eq:static_dyonic_non-susy}, we read off
\begin{align}
  u(z) ={}& U(r(z)) \frac{z^2}{\ell_4^2} \,, \quad v(z) = r(z)^2 \frac{z^2}{\ell_4^2} \,, \quad \widetilde{a}_t(z) = - \frac{q}{r(z)} - \Phi \,. 
\end{align}
Given this, $ \qty{u(z), v(z), \widetilde{a}_t(z)}$ can be determined to any desired order. Here we report their expansions up to 5th order
\begin{align}
  \begin{aligned}
    u(z) &= \frac{1}{\ell_4^2} + \frac{\kappa}{2 \ell_4^2} z^2 + \frac{2\mu}{3\ell_4^3} z^3 + \frac{12(\mathfrak{p}^2 + q^2) + \ell_4^2 \kappa^2}{16 \ell_4^4} z^4 + \frac{3\mu\kappa}{10 \ell_4^3} z^5 + \dots \,, \\
    v(z) &= 1 - \frac{\kappa}{2} z^2 - \frac{\mu}{3 \ell_4} z^3 - \frac{4(\mathfrak{p}^2 + q^2) - \ell_4^2 \kappa^2}{16 \ell_4^2} z^4 + \frac{\kappa\mu}{20 \ell_4} z^5 + \dots \,, \\
    \widetilde{a}_t(z) &= - \Phi - \frac{q}{\ell_4} z - \frac{q\kappa}{4 \ell_4} z^3 - \frac{q\mu}{6 \ell_4^2} z^4 - \frac{q(2(\mathfrak{p}^2 + q^2) + \ell_4^2 \kappa^2)}{16 \ell_4^3} z^5 + \dots \,. 
  \end{aligned}
\end{align}
Thus far, the conformal boundary metric and its background gauge field are
\begin{align}\label{eq:conf_bdy_4d}
  \begin{aligned}
  \dd[]{s}^2_\partial = - \frac{\dd[]{t}^2}{\ell_4^2} + \dd[]{s}^2_\Sigma \,, \quad A_\partial = - \Phi \dd[]{t} - \mathfrak{p} \, \omega_\Sigma \,. 
  \end{aligned}
\end{align}
Next, we supplement the bulk action in \eqref{eq:action_4d} with the standard set of (divergent) holographic boundary counterterms
\begin{align}
  S_{\text{bdy}} ={}& \frac{2}{16\pi G_4} \int_{\partial M_4} \qty(K - \frac{2}{\ell_4} - \frac{\ell_4}{2} R_h) \, \star_h \,,
\end{align}
where $K$ is the extrinsic curvature
\begin{align}
  K ={}& h^{ij} K_{ij} \,, \quad K_{ij} = - \frac{z}{2 \ell_4} \partial_z h_{ij} \,,
\end{align}
and $R_h$ is the Ricci scalar of the induced metric. The holographic stress-energy tensor is obtained as follows
\begin{align}
  T_{ij} ={}& - \frac{2}{\sqrt{-h}} \frac{\delta}{\delta h^{ij}} \qty[\sqrt{-h} \qty(K - \frac{2}{\ell_4} - \frac{\ell_4}{2} R_h)] \notag \\
  ={}& - K_{ij} + \ell_4 (R_h)_{ij} + h_{ij} \qty(K - \frac{2}{\ell_4} - \frac{\ell_4}{2} R_h) \,. 
\end{align}
Using this we can find the energy which reads
\begin{align}
  E ={}& - \frac{2}{16 \pi G_4} \int_{\Sigma} \sqrt{-h} \, T\indices{^t_t} = - \frac{\eta_\Sigma}{4G_4} \mu \,. 
\end{align}
The bulk part of the Euclidean on-shell action evaluates to
\begin{align}
  I ={}& - \iu \eval{S}_{t \rightarrow -\iu t_E} = \frac{\beta \eta_\Sigma}{4G_4} \qty[\frac{r_c^3}{\ell_4^2} - \frac{r_+^3}{\ell_4^2} + \frac{\mathfrak{p}^2 - q^2}{r_+} ] - \beta \eta_\Sigma \frac{\vartheta}{2\pi} \frac{\mathfrak{p} q}{r_+} \,,
\end{align}
where $r_c$ is the radial cutoff. Similarly, the boundary part of the Euclidean on-shell action evaluates to
\begin{align}
  I_{\text{bdy}} ={}& -\iu \eval{S_{\text{bdy}}}_{t \rightarrow -\iu t_E} = \frac{\beta \eta_\Sigma}{4 G_4} \qty[ - \frac{\ell_4}{z_c^3} + \frac{3 \ell_4 \kappa}{4 z_c}] \,. 
\end{align}
Using the relation between the two radial coordinates~\eqref{eq:radial_to_fg_radial} the divergent terms cancel and a finite piece remains. Overall, the renormalized Euclidean on-shell action evaluates to
\begin{align}\label{eq:osa_non-susy_4d}
  \widehat{I} = I + I_{\text{bdy}} ={}& \frac{\beta \eta_\Sigma}{4G_4} \qty[- \frac{r_+^3}{\ell_4^2} + \frac{\mathfrak{p}^2 - q^2}{r_+} - \frac{\mu}{2}] - \beta \eta_\Sigma \frac{\vartheta}{2\pi} \frac{\mathfrak{p} q}{r_+} \,.
\end{align}
For the electric charge, we use the Page definition, see \cite{Marolf:2000cb} 
\begin{align}
  Q ={}& \frac{1}{4\pi G_4} \int_{\Sigma} \star_4 F = \frac{\eta_\Sigma}{G_4} \frac{q}{2} \,.
\end{align}
For the magnetic charge, we find
\begin{align}
  P ={}& \frac{1}{2\pi} \int_{\Sigma} F = - \eta_\Sigma \mathfrak{p} \,.
\end{align}
The entropy is obtained from the horizon area as
\begin{align}
  \mathcal{S} ={}& \frac{\text{Area}_H}{4G_4} = \frac{1}{4G_4} \int_{\Sigma} \sqrt{g|_H} = \frac{\eta_\Sigma}{4G_4} 2\pi r_+^2 \,. 
\end{align}
%

%%%%%%%%%%%
\subsection{Thermodynamic relations}
\label{sec:4d_thermo}
%%%%%%%%%%%

The on-shell action and the thermodynamic quantities derived above are linked through the quantum statistical relation (QSR)
\begin{align}\label{eq:qsr_4d}
  \widehat{I} ={}& - \mathcal{S} + \beta E + \beta\Phi Q - \beta\Phi \frac{\vartheta}{2\pi} P \,. 
\end{align}
Observe that the QSR is invariant under the combined $\vartheta$-angle shift and electric-charge redefinition
\begin{align}
  \begin{aligned}
    \vartheta & &\rightarrow& & &\vartheta + 2\pi n \,, \\
    Q & &\rightarrow& & & Q + n P \,,
  \end{aligned}
\end{align}
where $n \in \mathbb{Z}$. As discussed in~\cite{Heydeman:2024fgk}, this is an avatar of the Witten effect \cite{Witten:1979ey}: a $\vartheta$-angle shift transmutes magnetic charge into electric charge. It is important to emphasize that the
expression for the on-shell action \eqref{eq:osa_non-susy_4d} that satisfies the QSR \eqref{eq:qsr_4d} is evaluated in an ensemble where 
$\qty{\beta, \Phi, \mathfrak{p}}$ are held fixed: this is indeed the data held fixed by the boundary field configuration \eqref{eq:conf_bdy_4d}.

%%%%%%%%%%%%%%%%
\subsection{Supersymmetric limits}
%\subsection{Thermal black holes $\rightarrow$ supersymmetric black saddles $\rightarrow$ BPS black holes}
\label{sec:4d_limits}
%%%%%%%%%%%%%%%%

The study of supersymmetric black hole solutions of 4d minimal gauged supergravity has a long history, starting with \cite{Romans:1991nq} and later on \cite{Caldarelli:1998hg}. See also \cite{Genolini:2021qbi, Heydeman:2024fgk} for recent discussions that explicitly include the $\vartheta$ term. Here we will borrow results from \cite{Bobev:2020pjk}, where the Euclidean Killing spinor
equations for arbitrary-genus, static, dyonic, Euclidean solutions --- dubbed \textit{black saddles} --- are solved directly. 
Converted to our conventions\footnote{$g^{\text{there}} = {1}/{(\sqrt{2} \ell_4)}$, $A^{\text{there}} = {A}/{\sqrt{2}}$, $Q^{\text{there}} = {q_E}/{(2 \sqrt{2})}$, $\xi^{\text{there}} = -1$.}, the supersymmetric black saddle configuration reads
\begin{align}\label{eq:static_dyonic_susy}
  \begin{aligned}
    \dd[]{s}^2_4 ={}& U(r) \dd[]{t_E}^2 + \frac{\dd[]{r}^2}{U(r)} + r^2 \dd[]{s}^2_\Sigma \,, \\
    U(r) ={}& \qty(\frac{r}{\ell_4} + \frac{\ell_4 \kappa}{2r})^2 - \frac{q_E^2}{r^2} \,, \\
    F ={}& \frac{q_E}{r^2} \dd[]{t_E} \wedge \dd[]{r} + \frac{\ell_4 \kappa}{2} \qty(\star_\Sigma) \,.
  \end{aligned}
\end{align}
Notice that this is a non-extremal, i.e. finite $\beta$, supersymmetric background of the 4d gauged supergravity. Under the Wick rotation $t = -\iu t_E$, one obtains a complex gauge field. Optionally, one can work on a real Euclidean section by further continuing $q = -\iu q_E$, as done in~\cite{Bobev:2020pjk}. Here we will directly work with the Lorentzian parameter $q$. Comparing \eqref{eq:static_dyonic_non-susy} and \eqref{eq:static_dyonic_susy}, we see that the \textit{supersymmetry locus} is specified by
\begin{align}\label{eq:susy_locus_4d}
  \mu ={}& 0 \,, \quad \mathfrak{p} = - \frac{\ell_4 \kappa}{2} \quad \implies \quad q = \mathfrak{s}_Q \, \iu \qty(\frac{r_+^2}{\ell_4} + \frac{\ell_4 \kappa}{2}) \,, \quad \mathfrak{p} = - \frac{\ell_4 \kappa}{2} \,,
\end{align}
where $\mathfrak{s}_Q = \pm 1$. We can now evaluate the regularized on-shell action and thermodynamic quantities on the supersymmetric locus \eqref{eq:susy_locus_4d} to find 
\begin{align}
  \begin{aligned}
    \widehat{I} ={}& \qty(\frac{\pi\ell_4^2}{4 G_4} + \mathfrak{s}_Q \frac{\iu \ell_4^2}{4} \vartheta) \eta_\Sigma \kappa \,, \\
    \beta ={}& \frac{2\pi r_+}{\frac{2 r_+^2}{\ell_4^2} + \kappa}\,, \\
    \Phi ={}& - \mathfrak{s}_Q \, \frac{\iu \ell_4}{2 r_+} \qty(\frac{2 r_+^2}{\ell_4^2} + \kappa)\,,
  \end{aligned} \quad 
  \begin{aligned}
    \mathcal{S} ={}& \frac{\pi r_+^2}{2 G_4} \eta_\Sigma \,, \\
    E ={}& 0 \,, \\
    Q ={}& \mathfrak{s}_Q \frac{\iu \ell_4}{4 G_4} \qty(\frac{2 r_+^2}{\ell_4^2} + \kappa) \eta_\Sigma \,, \\
    P ={}& \frac{\ell_4}{2} \eta_\Sigma \kappa \,. 
  \end{aligned}
\end{align}
Some notable features of the supersymmetric locus are:
\begin{itemize}
  \item The on-shell action becomes $\beta$-independent.
  \item The thermodynamic potentials satisfy a \textit{supersymmetric linear constraint} on the \textit{second sheet} in the language of \cite{Cassani:2021fyv}
    \begin{align}\label{eq:susy_linear_constraint_4d}
      \beta\Phi ={}& - \mathfrak{s}_Q \, 2 \pi \iu \, \frac{\ell_4}{2} \,. 
    \end{align}
  \item The entropy, written as a function of the charges, is
    \begin{align}\label{eq:entropy_4d_susy}
      \mathcal{S} = - \frac{\pi \ell_4}{2 G_4} P - \mathfrak{s}_Q \, \iu \, \pi \ell_4 Q \,. 
    \end{align}
\end{itemize}

Next, we move on to describe the \textit{BPS locus}, where supersymmetry and extremality, i.e. $\beta \to \infty$, are imposed simultaneously. It turns out that the BPS locus can be reached by demanding real entropy in \eqref{eq:entropy_4d_susy}
\begin{align}\label{eq:4d_Qzero_leads_to_extr}
  Q ={}& 0 \quad \implies \quad r_+ = \pm \iu \ell_4 \sqrt{\frac{\kappa}{2}} \quad \implies \quad \beta \rightarrow \infty \,.
\end{align}
On the BPS locus one has
\begin{align}
  \widehat{I}_{\text{BPS}} ={}& \qty(\frac{\pi\ell_4^2}{4 G_4} + \mathfrak{s}_Q \frac{\iu \ell_4^2}{4} \vartheta) \eta_\Sigma \kappa \,, \quad \mathcal{S}_{\text{BPS}} = - \frac{\pi \ell_4}{2 G_4} P \,. 
\end{align}
Note that the finite $\beta$ supergravity solution is regular in Euclidean signature for any genus~$\mathfrak{g}$. The $\mathfrak{g} = 0,1$ saddles are peculiar for different reasons:\footnote{Note that this puzzling behaviour of the $\mathfrak{g} = 0,1$ saddles can be modified in the presence of matter fields. Indeed, in the 4d $\mathcal{N}=2$ gauged supergravity STU model for a large range of values of the magnetic charges (subject to a supersymmetry constraint) the black saddles of \cite{Bobev:2020pjk} are the dominant saddles of the TTI on $S^1 \times \Sigma_{\mathfrak{g}}$ for any $\mathfrak{g}$ and their Euclidean on-shell actions are in perfect agreement with $ - \ln Z_{\text{CFT}}$ at large $N$. }
\begin{itemize}
  \item $\mathfrak{g} = 0$ ($\kappa = 1, \, \, \eta_\Sigma = 2$): The on-shell action\footnote{For the refinement of the $\mathfrak{g}=0$ case with rotation, 
  see also section 3.5.3 of \cite{Crisafio:2024fyc}.} at $\vartheta = 0$ is positive $\widehat{I}_{\text{BPS}} > 0$. Then $Z_{\text{grav}} \approx \eu^{- \widehat{I}_{\text{BPS}}}$ implies that the $\mathfrak{g} = 0$ saddle is subdominant in the path integral. Further, the entropy is negative $\mathcal{S}_{\text{BPS}} < 0$, signalling a non-regular geometry.
  \item $\mathfrak{g} = 1$ ($\kappa = 0, \, \, \eta_\Sigma = 1$): Both the on-shell action and the entropy vanish, signalling that the horizon disappears altogether. It will be very interesting to understand this gravitational solution and its field theory interpretation better.
\end{itemize}
When $\mathfrak{g} > 1$, the Euclidean black saddle described above is the dominant small-$G_4$ (equivalently, large $N$) saddle of the TTI and it admits a perfectly good $\beta \to \infty$ limit to a Lorentzian BPS configuration with positive, finite entropy. Concretely, for $\mathfrak{g} > 1$:
\begin{align}\label{eq:g>1_osa_and_entropy}
  \widehat{I}_{\text{BPS}} ={}& -\qty(\frac{\pi\ell_4^2}{2G_4} + \mathfrak{s}_Q \frac{\iu \ell_4^2}{2} \vartheta) (\mathfrak{g} - 1) \,, \quad \mathcal{S}_{\text{BPS}} = \frac{\pi \ell_4^2}{2 G_4} (\mathfrak{g} - 1) \,. 
\end{align}
%

%%%%%%%%%%%
\subsection{Holography and the TTI}\label{sec:TTI_4d}
%%%%%%%%%%%
The finite $\beta$ supersymmetric black saddle solutions of \cref{sec:4d_limits} are dual, via AdS$_4$/CFT$_3$, to a 3d $\mathcal{N}\geq 2$ SCFT on $S^1_\beta \times \Sigma_{\mathfrak{g}>1}$, with the abelian $\mathrm{U}(1)_R$ SCFT R-symmetry coupled to the background magnetic flux $\mathfrak{p}$ realizing the topological twist along $\Sigma$ \cite{Benini:2015noa,Benini:2015eyy}. We therefore expect that the supergravity partition function in the saddle-point approximation reproduces the \emph{topologically twisted index} (TTI)
\begin{align}\label{eq:tti_def}
  Z_{\text{TTI}}\bigl(\Sigma_{\mathfrak{g}>1}\bigr) ={}& \Tr_{\mathcal{H}_\Sigma} (-1)^F \eu^{-\beta H} \,, 
\end{align}
where $\mathcal{H}_\Sigma$ is the Hilbert space of the twisted theory on $\Sigma_{\mathfrak{g}>1}$. By supersymmetry, \eqref{eq:tti_def} is $\beta$-independent --- a feature precisely mirrored in the bulk by the $\beta$-independence of $\widehat{I}_{\text{BPS}}$ noted in \cref{sec:4d_limits}.
The holographic identification is
\begin{align}\label{eq:tti_grav_match}
  -\ln Z_{\text{TTI}}\bigl(\Sigma_{\mathfrak{g}>1}\bigr) ={}& \widehat{I}_{\text{BPS}} = -\qty(\frac{\pi \ell_4^2}{2 G_4} + \mathfrak{s}_Q \, \frac{\iu \ell_4^2}{2} \vartheta)(\mathfrak{g}-1) \,, 
\end{align}
which is the gravity-variable result \eqref{eq:g>1_osa_and_entropy}. In the remainder of this section we describe how to translate the result above from the gravity to the field theory side for two classes of SCFTs that arise from M-theory and for which one can employ a consistent truncation to 4d $\mathcal{N}=2$ minimal gauged supergravity.

Reducing 11d supergravity on $Y_4 \times SE_7$, with $SE_7$ a Sasaki-Einstein seven-manifold, yields 4d minimal gauged supergravity \cite{Gauntlett:2007ma} with $\vartheta=0$. The structural reason no $\vartheta$ angle is generated is topological and is discussed in some detail in~\cite{Genolini:2021qbi}. Employing M2-brane flux quantization one then finds the following holographic dictionary between the 4d supergravity parameters and the rank $N$ of the gauge group in the dual SCFT
\begin{align}\label{eq:dict_M2}
  \frac{\pi \ell_4^2}{2 G_4} ={}& N^{3/2}\sqrt{\frac{2 \pi^6}{27\,\mathrm{Vol}(SE_7)}} \, \equiv \, F_{S^3}^{(N)} \,, \quad \vartheta = 0 \,. 
\end{align}
Here $F_{S^3}^{(N)}$ is the large $N$ $S^3$ free energy of the dual 3d SCFT which for many examples of $SE_7$ can be computed at large $N$ via supersymmetric localization. For $SE_7 = S^7/\mathbb{Z}_k$ we find the $\rm{U}(N)\times \rm{U}(N)$ ABJM theory at level $k$, with $\mathrm{Vol}(S^7/\mathbb{Z}_k) = \pi^4/(3k)$, giving $F_{S^3}^{(N)} = \frac{\pi\sqrt{2k}}{3} N^{3/2}$. Substituting \eqref{eq:dict_M2} into \eqref{eq:tti_grav_match} yields the field-theory answer
\begin{align}\label{eq:tti_M2}
  -\ln Z_{\text{TTI}}^{\text{M2}}\bigl(\Sigma_{\mathfrak{g}>1}\bigr) = -\, F_{S^3}^{(N)} \, (\mathfrak{g}-1)\,,
\end{align}
which is purely real. The associated BPS black-hole entropy reads
\begin{align}\label{eq:entropy_M2}
  \mathcal{S}_{\text{BPS}}^{\text{M2}} ={}& F_{S^3}^{(N)} \, (\mathfrak{g}-1) \,. 
\end{align}
The expression \eqref{eq:tti_M2} matches the large $N$ TTI of 3d $\mathcal{N}\geq 2$ holographic SCFTs derived from supersymmetric localization \cite{Benini:2015eyy,Hosseini:2016tor,Azzurli:2017kxo}, providing the microscopic counting of the AdS$_4$ magnetic black-hole entropy.

The other route to 4d minimal gauged supergravity is the Pernici--Sezgin consistent truncation of 11d supergravity on $\Sigma_3 \times S^4$, with $\Sigma_3$ a closed hyperbolic three-manifold~\cite{Pernici:1984xx,Donos:2010ax}. The dual 3d $\mathcal{N}=2$ SCFT of class $\mathcal{R}$, denoted $T_N[\Sigma_3]$, is obtained from $N$ M5-branes wrapped on $\Sigma_3$ with the partial topological twist of~\cite{Gauntlett:2000ng,Dimofte:2011ju}. In contrast to the M2-brane case, the reduction of the 11d topological term $C\wedge G \wedge G$ on $\Sigma_3 \times S^4$ \emph{does} survive in four dimensions and produces a non-vanishing $\vartheta$. The leading-$N$ dictionary, derived in \cite{Choi:2020baw,Bobev:2020zov,Genolini:2021qbi}, is
\begin{align}\label{eq:dict_M5}
  \frac{\pi \ell_4^2}{2 G_4} ={}& \frac{2 N^3}{3\pi}\,\mathrm{vol}(\Sigma_3) \,, \quad \vartheta = \frac{2 N^3}{3\pi}\,\mathrm{cs}(\Sigma_3) \,, 
\end{align}
where $\mathrm{vol}(\Sigma_3)$ is the hyperbolic volume and $\mathrm{cs}(\Sigma_3)$ is the gravitational Chern-Simons invariant of $\Sigma_3$, well-defined modulo $2\pi$ for a fixed spin structure on $\Sigma_3$ \cite[App.\ A]{Genolini:2021qbi}. The two invariants combine into the \emph{complex hyperbolic volume},
\begin{align}\label{eq:complex_vol}
  \mathrm{Vol}_{\mathbb{C}}(\Sigma_3) ={}& \mathrm{vol}(\Sigma_3) + \iu \, \mathrm{cs}(\Sigma_3) \,, 
\end{align}
which is a topological invariant of the hyperbolic 3-manifold by Mostow rigidity. Inserting \eqref{eq:dict_M5} into \eqref{eq:tti_grav_match} produces the field-theory answer
\begin{align}\label{eq:tti_M5}
  -\ln Z_{\text{TTI}}^{\text{M5}}\bigl(\Sigma_{\mathfrak{g}>1};\, \Sigma_3\bigr) = -\, \frac{2 N^3}{3\pi}\,\bigl[\, \mathrm{vol}(\Sigma_3) + \mathfrak{s}_Q\, \iu\, \mathrm{cs}(\Sigma_3) \,\bigr]\,(\mathfrak{g}-1)
\end{align}
i.e.\ a multiple of $\mathrm{Vol}_{\mathbb{C}}(\Sigma_3)$ for $\mathfrak{s}_Q = +1$ and of its complex conjugate $\overline{\mathrm{Vol}_{\mathbb{C}}}(\Sigma_3)$ for $\mathfrak{s}_Q = -1$. The sign $\mathfrak{s}_Q$ selects between the two chiralities of the bulk Killing spinor at the bolt, which translates on the field-theory side to a choice between the geometric $SL(N,\mathbb{C})$ flat connection $A^{\text{geom}}$ and its conjugate $\overline{A^{\text{geom}}}$ in the dual complex Chern-Simons theory \cite{Genolini:2021qbi}. The corresponding BPS entropy is the real part,
\begin{align}\label{eq:entropy_M5}
  \mathcal{S}_{\text{BPS}}^{\text{M5}} ={}& \frac{2 N^3}{3\pi}\,\mathrm{vol}(\Sigma_3)\,(\mathfrak{g}-1) \,. 
\end{align}
Equation \eqref{eq:tti_M5} is the gravity prediction for the large $N$ TTI of $T_N[\Sigma_3]$ on $\Sigma_{\mathfrak{g}>1}$, and matches the 3d--3d correspondence \cite{Terashima:2011qi,Dimofte:2011ju} prediction that this partition function is computed by complex $SL(N,\mathbb{C})$ Chern-Simons theory on $\Sigma_3$ at large $N$, whose large $N$ limit is governed by the complex hyperbolic volume \cite{Gang:2014qla,Gang:2018wek}.

%%%%%%%
\section{The TTI on $T^2\times \Sigma_{\mathfrak{g}}$}\label{sec:TTI_5d}
%%%%%%%

The TTI is a supersymmetric partition function of a 4d $\mathcal{N}=1$ SCFT placed on $T^2\times \Sigma_{\mathfrak{g}}$ that was defined and studied in \cite{Benini:2015noa,Benini:2016hjo}, see also \cite{Closset:2017bse}. To avoid subtleties with singularities and boundary conditions we assume that $\Sigma_{\mathfrak{g}}$ is a smooth compact Riemann surface of genus $\mathfrak{g}$. Supersymmetry is preserved by turning on a specific value for the $\rm{U}(1)$ background gauge field that couples to the R-symmetry current such that it coincides with the spin-connection on $\Sigma_{\mathfrak{g}}$. This procedure implements Witten's partial topological twist on $\Sigma_{\mathfrak{g}}$ and effectively allows one to treat the 4d $\mathcal{N}=1$ SCFT as a 2d $\mathcal{N}=(0,2)$ theory on $T^2$. For this reason there are some close similarities between the 4d TTI and the (refined) elliptic genus of 2d supersymmetric QFTs. If the 4d SCFT has additional continuous global symmetries the TTI can be refined by turning on appropriate fugacities and/or background magnetic fluxes on $\Sigma_{\mathfrak{g}}$ for them. Here however we do not turn on such additional parameters and study the ``universal'' TTI, i.e. the one in which only the R-symmetry gauge field has a background magnetic flux and in which all flavor fugacities on $T^2$ are turned off.

The TTI is a complicated object which can be computed by supersymmetric localization and reduced to matrix integrals. These in turn can be solved by residues after an appropriate choice of contour and the problem of calculating the TTI can be organized into a set of algebraic ``Bethe Ansatz Equations''. It is subtle and involved to analyze the solutions of these equations for large $N$ QFTs. To make progress it is convenient to consider a Cardy-like limit in which the thermal circle on $T^2$ is very small. The leading term in this limit of the TTI appears to be controlled by the conformal anomalies of the 4d SCFT~\cite{Hosseini:2016cyf}. This is analogous to the situation for the SCI which exhibits an analogous simplification in the Cardy limit, see~\cite{GonzalezLezcano:2020yeb,Cassani:2021fyv,ArabiArdehali:2021nsx}. In the absence of a general treatment of the Cardy-like limit of the TTI analogous to the one in~\cite{Cassani:2021fyv} and valid for non-Lagrangian SCFTs we will proceed following~\cite{Hosseini:2016cyf} where it was shown that for a large class of 4d $\mathcal{N}=1$ Lagrangian QFTs the TTI takes the form
\begin{equation}\label{eq:logZQFT}
  \log Z_{T^2\times \Sigma_{\mathfrak{g}}}(\Delta_I,\mathfrak{n}_I) = \frac{\pi^2}{6 \hat{\beta}} c_{l}(\Delta_I,\mathfrak{n}_I)\,.
\end{equation}
Here $\mathfrak{n}_I$ are the background magnetic fluxes through $ \Sigma_{\mathfrak{g}}$ for the global symmetries of the SCFT which obey the topological twist constraint $\sum \mathfrak{n}_I=2-2\mathfrak{g}$. The chemical potentials $\Delta_{I}$ obey $\sum\Delta_I = 2\pi$. We work in conventions where the 2d $\mathcal{N}=(0,2)$ SCFT is supersymmetric in the right moving sector and thus the right-moving central charge $c_r(\Delta_I,\mathfrak{n}_I)$ is obtained by employing $c$-extremization~\cite{Benini:2012cz,Benini:2013cda}. The left-moving central charge appearing in~\eqref{eq:logZQFT} is determined by first extremizing $c_r$ and then using the gravitational anomaly relation $c_r-c_l=k$. For the universal twist of interest in this work both $c_r$ and $c_l$ are uniquely determined by the 4d $a$ and $c$ conformal anomalies. Here we have taken the modulus of the torus to be related to $\hat{\beta}$ as $\hat{\beta} = -2\pi \iu \tau$ and work in the Cardy limit $\hat{\beta} \to 0$. More generally, see~\cite{Hosseini:2020vgl}, the formula for the TTI reads
\begin{equation}\label{eq:logZQFTtau}
\log Z_{T^2\times \Sigma_{\mathfrak{g}}}(\Delta_I,\mathfrak{n}_I) = \frac{\iu\pi}{12\tau} c_{l}(\Delta_I,\mathfrak{n}_I)\,.
\end{equation}
We use conventions such that $2\pi\tau = \hat{\omega} + \iu \hat{\beta} $ where $\hat{\beta}$ is the conjugate variable to the 2d Hamiltonian and $\hat{\omega}$ is the conjugate to the angular momentum, i.e. if we decompactify the torus into a cylinder $\hat{\beta}$ is associated with translations along the non-compact direction and $\hat{\omega}$ with rotation around the circle\footnote{As emphasized in~\cite{Closset:2019ucb} the SCFT calculation of the TTI may suffer from scheme dependence. This will be mirrored in the supergravity analysis below where we find ambiguities in the evaluation of the gravitational on-shell action due to the presence of Chern-Simons terms in the action and potential finite counterterms in the holographic renormalization procedure. In our analysis we adopt the same scheme for the calculation of the TTI as in~\cite{Hosseini:2016cyf,Hosseini:2020vgl} and show how to reproduce the answer in~\eqref{eq:logZQFTtau} from an appropriate evaluation of the holographically dual on-shell action.}.

To define the unrefined limit of the TTI described above one only needs to specify the background values for the 4d metric and background gauge field that couples to the ${\rm U}(1)$ R-symmetry of the SCFT. For a holographic SCFT this implies that the result for the TTI in~\eqref{eq:logZQFTtau} should be equally universal and reproduced in the minimal 5d $\mathcal{N}=2$ gauged supergravity. Indeed, the supersymmetric Euclidean solutions we find in \cref{sec:susy_solutions} realize precisely this and their regularized on-shell action agrees with~\eqref{eq:logZQFTtau}.

For the universal twist\footnote{For the universal topological twist to be well-defined one must ensure that the R-charges of all gauge invariant operators are compatible with the quantization condition for the background R-symmetry gauge field on $\Sigma_{\mathfrak{g}}$. In our discussion we assume that the 4d $\mathcal{N}=1$ SCFT is such that this condition is obeyed.} the effective 2d central charges are related to the 4d conformal anomalies by, see~\cite{Benini:2015bwz,Bobev:2017uzs} 
\begin{equation}\label{eq:unitwistcrcl}
c_r = \frac{16}{3}(\mathfrak{g}-1)(5a-3c)\,, \qquad c_{l}=\frac{32}{3}(\mathfrak{g}-1)a\,.
\end{equation}
Due to superconformal Ward identities the 4d conformal anomalies can be expressed in terms of the cubic and linear 't Hooft anomalies of the 4d $\mathcal{N}=1$ R-symmetry as follows
\begin{equation}\label{eq:TrR2TrR}
k_{RRR} =\frac{16(5a-3c)}{9} \,, \qquad\qquad k_R = 16(a-c)\,.
\end{equation}

The results for the TTI summarized above clearly depend only on the conformal anomalies of the 4d $\mathcal{N}=1$ SCFT and are thus universal for all such theories that admit a weakly coupled bulk holographic description. It is nevertheless instructive to discuss examples of top-down holographic SCFTs to which these TTI results apply. We will consider here three classes of examples: D3-branes on conical singularities in IIB string theory, class $\mathcal{S}$ SCFTs arising from wrapped M5-branes on a Riemann surface, and 4d $\mathcal{N}=2$ SCFTs arising from D3-branes probing singularities in F-theory. 

We first consider the $Y^{p,q}$ SCFTs arising from D3-branes probing a cone over the respective Sasaki-Einstein manifolds~\cite{Gauntlett:2004yd}. The 't Hooft anomalies for these SCFTs are well-known, see \cite{Benvenuti:2004dy,Benini:2015bwz} and read
\begin{equation}
k_{RRR} = \frac{8p^2}{9q^4}(w^3+9pq^2-8p^3)N^2-2p\,, \qquad k_R = -2p\,,
\end{equation}
where $(p,q)$ are co-prime natural numbers and we have defined $w = \sqrt{4p^2-3q^2}$. In the special case of $\mathcal{N}=4$ SYM one has $k_R=0$ and $k_{RRR}=\frac{8}{9}d_{G}$ which in turn leads to the conformal anomalies $a=c=\frac{1}{4}d_{G}$. Here $d_{G}$ is the dimension of the gauge group, e.g. $d_G=N^{2}-1$ for $\rm{SU}(N)$. The partial topological twists of these 4d $\mathcal{N}=1$ SCFTs on $\Sigma_{\mathfrak{g}}$ have been studied in~\cite{Benini:2015bwz} (see also~\cite{Gauntlett:2006qw} for earlier work) and our results in this work can be viewed as a bulk derivation of the leading term in the TTI for this large class of models. Note that the consistent truncation of the 10d type IIB supergravity to the 5d minimal gauged supergravity for this class of D3-brane models has been established in~\cite{Gauntlett:2007ma}.

The 4d $\mathcal{N}=1$ theories of class $\mathcal{S}$ arise from M5-branes wrapped on a smooth compact Riemann surface $\Sigma_{\mathfrak{h}}$ of genus $\mathfrak{h}>1$. These models are characterized by a rational number $z$ that determines the partial topological twist of the 6d $\mathcal{N}=(2,0)$ SCFT on $\Sigma_{\mathfrak{h}}$ such that for $|z|=1$ one preserves 4d $\mathcal{N}=2$ supersymmetry, see~\cite{Maldacena:2000mw,Gaiotto:2009gz}, while for other values of $z$ one has a family of 4d $\mathcal{N}=1$ SCFTs, see~\cite{Benini:2009mz} and~\cite{Bah:2011vv,Bah:2012dg} for further details. The 4d 't Hooft anomalies for this class of SCFTs were computed in~\cite{Bah:2011vv,Bah:2012dg} and the result to leading order at large $N$ reads
\begin{equation}\label{eq:Mthanom}
\begin{split}
k_{RRR} &= \frac{2(\mathfrak{h}-1)}{27z^2}\left[9z^2-1+(3z^2+1)^{\frac{3}{2}}\right]N^3+\ldots\,,\\
k_R &= \frac{\mathfrak{h}-1}{3}\left[4-\sqrt{3z^2+1}\right]N+\ldots\,,
 \end{split}
\end{equation}
where $N$ is the number of M5-branes. As discussed in~\cite{Benini:2013cda} one can further compactify the 4d $\mathcal{N}=1$ class $\mathcal{S}$ theories with a partial topological twist on $\Sigma_{\mathfrak{g}}$ to obtain 2d $\mathcal{N}=(0,2)$ SCFTs. Our universal TTI results discussed above are also valid for this class of models and one can use the 't Hooft anomalies in~\eqref{eq:Mthanom} together with~\eqref{eq:TrR2TrR},~\eqref{eq:unitwistcrcl}, and~\eqref{eq:logZQFTtau} to obtain the leading term in the TTI. The consistent truncation from 11d supergravity to the 5d minimal gauged supergravity for these wrapped M5-brane models was established in~\cite{Gauntlett:2007sm,Szepietowski:2012tb,MatthewCheung:2019ehr,Faedo:2019cvr,Cassani:2020cod,Bobev:2022ocx}.

The final class of models we discuss are the 4d $\mathcal{N}=2$ SCFTs studied in~\cite{Aharony:1998xz,Aharony:2007dj} which arise from D3-branes at F-theory singularities. The $\text{U}\qty(1)_{\rm R}$ 't Hooft anomalies for these models are
\begin{equation}\label{eq:Fthanom}
\begin{split}
k_{RRR} &= \frac{8\Delta}{9} N^2+\frac{4(\Delta-1)}{9} N + \frac{2}{27}\,,\\
k_R &= 4(1-\Delta) N + \frac{2}{3}\,,
\end{split}
\end{equation}
where $N$ is the number of D3-branes and the rational number $\Delta$ specifies the global flavor symmetry of the SCFT and is related to the number of D7-branes, $n_7$, used in the construction. It is given by $\Delta = \frac{12}{12-n_7}$ with $n_7$ taking values in the set $\{2,3,4,6,8,9,10\}$. To the best of our knowledge, the consistent truncation to 5d minimal gauged supergravity for such F-theory models has not been rigorously established in the literature. We find it very likely to be true since the 5d supergravity fields of interest arise from the closed string modes not associated with the D7-brane singularity locus. It will be interesting to make this more rigorous by building on the analysis in \cite{Gauntlett:2007ma} and \cite{Cassani:2019vcl}.

%%%%%%%%%%%%%
\section{Thermal black string with $\mathbb{R} \times S^1 \times \Sigma_{\mathfrak{g}}$ boundary}
\label{sec:thermo_bs}
%%%%%%%%%%%%%

As explained above, the ``universal'' TTI on $T^2 \times \Sigma_{\mathfrak{g}}$ is supported by a background R-symmetry gauge field and no other gauge fields. Without further flavor symmetry refinements, the supergravity theory hosting the holographically dual solutions is 5d $\mathcal{N}=2$
minimal gauged supergravity. A supergravity theory admits non-supersymmetric thermal Lorentzian solutions --- the gravity duals
of complicated supersymmetry breaking, conformality breaking, finite temperature states in the dual QFT. In this section we construct a family of such asymptotically $\text{AdS}_5$ solutions, restricted to having an $\mathbb{R} \times S^1 \times \Sigma_{\mathfrak{g}}$ boundary in Lorentzian signature. This new family of solutions 
is important in its own right, irrespectively of the TTI.
First, it
provides the thermal completion of the twisted holographic RG flows across dimensions: at
$\mathfrak p=\mathfrak p_{\text{top. twist}}$ the backgrounds are dual to the 4d $\mathcal N=1$ SCFT
compactified on $\Sigma_{\mathfrak g}$ with the topological twist, at \emph{finite temperature and
angular velocity} on the residual $\mathbb{R}\times S^1$ --- i.e.\ to genuinely thermal states of
the effective 2d theory, of which the supersymmetric index saddles are a measure-zero corner.
As such the family is the natural arena for questions --- Hawking--Page-type transitions, the
approach to the 2d Cardy regime, hydrodynamics of the compactified theory --- that cannot even be
posed in the extremal or supersymmetric limits. Away from the twisted value of $\mathfrak p$ the
solutions describe conformality- and supersymmetry-breaking deformations in which the R-symmetry
background is misaligned with the spin connection, a setting about which very little is known
holographically. Finally, as we show below, the family exhibits structural surprises --- five
radially conserved charges and an on-shell action that localizes to the boundaries of the flow
without any supersymmetry --- which we expect to generalize to other cohomogeneity one settings.

%%%%%%%%%%
\subsection{Integrals of motion and non-supersymmetric localization}\label{sec:ioms}
%%%%%%%%%%
The bosonic action of 5d minimal gauged supergravity in Lorentzian signature reads
\begin{align}\label{eq:action_5d}
  S ={}& \frac{1}{16\pi G_5} \int_{\mathcal{M}} \qty[\qty(R_5 + \frac{12}{\ell_5^2}) \star_5 - \frac{3}{2} \mathcal{F} \wedge \star_5 \mathcal{F} - \mathcal{A} \wedge \mathcal{F} \wedge \mathcal{F}] \,, 
\end{align}
where $\ell_5$ is the scale of $\text{AdS}_5$. The equations of motion derived from \eqref{eq:action_5d} read
\begin{align}\label{eq:eom_5d}
  \begin{aligned}
  0 ={}& (R_5)_{\mu\nu} + \frac{4}{\ell_5^2} g_{\mu\nu} - \frac{3}{2} \mathcal{F}\indices{_\mu^\rho} \mathcal{F}_{\nu\rho} + \frac{1}{4} g_{\mu\nu} \mathcal{F}_{\rho\sigma} \mathcal{F}^{\rho\sigma} \,, \\
  0 ={}& \dd[]{\qty(\star_5 \mathcal{F})} + \mathcal{F} \wedge \mathcal{F} \,. 
  \end{aligned}
\end{align}
As anticipated above, we shall consider dyonic solutions that retain a $\Sigma_{\mathfrak{g} > 1}$ factor throughout the entire flow from the UV to the IR, with rotation in the direction transverse to the Riemann surface. Thus, topologically, we take $\mathcal{M} \cong \mathbb{R}_t \times [0,\infty] \times S^1_\varphi \times \Sigma_{\mathfrak{g} > 1}$, with $t$ denoting time, $[0,\infty]$ denoting the radial direction, and $\varphi$ denoting the compactified string direction.

Before proceeding to the ansatz that we will employ, we will first discuss how to treat correctly the Chern-Simons terms, as this will play a prominent role throughout the paper. In order to properly define Chern-Simons terms one needs to cover the space-time manifold $\mathcal{M}$ with patches $\{\mathcal{M}_a\}$, such that $\mathcal{M} =\cup_a \mathcal{M}_a$ and on each $\mathcal{M}_a$ 
 the gauge field $\mathcal{A}_a$ is a regular local one-form. Then the Chern-Simons integral is decomposed 
 as the sum of integrals on each patch, plus the contributions of lower-dimensional integrals arising from intersections of the overlapping patches. 
 We refer the reader to \cref{sec:patchwork} and 
 Appendix A.1 of \cite{Colombo:2025yqy} for details. In the present case, the space-time manifold is $\mathcal{M} \cong \mathbb{R}_t \times [0,\infty] \times S^1_\varphi \times \Sigma_\mathfrak{g}$ and to define the gauge field 
 we need \emph{two} patches $\mathcal{M}_a = \mathbb{R}_t \times [0,\infty] \times S^1_\varphi \times U_a$, where $U_a$ are two patches covering the Riemann surface\footnote{Globally, the compact Riemann surface is obtained as a quotient $\Sigma_{\mathfrak{g}} =\mathbb{H}^2/\Gamma$, where $\Gamma\subset $ PSL$(2,\mathbb{R})$. A closed orientable surface of arbitrary genus ${\mathfrak{g}}$ can be covered by two coordinate charts. A standard construction is to pick a point $p\in \Sigma_{\mathfrak{g}}$ and to define the two patches as $U_N\equiv \Sigma_{\mathfrak{g}} \setminus \{p\}$
and $U_S\equiv D$, where $D$ is a small disk around $p$. Then the overlap is an annulus $U_N \cap U_S\simeq S^1 \times (0,1)$. This extends the familiar ``North'' and ``South'' patches of the two-sphere.}, that we will refer to as the ``north'' and ``south'' patches $U_N$ and $U_S$. Notice that, differently from the assumptions made in \cite{Colombo:2025yqy}, our two patches $\qty{\mathcal{M}_N, \mathcal{M}_S}$ intersect the conformal boundary. With these preliminaries, the correct definition of the Chern-Simons integral is 
\begin{align}
  \int_{\mathcal{M}} \mathcal{A} \wedge \mathcal{F} \wedge \mathcal{F} \equiv{}& \int_{\mathcal{M}_N} \mathcal{A}_N \wedge \mathcal{F} \wedge \mathcal{F} + \int_{\mathcal{M}_S} \mathcal{A}_S \wedge \mathcal{F} \wedge \mathcal{F} + \int_{\Gamma_4} \mathcal{A}_N \wedge \mathcal{A}_S \wedge \mathcal{F} \,,
\end{align}
where $\Gamma_4 \equiv \mathbb{R}_t \times [0,\infty] \times S^1_\varphi \times S^1_\phi$ and $S^1_\phi$ is a circle inside the annulus $U_N \cap U_S$. We will also employ the short-hand notation 
 \begin{align}
\fint_{\mathcal{M}} \mathcal{A} \wedge \mathcal{F}\wedge \mathcal{F} \equiv {}& \int_{\mathcal{M}_N} \mathcal{A}_N \wedge \mathcal{F}\wedge \mathcal{F} + \int_{\mathcal{M}_S} \mathcal{A}_S \wedge \mathcal{F} \wedge \mathcal{F}\,, 
\end{align}
so that
\begin{align}\label{defbulkCS}
  \begin{aligned}
    \int_{\mathcal{M}} \mathcal{A} \wedge \mathcal{F} \wedge \mathcal{F} ={}& \fint_{\mathcal{M}} \mathcal{A} \wedge \mathcal{F}\wedge \mathcal{F} + \int_{\Gamma_4} \mathcal{A}_N \wedge \mathcal{A}_S \wedge \mathcal{F} \,. 
   \end{aligned}
\end{align}

Let us now move on to describing the local form of the ansatz that we shall employ. We shall assume that locally the $\Sigma_{\mathfrak{g} > 1}$ is equipped with the standard constant-curvature hyperbolic metric $ \dd[]{s}^2_\Sigma$, namely
\begin{align}
 \dd[]{s}^2_\Sigma ={}& \dd[]{\theta}^2 + \sinh^{2}{\theta} \dd[]{\phi}^2 \,, 
\end{align}
and that there exist a timelike Killing vector $\partial_t$ generating the (non-extremal) horizon and a spacelike Killing vector $\partial_\varphi$ associated to the angular momentum along the string, which we take to be a compact $S^1_\varphi$. Thus, locally the metric must be invariant under $\mathbb{R}_t\times \text{U}\qty(1)_\varphi \times \text{SL}\qty(2,\mathbb{R})$ and in particular it must have cohomogeneity one. Such configurations are described by the ansatz\footnote{
  Here $\mathcal{A}$ is viewed abstractly as the connection one-form of the $\text{U}\qty(1)$-bundle over $\mathcal{M}$ with representatives $\qty{\mathcal{A}_N, \mathcal{A}_S}$ in the respective patches. Similarly, $\omega_\Sigma$ is viewed abstractly as the connection one-form of the $\text{U}\qty(1)$-bundle over $\Sigma_{\mathfrak{g}}$ with representatives $\qty{\omega_N, \omega_S}$.}
\begin{align}\label{pre_ansatz_5d}
  \begin{aligned}
  \dd[]{s}^2 ={}& - g_{tt} (\dd[]{t} + b_r \dd[]{r})^2 + g_{rr} \dd[]{r}^2 + g_{\varphi\varphi} \qty(\dd[]{\varphi} + w \dd[]{t} + a_r \dd[] {r})^2 + g_{\Sigma} \dd[]{s}^2_\Sigma \,, \\
  \mathcal{A} ={}& a_t \dd[]{t} + a_\varphi \qty(\dd[]{\varphi} + w \dd[]{t}) - p \, \omega_\Sigma \,,
  \end{aligned}
\end{align}
where $\qty{g_{tt}, g_{rr}, g_{\varphi\varphi}, a_r, b_r,g_\Sigma, w, a_t, a_\varphi,p}$ are functions of the radial coordinate $r$ only. A possible fibration term proportional to $(\dd []{\varphi}+ c_r \omega_\Sigma)^2$
 should have constant $c_r$ in order to be globally defined, and therefore requiring $S^1\times \Sigma_{\mathfrak{g}}$ topology of the spatial section of the horizon fixes $c_r=0$. Similarly, demanding that the flux $\int_{\Sigma_r} \dd \mathcal{A}$ is correctly quantized for any fixed value of $r$ implies that $p= \text{const} \equiv \mathfrak{p}$. The metric ansatz may be simplified further as follows:
 defining the new coordinate $\tilde t= t + \int b_r \dd[]{r}$ removes the terms proportional to $b_r$. Then defining the new coordinate $\tilde \varphi= \varphi + \int (a_r - b_r w) \dd[]{r}$ and dropping tildes, we arrive at\footnote{We also ignore any terms of the type $\texttt{func}(r) \dd[]{r}$ in $\mathcal{A}$ since those are pure gauge along a non-compact direction.}
\begin{align}\label{eq:ansatz_5d}
  \begin{aligned}
  \dd[]{s}^2 ={}& - g_{tt} \dd[]{t}^2 + g_{rr} \dd[]{r}^2 + g_{\varphi\varphi} \qty(\dd[]{\varphi} + w \dd[]{t})^2 + g_{\Sigma} \dd[]{s}^2_\Sigma \,, \\
  \mathcal{A} ={}& a_t \dd[]{t} + a_\varphi \qty(\dd[]{\varphi} + w \dd[]{t}) - \mathfrak{p} \, \omega_\Sigma \,,
  \end{aligned}
\end{align}
where possible gauge shifts in $ \mathcal{A}$ are contained in the functions $\qty{a_t, a_\varphi}$. Note that setting $\qty{w, a_t, a_\varphi}=0$ we fall into the ansatz of \cite{Bernamonti:2007bu}, which, for the topological twist value of the magnetic charge, they manage to solve analytically. For the moment we will work in an arbitrary radial gauge; later we will find it convenient to fix the gauge as $g_{rr} = 1$. For future reference let us also record the following relations: 
\begin{align}\label{eq:Sigma_explicit}
  \begin{aligned}
  \dd[]{\omega_\Sigma} = {}& \Omega_\Sigma = \sinh^{}{\theta} \dd[]{\theta} \wedge \dd[]{\phi} \,, \\
  \int_\Sigma \Omega_\Sigma ={}& 4\pi(\mathfrak{g} - 1) \equiv 2\pi \eta_\Sigma \,,
  \end{aligned}
\end{align}
where we have defined $\eta_\Sigma$ as the $2\pi$-normalized volume of the unit Riemann surface.

For the ansatz \eqref{eq:ansatz_5d}, the equations of motion \eqref{eq:eom_5d} reduce to six second-order evolution equations for $\qty{g_{tt}, g_{\varphi\varphi}, g_\Sigma, w, a_t, a_\varphi}$:
\begin{align}\label{eq:evolution}
  \begin{aligned}
  0 ={}& \frac{g_{tt}''}{g_{tt}} - \frac{g_{tt}'}{g_{tt}} \qty(\frac{1}{2} \frac{g_{tt}'}{g_{tt}} + \frac{1}{2} \frac{g_{rr}'}{g_{rr}} + \frac{1}{2} \frac{g_{\varphi\varphi}'}{g_{\varphi\varphi}} + \frac{g_{\Sigma}'}{g_{\Sigma}}) - \frac{g_{\Sigma}'}{g_{\Sigma}} \qty(2 \frac{g_{\varphi\varphi}'}{g_{\varphi\varphi}} + \frac{g_{\Sigma}'}{g_{\Sigma}}) \\
       &+ 2 \frac{(a_\varphi')^2}{g_{\varphi\varphi}} - 2 \frac{g_{\varphi\varphi} (w')^2}{g_{tt}} - 5 \frac{(a_t' + a_\varphi w')^2}{g_{tt}} + g_{rr} \qty(\frac{16}{\ell_5^2} - \frac{4 \mathfrak{p}^2}{g_\Sigma^2} - \frac{4}{g_\Sigma}) \,, \\
  0 ={}& \frac{g_{\varphi\varphi}''}{g_{\varphi\varphi}} - \frac{g_{\varphi\varphi}'}{g_{\varphi\varphi}} \qty( - \frac{1}{2} \frac{g_{tt}'}{g_{tt}} + \frac{1}{2} \frac{g_{rr}'}{g_{rr}} + \frac{1}{2} \frac{g_{\varphi\varphi}'}{g_{\varphi\varphi}} - \frac{g_{\Sigma}'}{g_{\Sigma}}) \\
       &+ 2 \frac{(a_\varphi')^2}{g_{\varphi\varphi}} + \frac{g_{\varphi\varphi} (w')^2}{g_{tt}} + \frac{(a_t' + a_\varphi w')^2}{g_{tt}} + g_{rr} \qty( - \frac{8}{\ell_5^2} - \frac{\mathfrak{p}^2}{g_\Sigma^2}) \,, \\
   0 ={}& \frac{g_{\Sigma}''}{g_\Sigma} - \frac{g_\Sigma'}{g_\Sigma} \qty( - \frac{1}{2} \frac{g_{tt}'}{g_{tt}} + \frac{1}{2} \frac{g_{rr}'}{g_{rr}} - \frac{1}{2} \frac{g_{\varphi\varphi}'}{g_{\varphi\varphi}}) \\
        &- \frac{(a_\varphi')^2}{g_{\varphi\varphi}} + \frac{(a_t' + a_\varphi w')^2}{g_{tt}} + g_{rr} \qty( - \frac{8}{\ell_5^2} + \frac{2 \mathfrak{p}^2}{g_\Sigma^2} + \frac{2}{g_\Sigma}) \,, \\
   0 ={}& w'' - w' \qty(\frac{1}{2} \frac{g_{tt}'}{g_{tt}} + \frac{1}{2} \frac{g_{rr}'}{g_{rr}} - \frac{3}{2} \frac{g_{\varphi\varphi}'}{g_{\varphi\varphi}} - \frac{g_\Sigma'}{g_\Sigma}) + 3 \frac{a_\varphi' (a_t' + a_\varphi w')}{g_{\varphi\varphi}} \,, \\
   0 ={}& a_t'' - a_t' \qty(\frac{1}{2} \frac{g_{tt}'}{g_{tt}} + \frac{1}{2} \frac{g_{rr}'}{g_{rr}} - \frac{1}{2} \frac{g_{\varphi\varphi}'}{g_{\varphi\varphi}} - \frac{g_\Sigma'}{g_\Sigma}) \\
        &+ a_\varphi' \qty(w' - 3 \frac{a_\varphi (a_t' + a_\varphi w')}{g_{\varphi\varphi}} - \frac{2 \mathfrak{p} \sqrt{g_{tt} g_{rr} g_{\varphi\varphi}}}{g_{\varphi\varphi} g_\Sigma}) - a_\varphi \frac{w' g_{\varphi\varphi}'}{g_{\varphi\varphi}} \,, \\
   0 ={}& a_\varphi'' - a_\varphi' \qty( - \frac{1}{2} \frac{g_{tt}'}{g_{tt}} + \frac{1}{2} \frac{g_{rr}'}{g_{rr}} + \frac{1}{2} \frac{g_{\varphi\varphi}'}{g_{\varphi\varphi}} - \frac{g_\Sigma'}{g_\Sigma}) \\
        &+ (a_t' + a_\varphi w') \qty(\frac{g_{\varphi\varphi} w'}{g_{tt}} - \frac{2 \mathfrak{p} \sqrt{g_{tt} g_{rr} g_{\varphi\varphi}}}{g_{tt}g_\Sigma}) \,,
  \end{aligned}
\end{align}
together with a first-order constraint
\begin{align}\label{eq:constraint}
  0 ={}& \frac{1}{2} \frac{g_{tt}' g_{\varphi\varphi}'}{g_{tt} g_{\varphi\varphi}} + \frac{g_\Sigma'}{g_\Sigma} \qty(\frac{g_{tt}'}{g_{tt}} + \frac{g_{\varphi\varphi}'}{g_{\varphi\varphi}} + \frac{1}{2} \frac{g_\Sigma'}{g_\Sigma}) \notag \\
       &- \frac{3}{2} \frac{(a_\varphi')^2}{g_{\varphi\varphi}} + \frac{1}{2} \frac{g_{\varphi\varphi} (w')^2}{g_{tt}} + \frac{3}{2} \frac{(a_t' + a_\varphi w')^2}{g_{tt}} + g_{rr} \qty( - \frac{12}{\ell_5^2} + \frac{3 \mathfrak{p}^2}{2 g_\Sigma^2} + \frac{2}{g_{\Sigma}}) \,. 
\end{align}
Above and in what follows, primes denote $\partial_r$. This coupled system appears unwieldy but, remarkably, we will show that it admits a number of 
 \emph{radially conserved} integrals of motion (IOMs). The first three IOMs are relatively simple to construct, although they are somewhat non-standard and their specific form is related to the Chern-Simons 
 coupling in the theory.

One can define the following three forms
\begin{align}
  \begin{aligned}
    \mathcal{Q}_{\text{Page}} \equiv {}& \star_5 \mathcal{F} + \mathcal{A} \wedge \mathcal{F} \,, \\
    \mathcal{J}_{\text{Wald}} \equiv {}& \star_5 \dd[]{\qty((\partial_\varphi)^{\flat}) } + 3 \qty(\iota_{\partial_\varphi} \mathcal{A}) \qty(\star_5 \mathcal{F} + \tfrac{2}{3}\mathcal{A} \wedge \mathcal{F}) \,, 
  \end{aligned}
\end{align}
which are closed\footnote{$\mathcal{J}_{\text{Wald}}$ is closed only when pulled back to $S^1\times \Sigma_{\mathfrak{g}}$. See \cref{app:patchwise-angular-momentum}.} by virtue of the equations of motion \cite{Suryanarayana:2007rk,Hanaki:2007mb} and should lead to radially conserved integrals of motion. Indeed, employing the procedure to perform integrals on manifolds involving $\Sigma_{\mathfrak{g}}$ explained in \cref{sec:patchwork}, we have that
\begin{align}\label{eq:qs_patchwise}
  \begin{aligned}
  \frac{1}{4\pi^2 \eta_\Sigma}\int_{S^1_\varphi \times \Sigma} \mathcal{Q}_{\text{Page}} ={}& \frac{1}{4\pi^2 \eta_\Sigma} \fint_{S^1_\varphi \times \Sigma} \qty(\star_5 \mathcal{F} + 2 \mathcal{A} \wedge \mathcal{F}) \equiv q \,, \\
  \frac{1}{L(\mathbb{R}_t) 2 \pi \eta_\Sigma}\int_{\mathbb{R}_t \times \Sigma} \mathcal{Q}_{\text{Page}} ={}& \frac{1}{L(\mathbb{R}_t) 2 \pi \eta_\Sigma} \fint_{\mathbb{R}_t \times \Sigma} \qty(\star_5 \mathcal{F} + 2 \mathcal{A} \wedge \mathcal{F}) \equiv \widetilde{q} \,,
  \end{aligned}
\end{align}
where we have used the identity
\begin{align}\label{pippo}
  \int \mathcal{A} \wedge \mathcal{F} \equiv{}& \fint \mathcal{A} \wedge \mathcal{F} + \int_{\Gamma_2} \mathcal{A}_N \wedge \mathcal{A}_S = 2 \fint \mathcal{A} \wedge \mathcal{F} \,, 
\end{align} 
with $\Gamma_2 \cong \mathbb{R}_t \times S^1_\phi$ or $ \Gamma_2 \cong S^1_\varphi \times S^1_\phi$ being the 2d interface integral. The fact that the charges $q$ and $\tilde q$ are radially conserved can be seen a priori in two slightly different ways: either as conservation of the standard Page charge, where the integral on the 3d hypersurface has been supplemented by the 2d interface, or as conservation of a non-closed modified charge,
\begin{align}
\begin{aligned}
 \mathcal{\hat Q}_{\text{Page}}\equiv {}& \star_5 \mathcal{F} + 2 \mathcal{A} \wedge \mathcal{F} \,, 
  \end{aligned}
\end{align}
 where the integral on the 3d hypersurface has not been supplemented by the 2d interface. The equivalence of these two definitions follows from the identity \eqref{pippo} and the calculations proving radial conservation of these charges, using both points of view, have been relegated to \cref{sec:patchwork}. Of course, one can simply check using the (local) equations of motion that the explicit expressions for $q$ and $\widetilde{q}$ are radially conserved in the sense that $q' = 0 = \widetilde{q}'$. Note that $\widetilde{q}$ should really be thought of as charge density distributed along the time line $\mathbb{R}_t$. Similarly, $q$ can be thought of as a charge density integrated over the compact string direction $S^1_\varphi$.
 In \cref{sec:patchwork} we also discuss the following non-closed 
 modification of $\mathcal{J}_{\text{Wald}}$
 \begin{align}
  \begin{aligned}
  \mathcal{\hat J}_{\text{Wald}} \equiv {}& \star_5 \dd[]{\qty((\partial_\varphi)^{\flat}) } + 3 \qty(\iota_{\partial_\varphi} \mathcal{A}) \qty(\star_5 \mathcal{F} + \mathcal{A} \wedge \mathcal{F}) \,,
  \end{aligned}
\end{align}
that leads to another radially conserved quantity
\begin{align}
  \begin{aligned}
\frac{1}{4\pi^2 \eta_\Sigma}\fint_{S^1_\varphi \times \Sigma} \mathcal{\hat J}_{\text{Wald}} \equiv j \,.
  \end{aligned}
\end{align}
Note that, contrary to $ \mathcal{\hat Q}_{\text{Page}}$ that gives a conserved quantity when $\fint$ integrated
over either $\mathbb{R}_t \times \Sigma$ or $S^1_\varphi \times \Sigma$, the $\fint$ integral of $\mathcal{\hat J}_{\text{Wald}}$ over $\mathbb{R}_t \times \Sigma$ \textit{does not} lead to a conserved quantity. The explicit expressions for these IOMs, evaluated on our ansatz \eqref{eq:ansatz_5d} are
\begin{align}\label{eq:iom_qj}
  \begin{aligned}
  q ={}& - 2 \mathfrak{p} a_\varphi + \frac{g_\Sigma \sqrt{g_{tt} g_{rr} g_{\varphi\varphi}}}{g_{rr}} \frac{\qty(a_t' + a_\varphi w')}{g_{tt}} \,, \\
  \widetilde{q} ={}& - 2 \mathfrak{p} \qty(a_t + a_\varphi w) + \frac{g_\Sigma \sqrt{g_{tt} g_{rr} g_{\varphi\varphi}}}{g_{rr}} \qty( \frac{a_\varphi'}{g_{\varphi\varphi}} + \frac{w \qty(a_t' + a_\varphi w')}{g_{tt}}) \,, \\
  j ={}& -3 \mathfrak{p} a_\varphi^2 + \frac{g_\Sigma \sqrt{g_{tt} g_{rr} g_{\varphi\varphi}}}{g_{rr}} \qty(3 \frac{a_\varphi(a_t' + a_\varphi w')}{g_{tt}} + \frac{g_{\varphi\varphi} w'}{g_{tt}}) \,,
  \end{aligned}
\end{align}
with $q' = \widetilde{q}' = j' = 0$, as can be directly verified by using the evolution equations \eqref{eq:evolution}.

Somewhat surprisingly, it turns out that two additional IOMs exist. To see how they arise, consider the $t$--$\varphi$ slices of the metric and the gauge field
\begin{align}
  \mathbb{G} ={}& \mqty(
  - g_{tt} + g_{\varphi\varphi} w^2 & g_{\varphi\varphi} w \\
  g_{\varphi\varphi} w & g_{\varphi\varphi}
  ) \,, \quad \mathbb{A} = \mqty(
 \mathcal{A}_t \\
  \mathcal{A}_\varphi
  ) \,,
\end{align}
where $\mathcal{A}_t = a_t + a_\varphi w$ and $\mathcal{A}_\varphi = a_\varphi$, together with the current
\begin{align}\label{eq:current_mu}
  \mu(T) ={}& - \frac{1}{2} \frac{g_\Sigma \sqrt{g_{tt} g_{rr} g_{\varphi\varphi}}}{g_{rr}} \qty[\Trace(T \, \mathbb{G}^{-1} \, \partial_r \mathbb{G}) + 3 \, \mathbb{A} ^{\mathsf{T}} \, T \, \mathbb{G}^{-1} \, \partial_r \mathbb{A} ] + \frac{3 \mathfrak{p}}{2} \, \mathbb{A} ^{\mathsf{T}} \, \varepsilon \, T \, \mathbb{A} \,, 
\end{align}
where $T \in \text{GL}\qty(2, \mathbb{R})$ and $\varepsilon$ is the antisymmetric symbol with $\varepsilon_{12} = 1$. Computing $\mu(T)'$ and substituting the evolution equations~\eqref{eq:evolution}, the result is proportional to the constraint~\eqref{eq:constraint} when $T$ takes the form
\begin{align}
  T ={}& c_t T_t + c_s T_s \,, \quad T_t = \mqty( - 1 & 0 \\ 0 & 1) \,, \quad T_s = \mqty( 0 & 1 \\ 1 & 0) \,,
\end{align}
where the subscripts $t$ and $s$ stand for ``traceless'' and ``symmetric'', respectively. This yields two further IOMs, $\mu_t = \mu(T_t)$ and $\mu_s = \mu(T_s)$, explicitly given by
\begin{align}\label{eq:iom_mu}
  \begin{aligned}
  \mu_t ={}& 3 \mathfrak{p} a_\varphi (a_t + a_\varphi w) \\
           &- \frac{g_\Sigma \sqrt{g_{tt} g_{rr} g_{\varphi\varphi}}}{g_{rr}} \Bigg[\frac{3}{2} \frac{a_\varphi a_\varphi'}{g_{\varphi\varphi}} - \frac{1}{2} \qty(\frac{g_{tt}'}{g_{tt}} - \frac{g_{\varphi\varphi}'}{g_{\varphi\varphi}}) + \frac{g_{\varphi\varphi} w w'}{g_{tt}} \\
           & \qquad \qquad \qquad \qquad + \frac{3}{2} \frac{(a_t + 2 a_\varphi w) (a_t' + a_\varphi w')}{g_{tt}} \Bigg] \,, \\
    \mu_s ={}& \frac{3\mathfrak{p}}{2} \qty(a_t + a_\varphi(w - 1)) \qty(a_t + a_\varphi(w + 1)) \\
             &- \frac{g_\Sigma \sqrt{g_{tt} g_{rr} g_{\varphi\varphi}}}{g_{rr}} \Bigg[\frac{3}{2} \frac{(a_t + a_\varphi w) a_\varphi'}{g_{\varphi\varphi}} \\
             & \qquad \qquad \qquad \qquad - \frac{w}{2} \qty(\frac{g_{tt}'}{g_{tt}} - \frac{g_{\varphi\varphi}'}{g_{\varphi\varphi}}) + \frac{1}{2} \frac{\qty(g_{tt} + g_{\varphi\varphi}(w^2 - 1)) w'}{g_{tt}} \\
             & \qquad \qquad \qquad \qquad + \frac{3}{2} \frac{\qty(a_t w + a_\varphi (w^2 - 1)) (a_t' + a_\varphi w')}{g_{tt}} \Bigg] \,.
  \end{aligned}
\end{align}
Indeed, one can directly verify that $\mu_t' = \mu_s' = 0$ by using the evolution equations \eqref{eq:evolution} and the constraint \eqref{eq:constraint}. The IOM $\mu_t$ is a generalization of (2.12) of \cite{Bernamonti:2007bu} for a rotating ansatz and non-zero electric charge. We are not aware of a paper that has discussed $\mu_s$. The emergence of five IOMs, which are independent in the sense that the expressions \cref{eq:iom_qj,eq:iom_mu}, schematically given by $\texttt{iom}(g, g', a, a')$, 
are clearly not linear combinations of each other, might seem surprising given the expected number of physical charges: mass, angular momentum, and electric charge. As we shall see below, on a physical solution there are non-trivial relations among the \textit{evaluated expressions of these functionals} that render the number of independent constants of integration equal to the number of physical charges. 

Remarkably, the entire action, put on-shell on our non-supersymmetric ansatz \eqref{eq:ansatz_5d}, with the Chern-Simons integral treated carefully, ``localizes'' to boundary contributions. First note that in \cref{sec:patchwork} we show that the net effect of incorporating the interface integral over $\Gamma_4 \cong \mathbb{R}_t \times [0,\infty] \times S^1_\varphi \times S^1_\phi$ is
\begin{align}
  \int_{\mathcal{M}} \mathcal{A} \wedge \mathcal{F} \wedge \mathcal{F} ={}& \frac{3}{2} \fint_{\mathcal{M}} \mathcal{A} \wedge \mathcal{F} \wedge \mathcal{F} \,,
\end{align}
where as a reminder
\begin{align}
\fint_{\mathcal{M}} \mathcal{A} \wedge \mathcal{F} \wedge \mathcal{F} ={}& \int_{\mathcal{M}_N} \mathcal{A}_N \wedge \mathcal{F} \wedge \mathcal{F} + \int_{\mathcal{M}_S} \mathcal{A}_S \wedge \mathcal{F} \wedge \mathcal{F} \, .
\end{align}
Then, on our specific ansatz the $\omega_{N,S}$ terms in $\mathcal{A}_{N,S}$ do not appear ``naked'' in the above integrals; for example
\begin{align}
  \mathcal{A}_N \wedge \mathcal{F} \wedge \mathcal{F} ={}& \qty(\mathcal{B} - \mathfrak{p} \, \omega_N) \wedge \qty(\dd[]{\mathcal{B}} - \mathfrak{p} \, \Omega_\Sigma)^2 = - 2 \mathfrak{p} \, \mathcal{B} \wedge \dd[]{\mathcal{B}} \wedge \Omega_\Sigma \,,
\end{align}
with $\mathcal{B} = \mathcal{A}_t \dd[]{t} + \mathcal{A}_\varphi \dd[]{\varphi}$. Thus, the $\fint$ integral collapses to the na\"{\i}ve integral on $\mathcal{M}$ and the only modification of the original Chern-Simons integral in the bulk action is the effective replacement of its coefficient $ (- 1) \rightarrow (- 3/2)$. That is the exact same logic that replaced the coefficient $( + 1) \rightarrow ( + 2)$ of the $\mathcal{A} \wedge \mathcal{F}$ integral to show the conservation of the Page charge. The end result of the above discussion can be neatly summarized as
\begin{align}
  S ={}& \frac{1}{16\pi G_5} \fint_{\mathcal{M}} \qty[\qty(R_5 + \frac{12}{\ell_5^2}) \star_5 - \frac{3}{2} \mathcal{F} \wedge \star_5 \mathcal{F} - \frac{3}{2} \mathcal{A} \wedge \mathcal{F} \wedge \mathcal{F}] \,.
\end{align}
The integrand above should be viewed as a local expression that one can integrate na\"{\i}vely to obtain the correct answer for the on-shell action\footnote{Note that 
we are not suggesting that the local Lagrangian gets modified. Equations of motion are still derived from the standard action of 5d gauged minimal supergravity \eqref{eq:action_5d}. The $\fint$ expression is merely the practical recipe to evaluate the on-shell action in the presence of an interface.}. The integrand reads
\begin{align}
  \texttt{integrand} ={}& \sqrt{-g} \qty[R_5 + \frac{12}{\ell_5^2} - \frac{3}{2} \frac{1}{2!} \mathcal{F}_{\mu\nu} \mathcal{F}^{\mu\nu} - \frac{3}{2} \frac{1}{(2!)^2} \varepsilon^{\mu\nu\rho\sigma\lambda} \mathcal{A}_\mu \mathcal{F}_{\nu\rho} \mathcal{F}_{\sigma\lambda}] \notag \\
  ={}& \Bigg[- \sqrt{g_{tt} g_{rr} g_{\varphi\varphi}} \qty(\frac{\mathfrak{p}^2}{g_\Sigma} + \frac{8 g_\Sigma}{\ell_5^2} + \frac{g_\Sigma}{g_{rr}} \qty(\frac{(a_\varphi')^2}{g_{\varphi\varphi}} - \frac{(a_t' + a_\varphi w')^2}{g_{tt}}) ) \notag \\
     & \quad + 3 \mathfrak{p} \qty(\mathcal{A}_t \mathcal{A}_\varphi' - \mathcal{A}_\varphi \mathcal{A}_t') \Bigg] \sinh^{}{\theta} \,, 
\end{align}
where to reach the second line we have used both the evolution equations \eqref{eq:evolution} and the constraint \eqref{eq:constraint}. Now, the non-trivial observation is that the expression can be written as
\begin{align}
  \texttt{integrand} ={}& \mu_{\text{id}}' \sinh^{}{\theta} \,, \quad \mu_{\text{id}} \equiv \mu(T = \text{diag}(1,1)) \,. 
\end{align}
That is: upon feeding the identity matrix to the current \eqref{eq:current_mu} we obtain an expression whose radial derivative is precisely the on-shell integrand. The explicit expression for $\mu_{\text{id}}$ is
\begin{align}
  \mu_{\text{id}} ={}& \frac{g_\Sigma \sqrt{g_{tt} g_{rr} g_{\varphi\varphi}}}{g_{rr}} \qty[ - \frac{3}{2} \frac{a_\varphi a_\varphi'}{g_{\varphi\varphi}} - \frac{1}{2} \qty(\frac{g_{tt}'}{g_{tt}} + \frac{g_{\varphi\varphi}'}{g_{\varphi\varphi}}) + \frac{3}{2} \frac{a_t \qty(a_t' + a_\varphi w')}{g_{tt}}] \,.
\end{align}
Then the action
\begin{align}
  S ={}& \frac{1}{16 \pi G_5} \int_{\mathcal{M}} \mu_{\text{id}}' \dd[]{t} \wedge \dd[]{r} \wedge \dd[]{\varphi} \wedge \Omega_\Sigma = \frac{L(\mathbb{R}_t) 4\pi^2 \eta_\Sigma}{16 \pi G_5} \qty[\mu_{\text{id}}(r_{\text{UV}}) - \mu_{\text{id}}(r_{\text{IR}})]
\end{align}
``localizes'' to boundary contributions: the asymptotic boundary $\mathbb{R}_t \times S^1_\varphi \times \Sigma_{\mathfrak{g}}$ in the UV and the \textit{bolt} $S^1_\varphi \times \Sigma_{\mathfrak{g}}$ in the IR. In Euclidean signature, upon $\mathbb{R}_t \rightarrow S^1_{t_E}$, both these surfaces are compact. Notice that on our concrete ansatz this localization requires no supersymmetry.

To close this section, we give some remarks on the existence of the five IOMs $\qty{q, \widetilde{q},j, \mu_t, \mu_s}$. It would be interesting to rederive them from the effective 1d Lagrangian approach intrinsic to cohomogeneity one metrics, see e.g. \cite{Ntokos:2021duk}. In fact, the structure of the current \eqref{eq:current_mu} strongly suggests such an origin.
Reducing the theory on the torus generated by $\qty{\partial_t,\partial_\varphi}$ yields a
one-dimensional effective mechanics for the matrix $\mathbb G$ of torus moduli, the doublet
$\mathbb A$ of Wilson lines and the breathing mode $g_\Sigma$, in which the global
$\mathrm{GL}(2,\mathbb R)$ acting on the torus (linearly on $\mathbb A$, by conjugation on
$\mathbb G$) is broken only by the periodic identifications of $(t_E,\varphi)$ and by the
Chern-Simons coupling. The five integrals of motion then organize into representations of this
$\mathrm{GL}(2,\mathbb R)$: the doublet $(q,\widetilde q)$ from shifts of the Wilson lines, and the
triplet $(\mu_t,\mu_s,j)$ --- the currents $\mu(T)$ with traceless $T$ --- from the
$\mathrm{SL}(2,\mathbb R)$ part, with the transformation laws \eqref{eq:frame_shift} and
\eqref{eq:gauge_shift} realizing the corresponding coadjoint action. From this perspective the
non-conservation of the trace part, $\mu_{\rm id}'\neq0$, is expected --- the trace of
$\mathrm{GL}(2,\mathbb R)$ acts as a scaling that is anomalous in the reduced mechanics --- and its
failure to be conserved is precisely what makes it the potential for the on-shell action.

\subsection{Boundary conditions}\label{sec:BC}
The central question is whether there exists a non-supersymmetric Lorentzian flow interpolating between
\begin{equation}
\begin{gathered}
\text{AlAdS}_5 \text{ with } \mathbb{R}_t \times S^1_\varphi \times \Sigma \text{ boundary} \\[4pt]
\rotatebox[origin=c]{-90}{\scalebox{1.8}[1]{$\rightsquigarrow$}} \\[4pt]
\mathbb{R}^{1,1}_{t,r} \times S^1_\varphi \times \Sigma \text{ near-horizon region}
\end{gathered}
\end{equation}
We dub such a flow a \textit{thermal black string} to distinguish it from
\begin{itemize}
  \item ``empty'' supersymmetric and extremal black strings which interpolate between $\mathbb{R}^{1,1} \times \Sigma$ boundary and $\text{AdS}_3 \times \Sigma$ and have non-compact string direction \cite{Chamseddine:1999xk,Klemm:2000nj,Maldacena:2000mw,Benini:2012cz,Benini:2013cda}
  \item supersymmetric and/or extremal generalizations of the above which compactify the string direction \cite{Bernamonti:2007bu,Hong:2021bzg}
\end{itemize} 
The boundary conditions of the thermal black string are straightforward to write down. Asymptotically (for $z \to 0$), the field configuration must take the form\footnote{Note that it is $\Omega_H$ (not $\Omega_\infty$) that appears in the asymptotic gauge field: this aligns the Wilson line term $\qty(\dd[]{\varphi} - \Omega_H \dd[]{t})$ with the horizon generator $V$ such that (ii) and (iii) below hold.} 
\begin{align}\label{eq:fg_z}
  \begin{aligned}
    \dd[]{s}^2 ={}& \frac{\ell_5^2}{z^2} \dd[]{z}^2 + \frac{\ell_5^2}{z^2} \qty[ - c_t^2 \frac{\dd[]{t}^2}{\ell_5^2} + c_\varphi^2 (\dd[]{\varphi} - \Omega_\infty \dd[]{t})^2 + c_\Sigma^2 \dd[]{s}^2_\Sigma] \,, \\
    \mathcal{A} ={}& - \Phi \dd[]{t} - \Psi_\infty \qty(\dd[]{\varphi} - \Omega_H \dd[]{t}) - \mathfrak{p} \, \omega_\Sigma \,,
  \end{aligned}
\end{align}
where $\qty{c_t, c_\varphi, c_\Sigma}$ are dimensionless constants. Near the horizon (for $R \to 0$), it must take the form
\begin{align}\label{eq:cap_R}
  \begin{aligned}
  \dd[]{s}^2 ={}& \dd[]{R}^2 - R^2 \qty(\frac{2\pi}{\beta})^2 \dd[]{t}^2 + L_\varphi^{2} \qty(\dd[]{\varphi} - \Omega_H \dd[]{t})^2 + L_\Sigma^2 \dd[]{s}^2_\Sigma \,, \\
  \mathcal{A} ={}& - \Psi_H \qty(\dd[]{\varphi} - \Omega_H \dd[]{t}) - \mathfrak{p} \, \omega_\Sigma \,,
  \end{aligned}
\end{align}
where $\qty{L_\varphi, L_\Sigma}$ are dimensionful constants
and an irrelevant positive constant has been set to one as it can be absorbed in the definition of $R$. The boundary conditions are chosen such that:
\begin{enumerate}[(i)]
  \item We allow a rotating frame at infinity $\Omega_\infty \neq 0$.
  \item The gauge field is regular on the horizon by virtue of
\begin{align}
  \eval{\iota_V \mathcal{A}}_{H} ={}& 0 \,, \quad V = \partial_t + \Omega_H \partial_\varphi \,,
\end{align}
where $V$ is the timelike Killing vector whose norm vanishes at the horizon.
  \item The constant $\Phi$ coincides with the physical electrostatic potential
\begin{align}
  \Phi_{\text{phys}} ={}& \eval{\iota_V \mathcal{A}}_{H} - \eval{\iota_V \mathcal{A}}_{\infty} = \Phi \,. 
\end{align}
  \item The constant $\beta$ coincides with the physical inverse temperature, where regularity of the Euclidean ($t = -\iu t_E$) near-horizon metric dictates the usual twisted periodicities
\begin{align}\label{eq:twisted_periodicity}
  \qty(t_E, \varphi) \sim \qty(t_E + \beta, \varphi - \iu \Omega_H \beta) \sim \qty(t_E, \varphi + 2\pi) \,. 
\end{align}
\end{enumerate}
The remaining constants $\qty{\Psi_\infty, \Psi_H}$ are, at this point, arbitrary. Within the ansatz \eqref{eq:ansatz_5d} we adopt the convenient radial gauge choice
\begin{align}
  g_{rr} ={}& 1 \,, 
\end{align}
such that both regions are described using the same radial coordinate $r$:
\begin{align}\label{eq:BC_UV}
  \begin{aligned}
    \dd[]{s}^2 ={}& \dd[]{r}^2 + \ell_5^2 \, \eu^{\frac{2r}{\ell_5}} \qty[ - c_t^2 \frac{\dd[]{t}^2}{\ell_5^2} + c_\varphi^2 (\dd[]{\varphi} - \Omega_\infty \dd[]{t})^2 + c_\Sigma^2 \dd[]{s}^2_\Sigma] \,, \\
    \mathcal{A} ={}& - \Phi \dd[]{t} - \Psi_\infty \qty(\dd[]{\varphi} - \Omega_H \dd[]{t}) - \mathfrak{p} \, \omega_\Sigma \,,
  \end{aligned}
\end{align}
and
\begin{align}\label{eq:BC_IR}
  \begin{aligned}
  \dd[]{s}^2 ={}& \qty(\dd[]{r}^2 - r^2 \qty(\frac{2\pi}{\beta})^2 \dd[]{t}^2) + L_\varphi^{2} \qty(\dd[]{\varphi} - \Omega_H \dd[]{t})^2 + L_\Sigma^2 \dd[]{s}^2_\Sigma \,, \\
  \mathcal{A} ={}& - \Psi_H \qty(\dd[]{\varphi} - \Omega_H \dd[]{t}) - \mathfrak{p} \, \omega_\Sigma \,.
  \end{aligned}
\end{align}
Note that $r$ is a bona fide Fefferman-Graham radial coordinate, related to $z$ in \eqref{eq:fg_z} by $z = \eu^{- r/\ell_5}$, with the conformal boundary located at $r = \infty$. The coordinate $R$ in \eqref{eq:cap_R}, defined up to trivial rescalings $R \rightarrow c_R R$, is now simply related to the unified radial coordinate as $R = r$. 

To provide further intuition on the UV boundary conditions, note that in Euclidean signature,
through the coordinate transformation
\begin{align}
  \curly{z} ={}& \varphi + \iu \qty(\frac{c_t}{c_\varphi \ell_5} + \Omega_\infty) t_E \,, \quad \bar{\curly{z}} = \varphi - \iu \qty(\frac{c_t}{c_\varphi \ell_5} - \Omega_\infty) t_E \,,
\end{align}
the conformal boundary takes the canonical form
\begin{align}\label{eq:boundary_metric_canonical}
  \begin{aligned}
  \dd[]{s}^2_\partial ={}& c_\varphi^2 \dd[]{\curly{z}} \dd[]{\bar{\curly{z}}} + c_\Sigma^2 \dd[]{s}^2_\Sigma \,.
  \end{aligned}
\end{align}
In the $\qty{\curly{z}, \bar{\curly{z}}}$ coordinates the twisted periodicity \eqref{eq:twisted_periodicity} translates to
\begin{align}
  (\curly{z}, \bar{\curly{z}}) \sim \qty(\curly{z} + 2\pi \tau, \bar{\curly{z}} + 2\pi \bar{\tau}) \sim (\curly{z} + 2\pi, \bar{\curly{z}} + 2\pi) \,,
\end{align}
with the modular parameter coming out as\footnote{The modular parameters $\tau$ and $\bar{\tau}$ are exactly the same as the ones used in the QFT, see \cref{sec:TTI_5d}. The QFT angular fugacity $\hat{\omega}$ and temperature $\hat{\beta}$ are related to the supergravity parameters as dictated by \eqref{eq:modular_parameter}.}
\begin{align}\label{eq:modular_parameter}
  \tau ={}& \frac{\beta}{2\pi\iu} \qty(\Omega - \frac{c_t}{c_\varphi \ell_5}) \,, \quad \bar{\tau} = \frac{\beta}{2\pi\iu} \qty(\Omega + \frac{c_t}{c_\varphi \ell_5}) \,,
\end{align}
where we have defined
\begin{align}
  \Omega \equiv {}& \Omega_H - \Omega_\infty \,. 
\end{align}
The boundary value of the gauge field in the $\qty{\curly{z}, \bar{\curly{z}}}$ coordinates reads
\begin{align}
  \mathcal{A}_{\partial} ={}& \frac{\frac{\beta \Phi}{2\pi\iu} - \tau \Psi_\infty}{c_\varphi (\tau - \bar{\tau})}\dd[]{\curly{z}} - \frac{\frac{\beta \Phi}{2\pi\iu} - \bar{\tau} \Psi_\infty}{c_\varphi (\tau - \bar{\tau})}\dd[]{\bar{\curly{z}}} - \mathfrak{p} \, \omega_\Sigma \,. 
\end{align}
Thus, the boundary data are specified by:
\begin{itemize}
  \item $c_\varphi$ --- breathing mode of the boundary $T^2$
  \item $c_\Sigma$ --- breathing mode of the boundary $\Sigma_{\mathfrak{g}}$
  \item $\qty{\tau, \bar{\tau}}$ or alternatively $\qty{\beta, \Omega, \frac{c_t}{c_\varphi \ell_5}}$ --- ``tilt'' of the $T^2$
  \item $\qty{\Phi, \Psi_\infty}$ --- independent gauge field parameters
\end{itemize}

\subsection{Asymptotic solution}\label{sec:asymptotics}
In this section we impose the equations of motion \cref{eq:evolution,eq:constraint} and the IOM identities \cref{eq:iom_qj,eq:iom_mu} on the coefficients of the Fefferman-Graham expansion. The ansatz \eqref{eq:ansatz_5d}, in the radial gauge $g_{rr} = 1$, reads
\begin{align}
  \begin{aligned}
  \dd[]{s}^2 ={}& - g_{tt} \dd[]{t}^2 + \dd[]{r}^2 + g_{\varphi\varphi} \qty(\dd[]{\varphi} + w \dd[]{t})^2 + g_{\Sigma} \dd[]{s}^2_\Sigma \,, \\
  \mathcal{A} ={}& a_t \dd[]{t} + a_\varphi \qty(\dd[]{\varphi} + w \dd[]{t}) - \mathfrak{p} \, \omega_\Sigma \,.
  \end{aligned}
\end{align}
Imposing $\text{AdS}_5$ asymptotics, the above functions admit the Fefferman-Graham expansion\footnote{We have taken the liberty to omit orders that are allowed a priori but vanish identically once the equations of motion are imposed on the ansatz \eqref{eq:ansatz_5d}.}
\begin{align}\label{eq:fg_expansion}
  \begin{aligned}
  g_{\bullet} ={}& \eu^{\frac{2r}{\ell_5}} \Bigg[g_\bullet^{(0,0)} + g_{\bullet}^{(2,0)} \eu^{-\frac{2r}{\ell_5}} + \qty(g_{\bullet}^{(4,0)} + g_{\bullet}^{(4,1)} \frac{r}{\ell_5}) \eu^{- \frac{4r}{\ell_5}} \\
                 & \qquad + \qty(g_{\bullet}^{(6,0)} + g_{\bullet}^{(6,1)} \frac{r}{\ell_5}) \eu^{- \frac{6r}{\ell_5}} + \qty(g_{\bullet}^{(8,0)} + g_{\bullet}^{(8,1)} \frac{r}{\ell_5} + g_{\bullet}^{(8,2)} \frac{r^2}{\ell_5^2}) \eu^{- \frac{8r}{\ell_5}} + \dots \Bigg] \,, \\
            w ={}& w^{(0,0)} + w^{(4,0)} \eu^{- \frac{4r}{\ell_5}} + w^{(6,0)} \eu^{- \frac{6r}{\ell_5}} + w^{(8,0)} \eu^{- \frac{8r}{\ell_5}} + \dots \,, \\
            a_\bullet ={}& a_\bullet^{(0,0)} + a_\bullet^{(2,0)} \eu^{- \frac{2r}{\ell_5}} + a_\bullet^{(4,0)} \eu^{- \frac{4r}{\ell_5}} \\
                         &+ \qty(a_\bullet^{(6,0)} + a_\bullet^{(6,1)} \frac{r}{\ell_5}) \eu^{- \frac{6r}{\ell_5}} + \qty(a_\bullet^{(8,0)} + a_\bullet^{(8,1)} \frac{r}{\ell_5}) \eu^{- \frac{8r}{\ell_5}} + \dots \,,
  \end{aligned}
\end{align}
where the bullet under $g$ stands for $\qty{tt, \varphi\varphi, \Sigma}$ and the bullet under $a$ stands for $\qty{t, \varphi}$. With 
the standard abuse of terminology, we refer to the terms polynomial in $r$ as \emph{log terms}, since they correspond to $\log z$ contributions in the $z$-coordinate of~\eqref{eq:fg_z}. The log terms reflect that the boundary $\mathbb{R} \times S^1 \times \Sigma$ is not conformally flat. According to the boundary conditions \eqref{eq:BC_UV}, the leading coefficients are
\begin{align}
  \begin{aligned}
  g_{tt}^{(0,0)} ={}& c_t^2 \,, & g_{\varphi\varphi}^{(0,0)} ={}& \ell_5^2 c_\varphi^2 \,, & g_{\Sigma}^{(0,0)} ={}& \ell_5^2 c_\Sigma^2 \,, \\
  w^{(0,0)} ={}& - \Omega_\infty \,, & a_t^{(0,0)} ={}& - \Phi + \Psi_\infty \Omega \,, & a_\varphi^{(0,0)} ={}& - \Psi_\infty \,,
  \end{aligned}
\end{align}
where, as a reminder, $\Omega \equiv \Omega_H - \Omega_\infty$. We now substitute the expansions \eqref{eq:fg_expansion} into the equations of motion \cref{eq:evolution,eq:constraint} and solve order by order. The $2^{\text{nd}}$ order metric coefficients are fully determined by the $0^{\text{th}}$ order ones
\begin{align}
  g_{tt}^{(2,0)} ={}& - \frac{c_t^2}{6 c_\Sigma^2} \,, \quad g_{\varphi\varphi}^{(2,0)} = - \frac{c_\varphi^2 \ell_5^2}{6 c_\Sigma^2} \,, \quad g_\Sigma^{(2,0)} = \frac{\ell_5^2}{3} \,.
\end{align}
The same holds for the $4^{\text{th}}$ order log coefficients
\begin{align}
  \begin{aligned}
  g_{tt}^{(4,1)} ={}& \frac{3 c_t^2}{8 c_\Sigma^4 \ell_5^2} \qty(\frac{\ell_5^2}{9} - \mathfrak{p}^2)\,, \\
  g_{\varphi\varphi}^{(4,1)} ={}& \frac{3 c_\varphi^2}{8 c_\Sigma^4} \qty(\frac{\ell_5^2}{9} - \mathfrak{p}^2) \,, \\
  g_\Sigma^{(4,1)} ={}& - \frac{3}{8 c_\Sigma^2} \qty(\frac{\ell_5^2}{9} - \mathfrak{p}^2) \,. 
  \end{aligned}
\end{align}
The above expressions illustrate a general pattern: whenever $\mathfrak{p}$ is fixed to its topological twist value 
\begin{align}
  \mathfrak{p}_{\text{top. twist}} ={}& \frac{\ell_5}{3} \,,
\end{align}
all log terms in \eqref{eq:fg_expansion} vanish identically at every order. Thus, despite the fact that $\mathbb{R} \times S^1 \times \Sigma$ is not conformally flat, carefully tuning the background gauge field $\mathcal{A}_\partial$ makes the boundary effectively behave as if it was conformally flat. We emphasize that setting $\mathfrak{p} = \mathfrak{p}_{\text{top. twist}}$ is necessary but not sufficient to enforce supersymmetry on the entire bulk solution. However, at leading order asymptotically, setting $\mathfrak{p} = \mathfrak{p}_{\text{top. twist}}$ is precisely the topological twist, and it ensures that the four-dimensional field configuration $\qty{\dd[]{s}^2_\partial, \mathcal{A}_\partial}$ \emph{is supersymmetric}. The results of \cite{Cassani:2013dba} (see section 3.1) then imply that the contributions of the metric and of the gauge field exactly cancel so that the conformal anomaly vanishes. From the physical standpoint, the full five-dimensional background represents a finite temperature deformation of the supersymmetric TTI background. This phenomenon was observed already in \cite{Bernamonti:2007bu}, for a non-supersymmetric, but non-rotating, black string solution, which was found analytically.

The equations of motion \cref{eq:evolution,eq:constraint} alone provide only partial information about the $4^{\text{th}}$ order coefficients:
\begin{align}
  \begin{aligned}
    g_{tt}^{(4,0)} ={}& \frac{c_t^2}{8 c_\Sigma^4 \ell_5^2} \qty(\frac{5 \ell_5^2}{9} + \mathfrak{p}^2) - \frac{2 c_t^2}{c_\Sigma^2 \ell_5^2} \, g_{\Sigma}^{(4,0)} - \frac{c_t^2}{c_\varphi^2\ell_5^2} \, g_{\varphi\varphi}^{(4,0)} \,, \\
    a_t^{(4,0)} ={}& - \frac{1}{6 c_\Sigma^2} \, a_t^{(2,0)} - \frac{c_t \mathfrak{p}}{2 c_\varphi c_\Sigma^2 \ell_5^2} \, a_{\varphi}^{(2,0)} + \Psi_\infty w^{(4,0)} \,, \\
    a_\varphi^{(4,0)} ={}& - \frac{1}{6 c_\Sigma^2} \, a_\varphi^{(2,0)} - \frac{c_\varphi \mathfrak{p}}{2 c_t c_\Sigma^2} \, a_t^{(2,0)} \,,
  \end{aligned}
\end{align}
where the $2^{\text{nd}}$ order gauge field coefficients are also left undetermined. It is at this point that the IOM identities \cref{eq:iom_qj,eq:iom_mu} become particularly useful, as they express almost all of these undetermined coefficients in terms of four of the radially conserved IOMs $\qty{q, j, \mu_t, \mu_s}$
\begin{align}
  \begin{aligned}
    a_t^{(2,0)} ={}& - \frac{c_t (q - 2 \mathfrak{p} \Psi_\infty)}{2 c_\varphi c_\Sigma^2 \ell_5^2} \,, \\
    a_\varphi^{(2,0)} ={}& \frac{c_\varphi \mathfrak{p} \qty(\Phi - \Omega \Psi_\infty)}{2c_t c_\Sigma^2} - \frac{c_t \qty(j + 3 (q - \mathfrak{p} \Psi_\infty)\Psi_\infty)}{6 c_\varphi c_\Sigma^2 \ell_5^2 \qty(\Phi - \Omega \Psi_\infty)} - \frac{c_\varphi \qty(2\mu_s + 2\mu_t \Omega_\infty - j(1 + \Omega_\infty^2))}{6c_t c_\Sigma^2 \qty(\Phi - \Omega\Psi_\infty)}\,, \\
    w^{(4,0)} ={}& - \frac{c_t \qty(j + 3(q - \mathfrak{p}\Psi_\infty)\Psi_\infty)}{4 c_\varphi^3 c_\Sigma^2 \ell_5^4} \,, \\
    g_{\varphi\varphi}^{(4,0)} ={}& \frac{c_\varphi^2}{16 c_\Sigma^4} \qty(\frac{5 \ell_5^2}{9} + \mathfrak{p}^2) - \frac{c_t (j + 3(q - \mathfrak{p} \Psi_\infty) \Psi_\infty) \Psi_\infty}{8 c_\varphi c_\Sigma^2 \ell_5^2 \qty(\Phi - \Omega \Psi_\infty)} - \frac{c_\varphi (2\mu_s + 2\mu_t \Omega_\infty - j(1 + \Omega_\infty^2)) \Psi_\infty}{8 c_t c_\Sigma^2 (\Phi - \Omega\Psi_\infty)} \\
    & - \frac{3 c_\varphi (\Phi - \Omega \Psi_\infty)(q - \mathfrak{p} \Psi_\infty)}{8 c_t c_\Sigma^2} + \frac{c_\varphi (\mu_t + j\Omega_\infty)}{4c_t c_\Sigma^2} - \frac{c_\varphi^2}{c_t^2} g_{\Sigma}^{(4,0)} \,,
  \end{aligned}
\end{align}
and the final IOM $\widetilde{q}$ also ends up being fixed in terms of $\qty{q, j, \mu_t, \mu_s}$ and the UV boundary data
\begin{align}\label{eq:qt_UV}
  \widetilde{q} ={}& \mathfrak{p} \qty(\Phi - \Omega \Psi_\infty) - q \Psi_\infty + \frac{c_t^2 (j + 3(q - \mathfrak{p} \Psi_\infty)\Psi_\infty)}{3 c_\varphi^2 c_\Sigma^2 \ell_5^2 (\Phi - \Omega \Psi_\infty)} + \frac{2\mu_s + 2\mu_t \Omega_\infty - j(1 + \Omega_\infty^2)}{3(\Phi - \Omega \Psi_\infty)} \,. 
\end{align}
Of course $\widetilde{q}$ is a constant throughout the entire solution. Soon we will provide a similar expression for it in terms of the IR boundary data and equate the two to obtain a non-trivial UV-IR relation.

In summary, once the IOM identities have been imposed the entire UV solution is determined by
\begin{itemize}
  \item The boundary data: $\qty{c_t,c_\varphi, c_\Sigma, \Omega_\infty, \Omega_H, \Phi, \Psi_\infty}$ 
  \item The IOMs: $\qty{q,j, \mu_t, \mu_s}$
  \item The non-boundary data: $\qty{g_{\Sigma}^{(4,0)}}$ 
\end{itemize}
The Fefferman-Graham coefficient $g_{\Sigma}^{(4,0)}$ is related to the v.e.v. of the holographic stress energy tensor components along the $\Sigma_{\mathfrak{g}}$ (pressure) and is generically non-zero away from the supersymmetric locus. In practice, we have gone to 10th order, including the corresponding log terms. The first non-trivial log term in $w(r)$ appears at order 10 through $w^{(10,1)}$. As expected from a Fefferman-Graham expansion new orders (above 4th) get completely determined from the previous orders and no new free parameters get introduced. 

\subsection{Near-horizon solution}\label{sec:near-horizon}
In the previous section we showed that the equations of motion \cref{eq:evolution,eq:constraint} can be solved for the ansatz~\eqref{eq:ansatz_5d} near the boundary at any desired order in $\eu^{ -r/\ell_5}$. The non-trivial question is whether this asymptotic solution can be smoothly connected to the IR boundary conditions~\eqref{eq:BC_IR}.

To set the stage, we first solve the equations of motion perturbatively near the horizon. In the radial gauge $g_{rr} = 1$, we can always locate the horizon at $r = 0$\footnote{This can always be achieved by an additive shift $r\to r+ r_h $. The effect of this shift in the UV asymptotics is simply to redefine the terms in the expansions by multiplicative constants that are powers of $e^{r_h/\ell_5}$.}. For a thermal horizon the fields admit a Taylor expansion about $r = 0$\footnote{Around an extremal horizon, the powers are generically rationals, as opposed to integers, see section 2.2 of \cite{Bernamonti:2007bu}.}
\begin{align}\label{eq:horizon_expansion}
  \begin{aligned}
  g_\bullet ={}& g_\bullet^{(0)} + g_\bullet^{(2)} r^2 + g_\bullet^{(4)} r^4 + \dots \,, \\
  w ={}& w^{(0)} + w^{(2)} r^2 + w^{(4)} r^{4} + \dots \,, \\
  a_\bullet ={}& a_\bullet^{(0)} + a_\bullet^{(2)} r^2 + a_\bullet^{(4)} r^4 + \dots \,,
  \end{aligned}
\end{align}
where, as before, the bullet stands for $\qty{tt, \varphi\varphi, \Sigma}$ in the metric sector and for $\qty{t,\varphi}$ in the gauge-field sector; no confusion with the coefficients in \eqref{eq:fg_expansion}, which carry two order-labels, can arise. According to the boundary conditions \eqref{eq:BC_IR}, the leading coefficients are
\begin{align}\label{eq:IR_boundary_data}
  \begin{aligned}
    g_{tt}^{(2)} ={}& \qty(\frac{2\pi}{\beta})^2 \,, & g_{\varphi\varphi}^{(0)} ={}& L_\varphi^2 \,, & g_{\Sigma}^{(0)} ={}& L_\Sigma^2 \,, \\
    w^{(0)} ={}& - \Omega_H \,, & a_\varphi^{(0)} ={}& - \Psi_H \,,
  \end{aligned}
\end{align}
and, crucially, $g_{tt}^{(0)} = 0 = a_t^{(0)}$ in order to have a smooth cap in Euclidean signature and a regular gauge field. At leading order the equations of motion \cref{eq:evolution,eq:constraint} dictate
\begin{align}\label{eq:IR_second_order_eom_coeff}
  \begin{aligned}
  g_{\varphi\varphi}^{(2)} ={}& \frac{2 L_\varphi^2}{\ell_5^2} + \frac{L_\varphi^2 \mathfrak{p}^2}{4 L_\Sigma^4} - L_\varphi^2 \qty(\frac{\beta}{2\pi})^2 \qty(\qty(a_t^{(2)} - \Psi_H w^{(2)})^2 + L_\varphi^2 (w^{(2)})^2) \,, \\
  g_\Sigma^{(2)} ={}& - \frac{1}{2} + \frac{2L_\Sigma^2}{\ell_5^2} - \frac{\mathfrak{p}^2}{2 L_\Sigma^2} - L_\Sigma^2 \qty(\frac{\beta}{2\pi})^2 \qty(a_t^{(2)} - \Psi_H w^{(2)})^2 \,, \\
  a_\varphi^{(2)} ={}& L_\varphi \frac{\beta}{2\pi} \qty(a_t^{(2)} - \Psi_H w^{(2)}) \qty(\frac{\mathfrak{p}}{L_\Sigma^2} - L_\varphi \frac{\beta}{2\pi} w^{(2)}) \,.
  \end{aligned}
\end{align}
Similarly, at subleading orders, $\qty{g_{\bullet}^{(4)}, a_{\bullet}^{(4)}}$ are determined by $\qty{a_t^{(2)}, w^{(2)}}$. At subleading order, the IOM identities \cref{eq:iom_qj,eq:iom_mu} in turn fix $\qty{a_t^{(2)}, w^{(2)}}$
\begin{align}\label{eq:IR_second_order_iom_coeff}
  \begin{aligned}
  a_t^{(2)} ={}& \frac{\pi}{\beta} \frac{\qty(j + 3 (q - \mathfrak{p} \Psi_H)\Psi_H)\Psi_H }{L_\Sigma^2 L_\varphi^3} + \frac{\pi}{\beta} \frac{\qty(q - 2 \mathfrak{p} \Psi_H)}{L_\Sigma^2 L_\varphi}\,, \\
  w^{(2)} ={}& \frac{\pi}{\beta} \frac{\qty(j + 3 (q - \mathfrak{p} \Psi_H)\Psi_H)}{L_\Sigma^2 L_\varphi^3} \,,
  \end{aligned}
\end{align}
and, in addition, relate the IOMs $\qty{\widetilde{q}, \mu_t, \mu_s}$ to the IR boundary data 
\begin{align}\label{eq:IR_iom}
  \begin{aligned}
  \widetilde{q} ={}& - q \Omega_H \,, \\
  \mu_t ={}& \frac{2\pi}{\beta} L_\Sigma^2 L_\varphi + j \Omega_H \,, \\
  \mu_s ={}& - \frac{2\pi}{\beta} L_\Sigma^2 L_\varphi \Omega_H + \frac{j(1 - \Omega_H^2)}{2} \,.
  \end{aligned}
\end{align}
Thus, algebraically, the quintet $\qty{q, \widetilde{q}, j, \mu_t, \mu_s}$ can be exchanged for the triplet $\qty{q,j, L_\Sigma^2 L_\varphi}$ and the IR variable $\Omega_H$. 
In addition, comparing the expressions for $\widetilde{q}$ in the IR \eqref{eq:IR_iom} and the UV \eqref{eq:qt_UV} we obtain a non-trivial UV-IR relation
\begin{align}\label{eq:q_UV_IR}
  3 \mathfrak{p} \qty(\Phi - \Omega \Psi_\infty)^2 + \frac{c_t^2}{c_\varphi^2 \ell_5^2} (j + 3(q - \mathfrak{p} \Psi_\infty) \Psi_\infty) ={}& \qty(\frac{4\pi}{\beta} L_\Sigma^2 L_\varphi - 3 q (\Phi - \Omega \Psi_\infty) + j \Omega)\Omega \,,
\end{align}
involving $\qty{q,j, L_\Sigma^2 L_\varphi}$ and the UV and IR boundary data. The IR parameter combination $L_\Sigma^2 L_\varphi$ is related to the geometric notion of entropy as
\begin{align}\label{eq:entropy}
  \mathcal{S} ={}& \frac{\text{Area}_H}{4 G_5} = \frac{1}{4 G_5} \int_{S^1_\varphi \times \Sigma} \sqrt{g|_H} = \frac{\pi^2 \eta_\Sigma}{G_5} L_\Sigma^2 L_\varphi \,,
\end{align}
where $\eta_\Sigma = 2(\mathfrak{g} - 1)$ for $\mathfrak{g} > 1$.

\subsection{Coordinate transformations, gauge transformations and rescalings}\label{sec:symmetries}
In this section we describe a set of symmetries enjoyed by the system of EOMs \cref{eq:evolution,eq:constraint} and IOMs \cref{eq:iom_qj,eq:iom_mu}. Under the below coordinate transformation we have
\begin{align}\label{eq:frame_shift}
  \varphi \rightarrow \varphi - \widehat{\alpha} t : & & &
  \begin{aligned}
      \Omega_H \rightarrow {}& \Omega_H - \widehat{\alpha} \,, \\
      \Omega_\infty \rightarrow {}& \Omega_\infty - \widehat{\alpha} \,, \\
      q \rightarrow {}& q \,, \\ 
      \widetilde{q} \rightarrow {}& \widetilde{q} + q \widehat{\alpha} \,, \\
      j \rightarrow {}& j \,, \\
      \mu_t \rightarrow {}& \mu_t - j \widehat{\alpha} \,, \\
      \mu_s \rightarrow {}& \mu_s + \mu_t \widehat{\alpha} - \frac{j}{2} \widehat{\alpha}^2 \,,
  \end{aligned}
\end{align}
while the difference $\Omega = \Omega_H - \Omega_\infty$ remains invariant. This transformation reflects the ability to choose any frame at infinity, rotating or not.

Under a gauge transformation along the spatial circle we have 
\begin{align}\label{eq:gauge_shift}
  \mathcal{A} \rightarrow \mathcal{A} + \alpha \qty(\dd[]{\varphi} - \Omega_H \dd[]{t}) : & & &
      \begin{aligned}
        \Psi_H \rightarrow {}& \Psi_H - \alpha \,, \\
        \Psi_\infty \rightarrow {}& \Psi_\infty - \alpha \,, \\
        q \rightarrow {}& q - 2 \mathfrak{p} \alpha \,, \\
        \widetilde{q} \rightarrow {}& \widetilde{q} + 2 \mathfrak{p} \Omega_H \alpha \,, \\
        j \rightarrow {}& j + 3q \alpha - 3 \mathfrak{p} \alpha^2 \,, \\
        \mu_t \rightarrow {}& \mu_t + \frac{3}{2} \qty(q \Omega_H - \widetilde{q}) \alpha - 3 \mathfrak{p} \Omega_H \alpha^2 \,, \\
        \mu_s \rightarrow {}& \mu_s + \frac{3}{2} \qty(q + \widetilde{q} \Omega_H) \alpha + \frac{3}{2} \mathfrak{p} \qty(\Omega_H^2 - 1) \alpha^2 \,, 
      \end{aligned}
\end{align}
in such a way that the regular gauge is preserved $0 = \iota_V \mathcal{A}|_H \rightarrow \iota_V \mathcal{A}|_H = 0$ and the difference $\Psi \equiv \Psi_H - \Psi_\infty$ remains invariant. We do not consider such gauge transformations along $\mathbb{R}_t$ (or $S^1_{t_E}$ in Euclidean), since those would take us out of the regular gauge.

Under spatial circle rescaling
\begin{align}\label{eq:spatial_circle_rescaling}
  \varphi \rightarrow \widehat{\lambda}^{-1} \varphi : & & &
  \begin{aligned}
    c_\varphi \rightarrow {}& \widehat{\lambda} c_\varphi \,, \\
    \Psi \rightarrow {}& \widehat{\lambda} \Psi \,, \\
    \Omega \rightarrow {}& \widehat{\lambda}^{-1} \Omega \,, \\
    q \rightarrow {}& \widehat{\lambda} q \,, \\
    j \rightarrow {}& \widehat{\lambda}^2 j \,,
  \end{aligned}
\end{align}
in such a way that the period of $\varphi$ is preserved.

Under the time rescaling
\begin{align}\label{eq:time_rescaling}
  t \rightarrow \lambda t: & & &
      \begin{aligned}
        c_t \rightarrow {}& \lambda^{-1} c_t \,, \\
        \beta \rightarrow {}& \lambda \beta \,, \\
        \Omega \rightarrow {}& \lambda^{-1} \Omega \,, \\
        \Phi \rightarrow {}& \lambda^{-1} \Phi \,,
      \end{aligned}
\end{align}
in such a way that the period of $t_E = \iu t$ is preserved.

The former two transformations \cref{eq:frame_shift,eq:gauge_shift} keep us within the same family of solutions. The latter two transformations \cref{eq:spatial_circle_rescaling,eq:time_rescaling} keep the same schematic form of the solution but have the effect of rescaling the physical charges. Among those transformations the one we will need in the upcoming discussion is the 
regular gauge preserving gauge shift \eqref{eq:gauge_shift}; it will help us identify gauge invariant definitions of the integrated electric charge density and the Euclidean on-shell action.

\subsection{Numerics}\label{sec:non_susy_numerics}
\begin{figure}[htbp]
  \centering
  \includegraphics[width=0.98\textwidth]{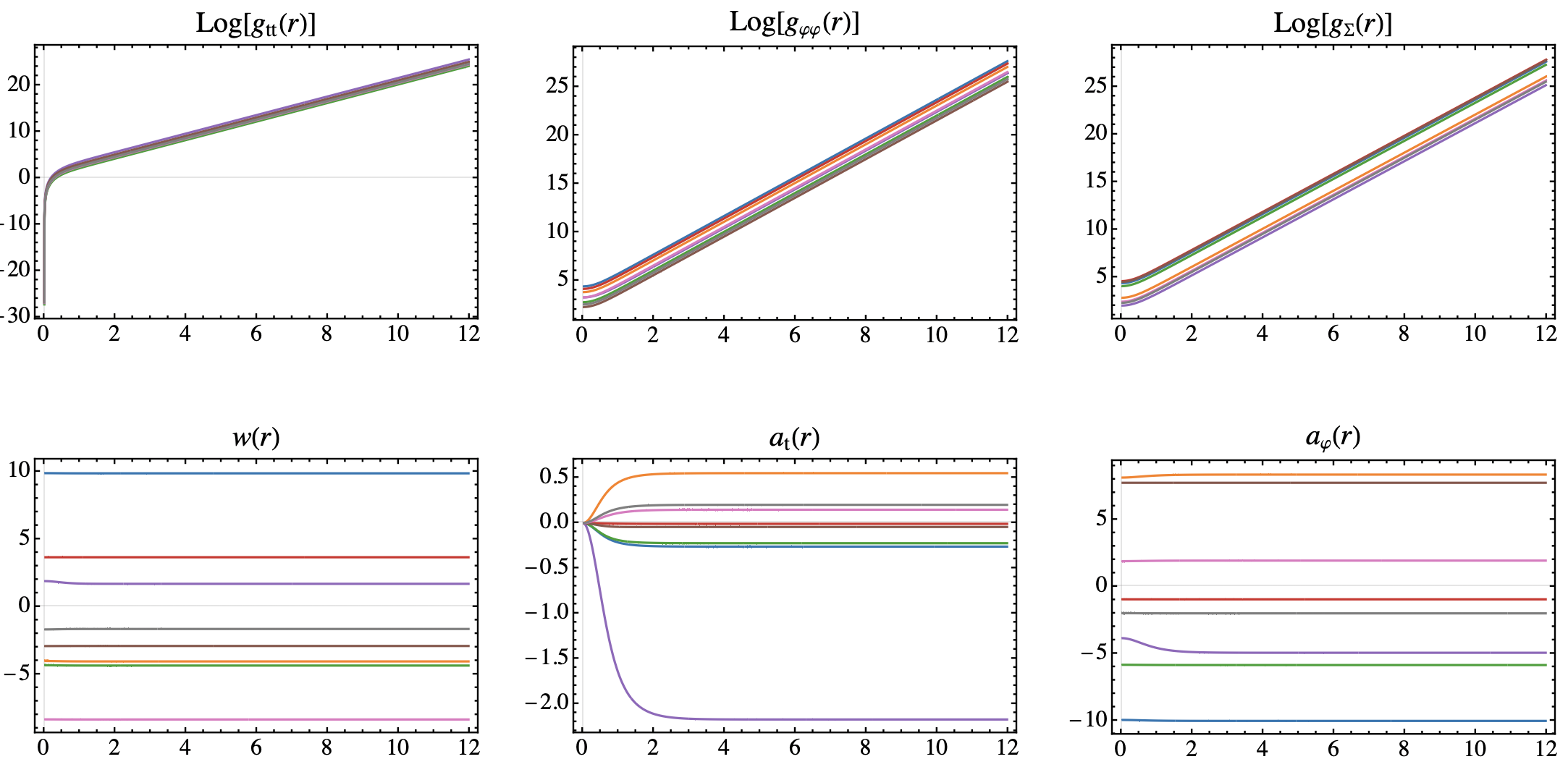}
  \caption{Radial profiles for a set of representative thermal black strings.}
  \label{fig:non_susy_plots}
\end{figure}

In this section we numerically solve the system of six evolution equations \eqref{eq:evolution} for the six functions $\mathcal{G}(r) = \qty{g_{tt}(r), g_{\varphi\varphi}(r), g_\Sigma(r), w(r), a_t(r), a_\varphi(r)}$. 
The strategy is to use the IR expansion \eqref{eq:horizon_expansion} to prepare the initial condition. Then we adopt a shooting method, implemented through \texttt{Mathematica}'s built-in \texttt{NDSolve}. Finally, we confirm the expected asymptotically locally $\text{AdS}_5$ asymptotics and read off the UV data. Since the EOMs are second-order ODEs, as initial conditions we supply
\begin{align}
  \mathcal{G}(\delta) \,, \quad \mathcal{G}'(\delta) \,, 
\end{align}
where $\delta$ is a small offset away from the horizon, which we have placed at $r = 0$. The ``positions'' $\mathcal{G}(\delta)$ involve the IR boundary data $\qty{\beta, L_\varphi, L_\Sigma, \Omega_H, \Psi_H}$ as in \eqref{eq:IR_boundary_data} and the ``velocities'' $\mathcal{G}'(\delta)$, in addition, involve $\qty{\mathfrak{p}, q, j}$ as in \cref{eq:IR_second_order_eom_coeff,eq:IR_second_order_iom_coeff}. For a vast array of initial conditions the numerical solver converges, see \cref{fig:non_susy_plots},
and at some large $r$ the behaviour of the functions is
consistent with $\text{AdS}_5$ asymptotics
\begin{align}
  \begin{aligned}
  \log g_{tt} \sim{}& \frac{2r}{\ell_5} + b_{tt}^\infty \,, & \log g_{\varphi\varphi} \sim{}& \frac{2r}{\ell_5} + b_{\varphi\varphi}^\infty \,, & \log g_{\Sigma} \sim{}& \frac{2r}{\ell_5} + b_\Sigma^\infty \,, \\
  w \sim {}& w^\infty \,, & a_t \sim{}& a_t^\infty \,, & a_\varphi \sim {}& a_\varphi^\infty \,,
  \end{aligned}
\end{align}
where the constants $\qty{b_{tt}^\infty, b_{\varphi\varphi}^\infty, b_{\Sigma}^{\infty}}$ are determined by numerical linear fits near the boundary, and the constants $\qty{w^\infty, a_{t}^\infty, a_{\varphi}^{\infty}}$ are simply numerically read off near the boundary. From there the UV boundary data is obtained as
\begin{align}
  \begin{aligned}
  c_t ={}& \eu^{\frac{b_{tt}^\infty}{2}} \,, & c_\varphi ={}& \eu^{\frac{b_{\varphi\varphi}^\infty}{2}} \,, & c_\Sigma ={}& \eu^{\frac{b_\Sigma^\infty}{2}} \,, \\
  \Omega_\infty ={}& - w^{\infty} \,, & \Phi ={}& - a_t^\infty - a_\varphi^\infty (\Omega_H + w^\infty) \,, & \Psi_\infty ={}& - a_\varphi^\infty \,. 
  \end{aligned}
\end{align}
Thus the shooter can be thought of as a black box that smoothly connects, see \cref{tab:non_susy_bb},
\begin{align}\label{eq:black_box}
  \qty{\mathfrak{p}, q, j, \beta, L_\varphi, L_\Sigma, \Omega_H, \Psi_H} \quad \longrightarrow \quad \qty{c_t, c_\varphi, c_\Sigma, \Omega_\infty, \Phi, \Psi_\infty} \,. 
\end{align}
\begin{table}[htbp]
  \centering
  \renewcommand{\arraystretch}{1.25}
  % tab10 palette (matches the notebook curves)
  \definecolor{solA}{RGB}{31,119,180} \definecolor{solB}{RGB}{255,127,14}
  \definecolor{solC}{RGB}{44,160,44} \definecolor{solD}{RGB}{214,39,40}
  \definecolor{solE}{RGB}{148,103,189} \definecolor{solF}{RGB}{140,86,75}
  \definecolor{solG}{RGB}{227,119,194} \definecolor{solH}{RGB}{127,127,127}
  \newcommand{\sw}[1]{\textcolor{#1}{\rule{2.6ex}{1.4ex}}}
  \resizebox{\textwidth}{!}{%
  \begin{tabular}{|c||c|c|c|c|c|c|c|c|c||c|c|c|c|c|c|}
    \hline
    \textbf{Color} & $\ell_5$ & $\mathfrak{p}$ & $q$ & $j$
    & $\beta$ & $\Omega_H$ & $\Psi_H$ & $L_\varphi$ & $L_\Sigma$
    & $c_t$ & $c_\varphi$ & $c_\Sigma$ & $\Omega_\infty$ & $\Phi$ & $\Psi_\infty$ \\
    \hline\hline
    \sw{solA} & $1$ & $5.375$ & $-2.393$ & $8.83$ & $1.746$ & $-9.894$ & $9.942$ & $9.075$ & $8.975$ & $1.279$ & $6.404$ & $6.327$ & $-9.881$ & $0.1354$ & $10.02$ \\ \hline
    \sw{solB} & $1$ & $2.735$ & $-9.238$ & $-3.379$ & $1.641$ & $4.005$ & $-8.15$ & $6.761$ & $4.168$ & $1.382$ & $4.755$ & $2.906$ & $4.035$ & $-0.2975$ & $-8.371$ \\ \hline
    \sw{solC} & $1$ & $2.464$ & $1.933$ & $0.4118$ & $1.992$ & $4.319$ & $5.826$ & $4.013$ & $7.66$ & $1.12$ & $2.833$ & $5.4$ & $4.342$ & $0.09018$ & $5.846$ \\ \hline
    \sw{solD} & $1$ & $8.15$ & $8.326$ & $1.033$ & $1.534$ & $-3.672$ & $0.9374$ & $7.974$ & $9.979$ & $1.45$ & $5.643$ & $7.042$ & $-3.672$ & $0.009009$ & $0.9394$ \\ \hline
    \sw{solE} & $1$ & $2.708$ & $-7.408$ & $-9.669$ & $1.111$ & $-1.907$ & $3.836$ & $5.135$ & $2.761$ & $2.258$ & $3.43$ & $1.853$ & $-1.706$ & $1.183$ & $4.928$ \\ \hline
    \sw{solF} & $1$ & $0.4951$ & $-5.36$ & $-3.801$ & $1.275$ & $2.899$ & $-7.76$ & $3.128$ & $9.593$ & $1.744$ & $2.212$ & $6.773$ & $2.892$ & $-0.009969$ & $-7.76$ \\ \hline
    \sw{solG} & $1$ & $1.46$ & $2.648$ & $4.788$ & $1.698$ & $8.323$ & $-1.911$ & $5.159$ & $3.391$ & $1.322$ & $3.658$ & $2.363$ & $8.331$ & $-0.1315$ & $-1.952$ \\ \hline
    \sw{solH} & $1$ & $0.05069$ & $5.246$ & $-0.4807$ & $1.61$ & $1.667$ & $1.973$ & $3.578$ & $3.192$ & $1.396$ & $2.532$ & $2.224$ & $1.635$ & $-0.1377$ & $1.987$ \\ \hline
  \end{tabular}%
  }
  \caption{Shooting data for the representative thermal black strings; row colors match the curves in \cref{fig:non_susy_plots}: the IR inputs
  $\{\ell_5,\mathfrak{p},q,j,\beta,\Omega_H,\Psi_H,L_\varphi,L_\Sigma\}$ (left of the double rule)
  and the UV data $\{c_t,c_\varphi,c_\Sigma,\Omega_\infty,\Phi,\Psi_\infty\}$ read off from the
  numerical flow (right).}
  \label{tab:non_susy_bb}
\end{table}
We verify numerically that the constraint equation \eqref{eq:constraint} holds $\qty(\text{to} \sim 10^{-14})$ pointwise\footnote{Here and in what follows when reporting such numerical accuracy we always have in mind that we have generated many example solutions where the initial conditions are $\order{1}$, see \cref{tab:non_susy_bb}, thus $10^{-n}$ accuracy is measured relative to $1$.}. Similarly, we test the IOM identities
\begin{align}
  \begin{aligned}
  q ={}& q(g,g',a,a') \,, \\
  - q \Omega_H ={}& \widetilde{q}(g,g',a,a') \,, \\
  j ={}& j(g,g',a,a') \,, \\
  \frac{2\pi}{\beta} L_\Sigma^2 L_\varphi + j \Omega_H ={}& \mu_t(g,g',a,a') \,, \\
  - \frac{2\pi}{\beta} L_\Sigma^2 L_\varphi \Omega_H + \frac{j(1 - \Omega_H^2)}{2} ={}& \mu_s(g,g',a,a') \,,
  \end{aligned}
\end{align}
where on the left hand sides $\qty{q,j}$ are part of the IR shooting data and $\qty{\widetilde{q}, \mu_t, \mu_s}$ are determined in terms of the IR shooting data from \eqref{eq:IR_iom}, and on the right hand sides we use the functional expressions \cref{eq:iom_qj,eq:iom_mu}. The above holds $(\text{to} \sim 10^{-8})$ pointwise when the \texttt{WorkingPrecision} of the solver is set to 25, and the accuracy increases when the working precision is increased. Finally, we verify the algebraic UV-IR relation \eqref{eq:q_UV_IR} $(\text{to} \sim 10^{-10})$. 
As such, we numerically establish the existence of the \textit{thermal black string} advertised in \cref{sec:BC}. We emphasize that nothing that follows directly relies on the explicit numerical data 
from the shooter. Concretely, the entire thermodynamics will be deduced analytically in the subsequent sections.

\subsection{Thermodynamic quantities}\label{sec:non_susy_thermodynamic_quantities}
In this section we compute the charges of the thermal black string using standard holographic renormalization.

As explained above, in the radial gauge $g_{rr} = 1$, the field configuration
\begin{align}
  \begin{aligned}
  \dd[]{s}^2 ={}& \dd[]{r}^2 + \dd[]{s}^2_h \,, \\
  \dd[]{s}^2_h ={}& - g_{tt} \dd[]{t}^2 + g_{\varphi\varphi} \qty(\dd[]{\varphi} + w \dd[]{t})^2 + g_\Sigma \dd[]{s}^2_\Sigma \,, \\
  \mathcal{A} ={}& a_t \dd[]{t} + a_\varphi \qty(\dd[]{\varphi} + w \dd[]{t}) - \mathfrak{p} \, \omega_\Sigma \,, 
  \end{aligned}
\end{align}
is already in Fefferman-Graham form. The functions are expanded for large $r$ as in \eqref{eq:fg_expansion}, and their coefficients are computed as in \cref{sec:asymptotics}. To the bulk action we add the standard set of (divergent) counterterms
\begin{align}\label{eq:counterterms}
  \begin{aligned}
  S_{\text{bdy}} ={}& \frac{2}{16 \pi G_5} \int_{\partial M_5} \dd[4]{x} \mathcal{L}_{\text{bdy}} \,, \\
  \mathcal{L}_{\text{bdy}} ={}& \sqrt{-h} \, \qty(K - \frac{3}{\ell_5} - \frac{\ell_5}{4} R_h - \frac{r}{\ell_5} \texttt{anomaly}) \,, \\
  \texttt{anomaly} ={}& \frac{1}{8} \qty(\ell_5^3 \qty(R_h)_{ij} \qty(R_h)^{ij} - \frac{\ell_5^3}{3} R_h^2 - 3 \ell_5 \mathcal{F}_{ij} \mathcal{F}^{ij}) \,,
  \end{aligned}
\end{align}
where $K$ is the extrinsic curvature
\begin{align}
  K ={}& h^{ij} K_{ij} \,, \quad K_{ij} = \frac{1}{2} \partial_r h_{ij} \,, 
\end{align}
and $\texttt{anomaly}$ is the conformal anomaly, multiplied by the log term $r/\ell_5$. The holographic stress energy tensor is defined by
\begin{align}
  T_{ij} ={}& - \frac{2}{\sqrt{-h}} \frac{\delta}{\delta h^{ij}} \mathcal{L}_{\text{bdy}} \,. 
\end{align}
The intrinsically finite up-down tensor is defined as $\mathcal{T}\indices{^i_j} \equiv \displaystyle{\lim_{r \rightarrow \infty}} 2 \sqrt{-h} \, h^{ik} T_{kj}$ and its explicit components read
\begin{align}\label{eq:stress-energy_explicit}
  \begin{aligned}
    \mathcal{T}\indices{^t_t} ={}& \qty[ - \mathcal{E} - \qty(\frac{2 \pi}{\beta} L_\Sigma^2 L_\varphi - \frac{3}{2} (q - \mathfrak{p} \Psi_\infty) \Phi) \frac{\Phi}{\Phi - \Omega \Psi_\infty} ] \sinh^{}{\theta} \\
                                 & \quad - \qty[\Omega + \frac{1}{2} \qty(\Omega^2 - \frac{c_t^2}{c_\varphi^2 \ell_5^2}) \frac{\Psi_\infty}{\Phi - \Omega \Psi_\infty} + \Omega_\infty] \mathcal{T}\indices{^t_\varphi} \,, \\
    \mathcal{T}\indices{^\varphi_\varphi} ={}& \qty[ - \mathcal{E} + \qty(\frac{2 \pi}{\beta} L_\Sigma^2 L_\varphi - \frac{3}{2} (q - \mathfrak{p} \Psi_\infty) \Phi) \frac{\Phi}{\Phi - \Omega \Psi_\infty} ] \sinh^{}{\theta} \\
                                 & \quad + \qty[\Omega + \frac{1}{2} \qty(\Omega^2 - \frac{c_t^2}{c_\varphi^2 \ell_5^2}) \frac{\Psi_\infty}{\Phi - \Omega \Psi_\infty} + \Omega_\infty] \mathcal{T}\indices{^t_\varphi} \,, \\
  \mathcal{T}\indices{^t_\varphi} ={}& \qty[j + 3 (q - \mathfrak{p} \Psi_\infty) \Psi_\infty] \sinh^{}{\theta} \,, \\
  \mathcal{T}\indices{^\theta_\theta} ={}& \qty[\mathcal{E} + \frac{3 c_t c_\varphi}{4 c_\Sigma^2} \qty(\frac{\ell_5^2}{9} - \mathfrak{p}^2)] \sinh^{}{\theta} \,, \\
  \mathcal{T}\indices{^\phi_\phi} ={}& \mathcal{T}\indices{^\theta_\theta} \,,
  \end{aligned}
\end{align}
where we have isolated the combination
\begin{align}\label{eq:casimir}
  \mathcal{E} \equiv {}& c_t c_\varphi \qty(\frac{\frac{\ell_5^2}{9} - \mathfrak{p}^2}{8 c_\Sigma^2} + 4 g_\Sigma^{(4,0)}) \,,
\end{align}
to be given an interpretation momentarily. Then the energy and angular momentum are obtained as\footnote{The slightly non-standard expression for the energy is due to allowing a general rotating frame at infinity --- which leads to the energy being associated to the Killing vector $\partial_t + \Omega_\infty \partial_\varphi$.}
\begin{align}
  \begin{aligned}
    E ={}& - \frac{1}{16 \pi G_5} \lim_{r \rightarrow \infty} \int_{S^1_\varphi \times \Sigma} \qty(\mathcal{T}\indices{^t_t} + \Omega_\infty \mathcal{T}\indices{^t_\varphi}) \,, \\
  J ={}& \frac{1}{16 \pi G_5} \lim_{r \rightarrow \infty} \int_{S^1_\varphi \times \Sigma} \mathcal{T}\indices{^t_\varphi} \,, 
  \end{aligned}
\end{align}
yielding the expressions
\begin{align}\label{eq:energy_ang_mom}
  \begin{aligned}
    E ={}& \frac{\pi \eta_\Sigma}{4 G_5} \Bigg[ \mathcal{E} + \qty(\frac{2\pi}{\beta} L_\Sigma^2 L_\varphi - \frac{3}{2} (q - \mathfrak{p} \Psi_\infty) \Phi)\frac{\Phi}{\Phi - \Omega \Psi_\infty} \\
    & \quad \quad \quad + \qty(j + 3(q - \mathfrak{p} \Psi_\infty) \Psi_\infty) \qty(\Omega + \frac{1}{2}\qty(\Omega^2 - \frac{c_t^2}{c_\varphi^2 \ell_5^2}) \frac{\Psi_\infty}{\Phi - \Omega \Psi_\infty})\Bigg] \,, \\
    J ={}& \frac{\pi \eta_\Sigma}{4 G_5} \qty(j + 3 \Psi_\infty(q - \mathfrak{p} \Psi_\infty)) \,.
  \end{aligned}
\end{align}

Next we turn to the holographic electric charge. First, we verify that the electric charge is non-anomalous by explicitly showing that
\begin{align}
  \nabla_i \mathcal{F}^{ij} ={}& 0 \,. 
\end{align}
The holographic electric charge is then given by \cite{Cassani:2014zwa}\footnote{Our gauge field is related to the one in \cite{Cassani:2014zwa} by $A^{\text{there}} = \frac{\sqrt{3}}{2} \mathcal{A}$.}
\begin{align}\label{eq:def_el_charge}
  Q ={}& \frac{3}{16 \pi G_5} \lim_{r \rightarrow \infty} \int_{S^1_\varphi \times \Sigma} \qty(\star_5 \mathcal{F} + \frac{2}{3} \mathcal{A} \wedge \mathcal{F}) \,,
\end{align}
where we have kept the relative coefficient between the two terms as in \cite{Cassani:2014zwa}. We evaluate the above integral following the procedure of \cref{sec:patchwork} as
\begin{align}\label{eq:el_charge}
  Q ={}& \frac{3}{16 \pi G_5} \lim_{r \rightarrow \infty} \fint_{S^1_\varphi \times \Sigma} \qty(\star_5 \mathcal{F} + \frac{4}{3} \mathcal{A} \wedge \mathcal{F}) \, \notag \\
  ={}& \frac{3\pi \eta_\Sigma}{4 G_5} \qty(q - \frac{2\Psi_\infty \mathfrak{p}}{3}) \,. 
\end{align}
Note that if we open the string direction through $S^1_\varphi \rightarrow \mathbb{R}_\varphi$ we will not be talking about total electric charge, but instead about electric charge density. In the present case $Q$ really should be thought of as \textit{integrated electric charge density}.

Finally, we compute the renormalized Euclidean on-shell action
\begin{align}
  \widehat{I} ={}& I + I_{\text{bdy}} \,,
\end{align}
where
\begin{align}
  \begin{aligned}
    I ={}& \eval{- \iu \, S}_{t \rightarrow - \iu t_E} = - \frac{1}{16 \pi G_5} \int_{\mathcal{M}} \mu_{\text{id}}' \dd[]{t_E} \wedge \dd[]{r} \wedge \dd[]{\varphi} \wedge \Omega_\Sigma \,, \\
  I_{\text{bdy}} ={}& \eval{- \iu \, S_{\text{bdy}}}_{t \rightarrow - \iu t_E} = - \frac{2}{16 \pi G_5} \int_{\partial \mathcal{M}} \mathcal{L}_{\text{bdy}}^E \dd[]{t_E} \wedge \dd[]{\varphi} \wedge \Omega_\Sigma \,. 
  \end{aligned}
\end{align}
The bulk part is
\begin{align}\label{eq:osa_naked}
  I ={}& - \beta \frac{\pi \eta_\Sigma}{4G_5} \qty(\mu_{\text{id}}(r_c) - \mu_{\text{id}}(0)) \,, 
\end{align}
where $r_c$ is the cutoff radius and we have used that $\Delta t_E = \beta$ and $\Delta \varphi = 2\pi$. The first term in \eqref{eq:osa_naked} contains both the non-log and the log divergences $\qty{\eu^{\frac{4 r_c}{\ell_5}}, \eu^{\frac{2 r_c}{\ell_5}}, \frac{r_c}{\ell_5}}$. These divergences are cancelled by $I_{\text{bdy}}$, leaving an additional finite contribution. Curiously, the IR contribution to the first term in \eqref{eq:osa_naked} evaluates to one-half of the (geometric) entropy
\begin{align}
  - \beta \frac{\pi \eta_\Sigma}{4 G_5} \mu_{\text{id}}(0) ={}& \frac{\pi^2 \eta_\Sigma}{2G_5} L_\Sigma^2 L_\varphi \equiv \frac{\mathcal{S}}{2} \,.
\end{align}
Combining all contributions, the renormalized Euclidean on-shell action reads 
\begin{align}\label{eq:osa}
    \widehat{I} ={}& \beta\frac{\pi \eta_\Sigma}{4 G_5} \Bigg[ c_t c_\varphi \qty(\frac{\frac{\ell_5^2}{9} - \mathfrak{p}^2}{8 c_\Sigma^2} + 4 g_{\Sigma}^{(4,0)}) + \qty(\frac{2\pi}{\beta} L_\Sigma^2 L_\varphi - \frac{3}{2} (q - \mathfrak{p} \Psi_\infty) \Phi) \qty(\frac{\Phi}{\Phi - \Omega \Psi_\infty} - 2) \notag \\
    & \quad \quad \quad + \frac{1}{2}\qty(j + 3(q - \mathfrak{p} \Psi_\infty) \Psi_\infty) \qty(\Omega^2 - \frac{c_t^2}{c_\varphi^2 \ell_5^2}) \frac{\Psi_\infty}{\Phi - \Omega \Psi_\infty} \Bigg] \,. 
\end{align}

\subsection{Thermodynamic relations}\label{sec:non_susy_thermodynamic_relations}
Collecting the thermodynamic quantities computed above --- the on-shell action $\widehat{I}$ \eqref{eq:osa}, the entropy $\mathcal{S}$ \eqref{eq:entropy}, the energy $E$ and angular momentum $J$ \eqref{eq:energy_ang_mom}, and the electric charge $Q$ \eqref{eq:el_charge} --- we verify that they satisfy the \textit{quantum statistical relation} (QSR)
\begin{align}
  \widehat{I} ={}& - \mathcal{S} + \beta \qty[E - \Omega J + \Phi \qty(Q - \Psi_\infty P) ] \,, \quad P \equiv \frac{\pi \eta_\Sigma}{4G_5} \mathfrak{p} \,. 
\end{align}
The UV-IR relation \eqref{eq:q_UV_IR} can be written in a more transparent fashion in terms of the charges: then it simply becomes the Smarr relation
\begin{align}\label{eq:smarr}
  E - \frac{\pi \eta_\Sigma}{4G_5}\mathcal{E} ={}& \frac{1}{2} \frac{\mathcal{S}}{\beta} + \Omega J - \frac{1}{2} \Phi \qty(Q - 4 \Psi_\infty P) \,.
\end{align}
All of the thermodynamic quantities are invariant under the coordinate transformation \eqref{eq:frame_shift}. Under the regular gauge shift \eqref{eq:gauge_shift} the ones that change are
\begin{align}
  \begin{aligned}
  \widehat{I} \rightarrow {}& \widehat{I} - 3 \beta \Phi P \alpha \,, \\
  Q \rightarrow {}& Q - 4 P \alpha \,, \\
  \Psi_\infty \rightarrow {}& \Psi_\infty - \alpha \,. 
  \end{aligned}
\end{align}
The above transformation of $\widehat{I}$ can be inferred from the slightly complicated expression~\eqref{eq:osa}. Alternatively, it can be obtained immediately by observing that the entire contribution to the transformation of $\widehat{I}$ comes from the Chern-Simons term, which can be written as
\begin{align}\label{effectiveCS}
  \begin{aligned}
  S_\mathrm{CS} ={}& -\frac{3}{2}\frac{1}{16\pi G_5} \fint_{\mathcal{M}} \mathcal{A} \wedge \mathcal{F}\wedge \mathcal{F} \, . 
   \end{aligned}
\end{align}
A straightforward direct calculation then yields
\begin{align}
  \begin{aligned}
 \Delta \widehat{I} = \Delta S_\mathrm{CS} ={}& -\frac{3}{2}\frac{1}{16\pi G_5} \fint_{\mathcal{M}}\Delta \mathcal{A} \wedge \mathcal{F}\wedge \mathcal{F} = - 3 \beta \Phi P \alpha \, . 
   \end{aligned}
\end{align}
The charge combination that appears in the Smarr relation is invariant under the regular gauge shift, as it should be given that $\qty{E, \mathcal{E}, \mathcal{S}, J}$ are invariant. It coincides with the Maxwell charge
\begin{align}
  Q - 4 \Psi_\infty P = {}& Q_{\text{Maxwell}} \,, \qquad Q_{\text{Maxwell}} \equiv \frac{3}{16 \pi G_5} \lim_{r \rightarrow \infty} \int_{S^1_\varphi \times \Sigma} \star_5 \mathcal{F} \,.
\end{align}
The regular gauge shift of $\widehat{I}$ is precisely compensated by the regular gauge shift of the charge combination appearing in the QSR
\begin{align}
  \begin{aligned}
  \widehat{I} \rightarrow {}& \widehat{I} - 3 \beta \Phi P \alpha \,, \\
  \qty(Q - \Psi_\infty P) \rightarrow {}& \qty(Q - \Psi_\infty P) - 3 P \alpha \,,
  \end{aligned}
\end{align}
such that the QSR is invariant under regular gauge shift.

\subsection{More convenient variables and a prelude to supersymmetry}\label{sec:convenient_vars}

Let us now give a succinct description of the thermodynamic quantities in more convenient variables. We define \textit{new thermodynamic potentials}\footnote{Note that $\qty{\omega, \widetilde{\omega}}$ are related to modular parameter $\qty{\tau, \bar{\tau}}$ \eqref{eq:modular_parameter} as $\omega = 2\pi\iu \tau$, $\widetilde{\omega} = 2\pi\iu \bar{\tau}$.}
\begin{align}
    \omega \equiv \beta \qty(\Omega - \frac{c_t}{c_\varphi \ell_5}) \,, \quad \widetilde{\omega} \equiv \beta \qty(\Omega + \frac{c_t}{c_\varphi \ell_5}) \,, \quad \Delta \equiv \frac{\beta \Phi}{\ell_5} \,.
\end{align}
We define $\ell_5$-rescaled parameters
\begin{align} 
    \ell_\Sigma \equiv \frac{L_\Sigma}{\ell_5} \,, \quad \ell_\varphi \equiv \frac{L_\varphi}{\ell_5} \,, \quad \psi_\infty \equiv \frac{\Psi_\infty}{\ell_5} \,, \quad \psi_H \equiv \frac{\Psi_H}{\ell_5} \,, \quad \widehat{\mathfrak{p}} \equiv \frac{\mathfrak{p}}{\ell_5} \,, 
\end{align}
and more convenient charges
\begin{align}
  \widehat{\mathcal{E}} \equiv{}& \frac{3}{16} \frac{c_\varphi \ell_5}{c_t} \frac{\mathcal{E}}{\ell_5^3} \,, \quad L \equiv \frac{1}{2} \qty(\frac{c_\varphi \ell_5}{c_t} E + J) \,, \quad \widetilde{L} \equiv \frac{1}{2} \qty(\frac{c_\varphi \ell_5}{c_t} E - J) \,, \quad \widehat{Q} \equiv Q \ell_5 \,. 
\end{align}
We also express the Newton constant $G_5$ suggestively in terms of a variable called $\mathfrak{c}$ that will turn out to be the already extremized 2d central charge at the end point of the field theory flow $T^2 \times \Sigma_{\mathfrak{g}} \longrightarrow T^2$, see details in \cref{sec:susy_thermo}:
\begin{align}
  \mathfrak{c} = \frac{2\pi \eta_\Sigma}{3} \frac{\ell_5^3}{G_5} \,. 
\end{align}
The definitions for $\{L, \widetilde{L}\}$ are inspired by 2d CFTs where
\begin{align}
  L ={}& L_0 - \frac{\mathfrak{c}}{24} \,, \quad \widetilde{L} = \bar{L}_0 - \frac{\mathfrak{c}}{24} \,, 
\end{align}
with $L_0$, $\bar{L}_0$ being the Virasoro zero modes and the subtraction of $\mathfrak{c}/24$ reflects a Casimir energy contribution due to the 2d theory being placed on $T^2$ instead of $\mathbb{R}^2$. 
Then the set of thermodynamic quantities for the \textit{fully non-supersymmetric thermal black string} can be compactly written as
\begin{align}\label{eq:full_thermo}
  \begin{aligned}
    \mathcal{S} ={}& \frac{3 \pi}{2} \ell_\Sigma^2 \ell_\varphi \mathfrak{c} \,, \\
    \widehat{I} - \frac{9}{8} \widehat{\mathfrak{p}} \Delta \psi_\infty \mathfrak{c} - (\widetilde{\omega} - \omega) \widehat{\mathcal{E}} \mathfrak{c} ={}&
    - \frac{9}{8} \frac{\widehat{\mathfrak{p}} \Delta^2}{\omega + \widetilde{\omega}} \mathfrak{c} 
    + \frac{\omega \widetilde{\omega}}{\omega + \widetilde{\omega}} J \,, \\
    L - \widehat{\mathcal{E}} \mathfrak{c} ={}& 
    - \frac{9}{8} \frac{\widehat{\mathfrak{p}} \Delta^2}{\omega^2 - \widetilde{\omega}^2} \mathfrak{c} 
    - \frac{\widetilde{\omega}^2}{\omega^2 - \widetilde{\omega}^2} J \,, \\
    \widetilde{L} - \widehat{\mathcal{E}} \mathfrak{c} ={}& 
    - \frac{9}{8} \frac{\widehat{\mathfrak{p}} \Delta^2}{\omega^2 - \widetilde{\omega}^2} \mathfrak{c} 
    - \frac{\omega^2}{\omega^2 - \widetilde{\omega}^2} J \,, \\
    \widehat{Q} - \frac{3}{2} \widehat{\mathfrak{p}} \psi_\infty \mathfrak{c} ={}& \frac{1}{\Delta} \qty(\frac{3 \pi}{2} \ell_\Sigma^2 \ell_\varphi - \frac{9}{4} \frac{\widehat{\mathfrak{p}} \Delta^2}{\omega + \widetilde{\omega}}) \mathfrak{c} + \frac{2}{\Delta} \frac{\omega \widetilde{\omega}}{\omega + \widetilde{\omega}} J \,.
  \end{aligned}
\end{align}
To reach the above expressions we have used the UV-IR relation \eqref{eq:q_UV_IR} to eliminate $q$ in favour of $\qty{J, \ell_\Sigma^2 \ell_\varphi}$. Note that the combinations
\begin{align}
   \widehat{I} - \frac{9}{8} \widehat{\mathfrak{p}} \Delta \psi_\infty \mathfrak{c} \equiv \widehat{I}_{\text{G.I.}} \,, \qquad \widehat{Q} - \frac{3}{2} \widehat{\mathfrak{p}} \psi_\infty \mathfrak{c} = Q_{\text{Maxwell}} \ell_5 \equiv \widehat{Q}_{\text{Maxwell}} \,, 
\end{align}
are gauge invariant under the regular gauge shift \eqref{eq:gauge_shift}. Redefining $\widehat{I}$ to make it gauge invariant can be motivated by including the following non-gauge invariant local counterterm\footnote{The same local expression, with a different numerical factor,
has been used in \cite{Ntokos:2021duk} to enforce invariance of the on-shell action for a class of (extremal) solutions having $S^1\times S^3_{\rm squashed}$ conformal boundary. In particular, a single global patch is sufficient for describing those solutions.}
 \begin{align}
 \begin{aligned}
 \Theta =\lim_{r\to \infty} -\frac{3\beta}{16 \pi G_5} \fint_{S^1_\varphi \times \Sigma_\mathfrak{g}} (\iota_V \mathcal{A}) \mathcal{A}\wedge \mathcal{F} \,,
  \end{aligned}
\end{align}
whose variation is $\Delta\Theta=3 \beta \Phi P \Psi_\infty$. It would be interesting to investigate whether this counterterm is compatible with supersymmetry, along the lines of the discussion in \cite{BenettiGenolini:2016qwm,BenettiGenolini:2016tsn,Papadimitriou:2017kzw}. Finally, we identify the combination
\begin{align}
  \widehat{\mathcal{E}} ={}& \frac{3}{128} \frac{c_\varphi^2}{c_\Sigma^2} \qty(\frac{1}{9} - \widehat{\mathfrak{p}}^2) + \frac{3}{4} c_\varphi^2 \frac{g_\Sigma^{(4,0)}}{\ell_5^2} \,,
\end{align}
as a Casimir energy contribution, additional to $\mathfrak{c}/24$, due to the full boundary theory being placed on $T^2 \times \Sigma_{\mathfrak{g}}$\footnote{ It will be of utmost interest to derive this quantity directly from a field theory calculation. For the time being we note that it is written in the correct variables: $c_\varphi$ --- breathing mode of the $T^2$, $c_\Sigma$ --- breathing mode of the $\Sigma_{\mathfrak{g}}$, $g_\Sigma^{(4,0)}$ --- v.e.v. of the pressure extending from $\Sigma_{\mathfrak{g}}$. We also give a word of caution that such quantities could be scheme dependent.
}. As such, we define
\begin{align}
  \vb{I} \equiv {}& \widehat{I}_{\text{G.I.}} - \qty(\widetilde{\omega} - \omega) \widehat{\mathcal{E}} \mathfrak{c} \,, \quad \vb{L} = L - \widehat{\mathcal{E}} \mathfrak{c} \,, \quad \widetilde{\vb{L}} = \widetilde{L} - \widehat{\mathcal{E}} \mathfrak{c} \,, \quad \vb{Q} = \widehat{Q}_{\text{Maxwell}} \,, 
\end{align}
as the \textit{gauge invariant and Casimir-energy-subtracted} thermodynamic quantities, in terms of which the quantum statistical relation takes the simple form
\begin{align}
  \vb{I} ={}& - \mathcal{S} - \omega \vb{L} + \widetilde{\omega} \widetilde{\vb{L}} + \Delta \vb{Q} \,,
\end{align}
and the Smarr relation \eqref{eq:smarr} becomes
\begin{align}
  0 ={}& \frac{1}{2} \mathcal{S} - \frac{1}{2} \Delta\vb{Q} + \omega \vb{L} - \widetilde{\omega} \widetilde{\vb{L}} \,. 
\end{align}
In particular, these relations can be used to express the entropy and the on-shell action in terms of the charges and potentials as
\begin{align}
\label{gatto}
   \mathcal{S} = -2 \vb{I} + \Delta\vb{Q} \,, \qquad \vb{I} = \omega \vb{L} - \widetilde{\omega} \widetilde{\vb{L}} \,. 
\end{align}

Taking inspiration from 2d CFTs we explore what happens upon demanding that the anti-holomorphic movers are in the vacuum state
\begin{align}\label{eq:anti_holo_constr}
  \widetilde{\vb{L}} ={}& \bar{L}_0 - \frac{\mathfrak{c}}{24} - \widehat{\mathcal{E}} \mathfrak{c} \overset{!}{=} 0 \,. 
\end{align}
This leads to $J = - \frac{9}{8} \frac{\widehat{\mathfrak{p}} \Delta^2}{\omega^2} \mathfrak{c}$, and the thermodynamic quantities simplify to
\begin{align}
  \begin{aligned}
    \vb{I} ={}& - \frac{9}{8} \frac{\widehat{\mathfrak{p}} \Delta^2}{\omega} \mathfrak{c} \,, \quad \vb{L} = - \frac{9}{8} \frac{\widehat{\mathfrak{p}} \Delta^2}{\omega^2} \mathfrak{c} \,, \quad \widetilde{\vb{L}} = 0 \,, \quad \vb{Q} = \frac{1}{\Delta}\qty(\frac{3 \pi}{2} \ell_\Sigma^2 \ell_\varphi - \frac{9}{4} \frac{\widehat{\mathfrak{p}} \Delta^2}{\omega}) \mathfrak{c} \,.
  \end{aligned}
\end{align}
The entropy can be obtained from the quantum statistical relation as
\begin{align}
  \mathcal{S} ={}& - \vb{I} - \omega \vb{L} + \Delta \vb{Q} = \mp \iu \, (3\Delta) (3 \widehat{\mathfrak{p}})^{1/2} \sqrt{\frac{\mathfrak{c}}{6} \vb{L}} + \Delta \vb{Q} \,,
\end{align}
where in the last equality we have written it, as much as possible, in terms of the charges. At this point we have not truly imposed supersymmetry. As we shall see below, \eqref{eq:anti_holo_constr} arises naturally once supersymmetry is imposed. However, it must be supplemented by:
\begin{align}
  \widehat{\mathfrak{p}} = \widehat{\mathfrak{p}}_{\text{top. twist}} = \frac{1}{3} \,, \quad \Delta = \pm \frac{2\pi\iu}{3} \,,
\end{align}
i.e.\ supersymmetry demands the topological twist, together with a \textit{supersymmetric linear constraint} on $\Delta$. These conditions are analogous to \eqref{eq:susy_linear_constraint_4d} and \eqref{eq:susy_locus_4d}. Note that, practically, there is no need for extremization in order to obtain the supersymmetric entropy; one simply plugs in the supersymmetric value for $\Delta$\footnote{The situation is vastly different for
  \begin{enumerate}[(a)]
    \item The flavor-full TTI on $T^2 \times \Sigma_{\mathfrak{g}}$, where there are multiple $\Delta_I$'s and the supersymmetric linear constraint is $\sum_{I} \Delta_I = \pm 2\pi\iu$. One then extremizes in order to obtain the correct R-symmetry in the IR \cite{Benini:2013cda,Benini:2012cz,Hosseini:2016cyf},
    \item The SCI on $S^1 \times S^3$, where the supersymmetric linear constraint links $\Delta$ and the rotational fugacities associated to the $S^3$, even in the flavor-less case \cite{Cabo-Bizet:2018ehj}.
  \end{enumerate}
  The lack of extremization above is a mere consequence of the fact that the supersymmetric linear constraint $\Delta = \pm 2\pi\iu/3$ involves only one variable and there is nothing to extremize over.
}.
We find it useful to already present the final supersymmetric thermodynamic quantities in the $\tau = \omega/(2\pi\iu)$ modular variable
\begin{align}
  \vb{I} ={}& -\frac{\pi\iu \mathfrak{c}}{12 \tau} \,, \quad \vb{L} = - \frac{\mathfrak{c}}{24 \tau^2} \,, \quad \widetilde{\vb{L}} = 0 \,, \quad \vb{Q} = \mp \frac{\mathfrak{c}}{4\tau} \mp \iu \frac{9 \mathfrak{c}}{4} \ell_\Sigma^2 \ell_\varphi \,, 
\end{align}
together with the QSR-obtained entropy
\begin{align}
  \begin{aligned}
  \mathcal{S} ={}& 2 \pi \sqrt{\frac{\mathfrak{c}}{6} \vb{L}} \pm \frac{2\pi\iu}{3} \vb{Q} \\
  ={}& \frac{\pi\iu \mathfrak{c}}{6\tau} \pm \frac{2\pi\iu}{3} \vb{Q} \\
  ={}& \cancel{\frac{\pi\iu \mathfrak{c}}{6\tau}} - \cancel{\frac{\pi\iu\mathfrak{c}}{6\tau}} + \frac{3\pi\mathfrak{c}}{2} \ell_\Sigma^2 \ell_\varphi \,,
  \end{aligned}
\end{align}
which agrees with the geometric definition \eqref{eq:entropy}. As a word of caution we want to emphasize that the geometric notion of area of the horizon fails to be directly computable on a general complex saddle. We will deal with this problem explicitly in \cref{sec:susy_numerics} when we construct a 
supersymmetric shooter in a particular way which allows for real IR shooting parameters (and thus a sensible geometric notion of horizon area).

\section{The gravity dual of the TTI on $T^2 \times \Sigma_{\mathfrak{g}}$}\label{sec:susy_solutions}
In this section we impose supersymmetry on the thermal black strings of \cref{sec:thermo_bs}. Rather than solving Killing spinor
equations on a numerical background, we use the classification of supersymmetric solutions of
minimal gauged supergravity \cite{Gauntlett:2003fk} to recast our ansatz in manifestly
supersymmetric form, and then demand compatibility with the UV and IR expansions of \cref{sec:thermo_bs}. The output is remarkably rigid: supersymmetry enforces the topological twist
$\widehat{\mathfrak p}=\tfrac13$, sets the Casimir-like charge $\mathcal E$ to zero, imposes the
linear constraint $\Delta=\pm\tfrac{2\pi\iu}{3}$ together with one new UV-IR relation, and
collapses the thermodynamics \eqref{eq:full_thermo} to the holomorphic form anticipated in \cref{sec:convenient_vars}, with on-shell action $\vb I=-\pi\iu\,\mathfrak c/(12\tau)$ reproducing (minus the log of) the TTI partition function at large $N$ \eqref{eq:TTItauintro}. The resulting saddles are supersymmetric but neither extremal nor, in
general, real --- finite-$\beta$ complex black saddles in the sense of
\cite{Bobev:2020pjk,Cabo-Bizet:2018ehj} --- and we exhibit a real supersymmetric sub-locus, with
real metric and imaginary Wilson line, on which we construct the solutions numerically. We
conclude by presenting two corners of the family where the solution is fully analytic.

\subsection{General supersymmetry constraints on the thermal black string}\label{sec:susy_generalities}
The most direct way to impose supersymmetry is to write down the Killing spinor equations (KSE) and derive the constraints on the solution parameters under which the KSE admit a non-trivial solution. In the present case, in the absence of an analytic solution, this procedure is difficult to implement in practice. Fortunately, supersymmetric solutions of 5d minimal gauged supergravity have been classified in~\cite{Gauntlett:2003fk} and their geometric structure characterized in a spinor-free manner in terms of bilinears.
In Lorentzian signature, there are two classes of supersymmetric solutions: \textit{null} and \textit{timelike}. They are distinguished by whether the Killing spinor bilinear
\begin{align}
  \overline{\epsilon^c} \cdot \epsilon = - \iu \epsilon^\dagger \Gamma^0 \epsilon \,,
\end{align}
where $\epsilon$ is the 5d Dirac Killing spinor, vanishes identically (null class) or not (timelike class)\footnote{In Euclidean signature, one must \textit{double the spinors}: treat the charge conjugate spinor as independent $\epsilon^c \rightarrow \widetilde{\epsilon}$, after which the following scalar bilinears can be formed
\begin{align*}
  \overline{\widetilde{\epsilon}} \cdot \epsilon \,, \quad \overline{\widetilde{\epsilon}} \cdot \widetilde{\epsilon} \,, \quad \overline{\epsilon} \cdot \epsilon \,. 
\end{align*}
One can still sensibly ask whether $\overline{\widetilde{\epsilon}} \cdot \epsilon$ vanishes, leading to a two-branch classification. Thus, the terminology \textit{null class} and \textit{timelike class} can in principle be extended to Euclidean signature. However, the full Euclidean classification has not been carried out in sufficient detail; see however \cite{Sabra:2016abd,BenettiGenolini:2016qwm,BenettiGenolini:2016tsn,BenettiGenolini:2025icr} for noteworthy attempts in 5d. For this reason we will remain in Lorentzian signature and rely on the intuition that a Lorentzian supersymmetric solution
can always be analytically continued to a Euclidean supersymmetric saddle.}.

Since the supersymmetric extremal black string solution of \cite{Klemm:2000nj} has been 
shown in \cite{Gauntlett:2003fk} to fall into the null class, it is reasonable to expect that its non-extremal deformation should fall in the same class. 
One can then recast our ansatz \eqref{eq:ansatz_5d} in the null form and impose the supersymmetry conditions. 
It is then simple to integrate the equation for the function $H$, which turns out to be unchanged with respect to that of the extremal 
solution of \cite{Klemm:2000nj}. The solution for the function ${\cal F}$ is modified, but one can show that regular solutions all have an extremal horizon. 
The analysis was performed in \cite{Bernamonti:2007bu} and we therefore do not present the details and refer to that paper. 
Since the most general supersymmetric solution in the null class possessing the symmetries of our ansatz 
is necessarily extremal, we take this as an indication that, if it exists, a Lorentzian supersymmetric solution 
with a smoothly capping thermal IR region of the type $\mathbb{R}^{1,1} \times S^1_\varphi \times \Sigma_\mathfrak{g}$ --- or $\mathbb{R}^2 \times S^1_\varphi \times \Sigma_\mathfrak{g}$ in Euclidean signature --- must belong to the timelike class. In this section we vindicate this hypothesis, by constructing such a solution, perturbatively. For a sub-family, physically characterized by real entropy and charges, we will also confirm the existence of the solution numerically.

Even though we mostly work in Lorentzian signature, in order to directly employ the machinery of \cite{Gauntlett:2003fk}, we want to emphasize that any Lorentzian supersymmetric configuration (which might be complex in itself and might contain CTCs) admits the analytic continuation $t \rightarrow -\iu t_E$ to a complex Euclidean background.

Thus, to set the stage we briefly recall the geometry of Lorentzian timelike supersymmetric solutions of 5d gauged minimal supergravity. Locally, the field configuration must take the form of a $\text{U}\qty(1)_y$ fibre over a \textit{K\"ahler} base $B$ 
\begin{align}
  \begin{aligned}
  \dd[]{s}^2 ={}& - f^2 \qty(\dd[]{y} + \varpi)^2 + f^{-1} \dd[]{s}^2_B \,, \\
  \mathcal{A} ={}& f(\dd[]{y} + \varpi) + \frac{\ell_5}{3} \mathcal{P} + \dd \Lambda \,, 
  \end{aligned}
\end{align}
where $\mathcal{P}$ is the potential for the Ricci form $\mathcal{R} = \dd[]{\mathcal{P}}$ on the base, $f$ and $\varpi$ are a function and a one-form on the base, respectively, and $\dd \Lambda$ is (locally) a gauge transformation. In these coordinates $\partial_y$ is referred to as the timelike ``supersymmetric Killing vector'' as it arises as a bilinear in the Killing spinors. 
The K\"ahler base may be conveniently described 
in terms of an almost hyper-K\"ahler structure, namely an $\text{SU}\qty(2)$ triplet of real two-forms $\mathcal{X}^i$, $i = 1,2,3$, satisfying
\begin{align}\label{eq:constraints_Kahler_geometry}
  \begin{aligned}
  (\mathcal{X}^{i})\indices{_m^p} (\mathcal{X}^{j})\indices{_p^n} ={}& - \delta^{ij} \delta\indices{_m^n} + \varepsilon^{ijk} (\mathcal{X}^{k})\indices{_m^n} \,, \\
  \star_B ={}& - \frac{1}{2} \mathcal{X}^1 \wedge \mathcal{X}^1 = - \frac{1}{2} \mathcal{X}^2 \wedge \mathcal{X}^2 = - \frac{1}{2} \mathcal{X}^3 \wedge \mathcal{X}^3 \,, \\
  \star_B \mathcal{X}^{i} ={}& - \mathcal{X}^{i} \,, \\
  \dd[]{\mathcal{X}^1} ={}& 0 \,, \\
  (\nabla_m + \iu \mathcal{P}_m) (\mathcal{X}^2 + \iu \mathcal{X}^3)\indices{_n_p} ={}& 0 \,, \\
  \mathcal{R}_{mn} ={}& \frac{1}{2} (R_B)_{mnpq} (\mathcal{X}^1)^{pq} \,, 
  \end{aligned}
\end{align}
where $R_B$ denotes the Ricci scalar on the base and $m,n, \dots $ are the indices on the base. Further, supersymmetry constrains the function $f$ and the one-form $\varpi$ as
\begin{align}\label{eq:susy_f_omega}
  f ={}& - \frac{24}{\ell_5^2 R_B} \,, \quad \dd[]{\varpi} + \star_B \dd[]{\varpi} = - \frac{\ell_5}{f} \qty(\mathcal{R} - \frac{1}{4} R_B \mathcal{X}^1) \,. 
\end{align}

Building on this general supersymmetric geometry, we now impose the presence of a $\Sigma_{\mathfrak{g}}$ factor throughout the flow, arriving at a refined supersymmetric ansatz compatible with~\eqref{eq:ansatz_5d}
\begin{align}\label{eq:refined_susy_ansatz}
  \begin{aligned}
  \dd[]{s}^2 ={}& - f^2 \qty(\dd[]{y} + \varpi_\zeta \dd[]{\zeta})^2 + f^{-1} \dd[]{s}_B^2 \,, \\
  \dd[]{s}^2_B ={}& H_{rr}^2 \dd[]{r}^2 + H_{\zeta \zeta}^2 \dd[]{\zeta}^2 + H_\Sigma^2 \qty(\dd[]{\theta}^2 + \sinh^{2}{\theta} \dd[]{\phi}^2) \,, \\
  \mathcal{A} ={}& f(\dd[]{y} + \varpi_\zeta \dd[]{\zeta}) + \frac{\ell_5}{3} \mathcal{P} + a_y \dd[]{y} + a_\zeta \dd[]{\zeta} \,,
  \end{aligned}
\end{align}
where $\qty{f,H_{rr},H_{\zeta\zeta},H_\Sigma, \varpi_\zeta}$ are now functions of the radial coordinate $r$ alone, $\qty{a_y, a_\zeta}$ are constant gauge shifts, and we employ explicit local coordinates on $\Sigma_{\mathfrak{g}}$ as in \eqref{eq:Sigma_explicit}. A not so tedious calculation reveals that the geometric constraints \eqref{eq:constraints_Kahler_geometry} are satisfied by
\begin{align}
  \begin{aligned}
  H_{\Sigma} ={}& C = \text{const} \,, \\
  \mathcal{X}^1 ={}& - H_{rr} H_{\zeta\zeta} \dd[]{r} \wedge \dd[]{\zeta} + C^2 \sinh^{}{\theta} \dd[]{\theta} \wedge \dd[]{\phi} \,, \\
  \mathcal{X}^2 ={}& - C \qty(H_{rr} \dd[]{r} \wedge \dd[]{\theta} + H_{\zeta \zeta} \sinh^{}{\theta} \dd[]{\zeta} \wedge \dd[]{\phi}) \,, \\
  \mathcal{X}^3 ={}& C \qty(H_{rr} \sinh^{}{\theta} \dd[]{r} \wedge \dd[]{\phi} - H_{\zeta\zeta} \dd[]{\zeta} \wedge \dd[]{\theta}) \,, \\
  \mathcal{P} ={}& \frac{H_{\zeta\zeta}'}{H_{rr}} \dd[]{\zeta} - \cosh^{}{\theta} \dd[]{\phi} \,. 
  \end{aligned}
\end{align}
Note that supersymmetry in the bulk automatically implements the topological twist
\begin{align}
  \mathfrak{p} ={}& \mathfrak{p}_{\text{top. twist}} = \frac{\ell_5}{3} \,, 
\end{align}
on the boundary. This follows from a direct comparison of the terms proportional to $\omega_\Sigma = \cosh^{}{\theta} \dd[]{\phi}$ in the gauge field in \eqref{eq:refined_susy_ansatz} with those in \eqref{eq:ansatz_5d}. Importantly, this identification simplifies the analysis of \cref{sec:asymptotics} across the entire bulk flow. Finally, the constraints on $f$ and $\varpi$ in \eqref{eq:susy_f_omega} become
\begin{align}\label{eq:susy_f_omega_explicit}
  f ={}& - \frac{24}{\ell_5^2 \qty( - \frac{2}{C^2} + \frac{2 H_{rr}' H_{\zeta\zeta}' - 2 H_{rr} H_{\zeta\zeta}''}{H_{rr}^3 H_{\zeta\zeta}})} \,, \quad \varpi_\zeta' = \frac{(- 6 C^2 + \ell_5^2 f) H_{rr} H_{\zeta\zeta}}{C^2 \ell_5 f^2} \,. 
\end{align}

The next step is to relate the non-supersymmetric parametrization of the thermal black string in \eqref{eq:ansatz_5d} to the supersymmetric parametrization in \eqref{eq:refined_susy_ansatz}. To that end we relate the coordinates $(t, \varphi)$ to $(y,\zeta)$ as
\begin{align}
  t ={}& y \,, \quad \varphi = \zeta + c_0 y \,,
\end{align}
where $c_0$ is a constant that will soon be fixed\footnote{
The most general coordinate transformation one can consider is
\begin{align*}
  \mqty(t \\ \varphi) ={}& \mqty(a_0 & b_0 \\ c_0 & d_0) \mqty(y \\ \zeta) \,,
\end{align*}
with $a_0 d_0 - b_0 c_0 \neq 0$. By rescaling $\qty(y, \zeta)$ one can always set $a_0 = 1 = d_0$. The fact that one can also set $b_0 = 0$ is non-trivial to see. We performed the calculation that follows with a generic $b_0 \neq 0$ and found no difference to the final result for the thermodynamic quantities and the geometry --- the upshot of that analysis is that there is a $b_0$-independent combination of $(q, a_\zeta)$ that appears in all relevant expressions, despite $q$ and $a_\zeta$ depending on $b_0$, separately. For brevity in the main text we choose to set $b_0 = 0$. Note that generically $(y, \zeta)$ do not have closed orbits, while $(t_E = \iu t, \varphi)$ do. 
}. Matching the metrics in \eqref{eq:ansatz_5d} and \eqref{eq:refined_susy_ansatz} leads to
\begin{align}\label{eq:met_fn_non-susy_2_susy}
  \begin{aligned}
    g_{tt} ={}& \frac{f^2 H_{\zeta\zeta}^2}{H_{\zeta\zeta}^2 - f^3 \varpi_\zeta^2} \,, & f ={}& \frac{C^2}{g_\Sigma} \,, \\
    g_{rr} ={}& \frac{H_{rr}^2}{f} \,, & H_{rr} ={}& C \sqrt{\frac{g_{rr}}{g_\Sigma}} \,, \\
    g_{\varphi\varphi} ={}& \frac{H_{\zeta\zeta}^2 - f^3 \varpi_\zeta^2}{f} \,, & H_{\zeta\zeta} ={}& C \sqrt{\frac{g_{\varphi\varphi}}{g_\Sigma} \qty(1 + \frac{g_\Sigma^2 g_{\varphi\varphi} (w + c_0)^2}{C^4})} \,, \\
    g_\Sigma ={}& \frac{C^2}{f} \,, & H_\Sigma ={}& C \,, \\
    w ={}& -\frac{c_0 H_{\zeta\zeta}^2 + f^3 \varpi_\zeta(1 - c_0 \varpi_\zeta)}{H_{\zeta\zeta}^2 - f^3 \varpi_\zeta^2} \,, & \varpi_\zeta ={}& - \frac{g_\Sigma^2 g_{\varphi\varphi}(w + c_0)}{C^4} \,.
  \end{aligned}
\end{align}
In the supersymmetric ansatz \eqref{eq:refined_susy_ansatz} one of the five functions $\qty{f,H_{rr},H_{\zeta\zeta}, H_\Sigma, \varpi_\zeta}$ is a constant: $H_\Sigma = C$. In the non-supersymmetric ansatz this is reflected by the following non-linear relation between the five functions $\qty{g_{tt}, g_{rr}, g_{\varphi\varphi}, g_\Sigma, w}$: 
\begin{align}\label{eq:susy_constraint_special}
  C^4 - g_\Sigma^2(g_{tt} - g_{\varphi\varphi}(w + c_0)^2) ={}& 0 \,.
\end{align}
The above provides a necessary but not sufficient algebraic constraint between the expansion coefficients, derived in the previous sections, if one wants to impose supersymmetry. Matching the gauge fields in \eqref{eq:ansatz_5d} and \eqref{eq:refined_susy_ansatz} leads to
\begin{align}\label{eq:vec_fn_non-susy_2_susy}
  \begin{aligned}
    a_t ={}& \frac{3 H_{rr} \qty[(a_y + f) H_{\zeta\zeta}^2 + f^3 \varpi_\zeta(a_\zeta - a_y \varpi_\zeta)] + \ell_5 f^3 \varpi_\zeta H_{\zeta\zeta}'}{3H_{rr}(H_{\zeta\zeta}^2 - f^3 \varpi_\zeta^2)} \,, \\
    a_\varphi ={}& a_\zeta + f \varpi_\zeta + \frac{\ell_5 H_{\zeta\zeta}'}{3 H_{rr}} \,, \\
    \mathfrak{p} ={}& \frac{\ell_5}{3} \,. 
  \end{aligned}
\end{align}
Setting $\mathfrak{p} = \ell_5/3$ and converting $\qty{g_{tt}, g_{rr}, g_{\varphi\varphi}, g_\Sigma, w, a_t, a_\varphi}$ to $\qty{f,H_{rr}, H_{\zeta\zeta},H_\Sigma, \varpi_\zeta}$ via \cref{eq:met_fn_non-susy_2_susy,eq:vec_fn_non-susy_2_susy}, the entire system --- (i) the equations of motion \cref{eq:evolution,eq:constraint}, (ii) the IOM identities \cref{eq:iom_qj,eq:iom_mu}, and (iii) the differential supersymmetry constraints \eqref{eq:susy_f_omega_explicit} --- is satisfied provided the following first-order differential equations hold,
\begin{align}\label{eq:susy_diff}
  \begin{aligned}
    0 ={}& - \qty(f')^2 + f^2 H_{rr}^2 \qty( - \frac{8}{3 C^2} + \frac{4 \ell_5^2 f}{9 C^4}) \\
         &+ \frac{f^2 H_{rr}^2}{H_{\zeta\zeta}^2} \qty( \frac{(4j\ell_5 + 9 q^2)f^2}{9 C^4} + \frac{4(2a_\zeta \ell_5 + 3q) f^2 \varpi_\zeta}{3 C^2 \ell_5} + \frac{4 f^2 \varpi_\zeta^2}{\ell_5^2}) \,, \\
    0 ={}& - \qty((2 a_\zeta \ell_5 + 3q) + \frac{6 C^2 \varpi_\zeta}{\ell_5} + \frac{2 \ell_5^2 H_{\zeta\zeta}'}{3 H_{rr}})^2 + (4j \ell_5 + 9 q^2) + \frac{4 H_{\zeta\zeta}^2}{f} \qty(\ell_5^2 - \frac{6 C^2}{f}) \\
         &+ \frac{12 C^2(2a_\zeta \ell_5 + 3 q) \varpi_\zeta}{\ell_5} + \frac{36 C^4 \varpi_\zeta^2}{\ell_5^2} \,, \\
    0 ={}& - \varpi_\zeta' + \frac{H_{rr} H_{\zeta\zeta}}{C^2 \ell_5 f} \qty(\ell_5^2 - \frac{6 C^2}{f}) \,,
  \end{aligned}
\end{align}
together with the algebraic constraints
\begin{align}\label{eq:susy_algebraic}
  \begin{aligned}
    \widetilde{q} ={}& - c_0 q - \frac{2 C^2}{\ell_5} - \frac{2 a_y \ell_5}{3} \,, \\
    \mu_t ={}& c_0 j + \frac{3 C^2 a_\zeta}{\ell_5} - \frac{3 a_y q}{2} \,, \\
    \mu_s ={}& \frac{3 C^2 (a_y - a_\zeta c_0)}{\ell_5} + \frac{a_y^2 \ell_5}{2} + \frac{3a_y c_0 q}{2} + \frac{j}{2}\qty(1 - c_0^2) \,. 
  \end{aligned}
\end{align}

To summarize: so far we have worked in an arbitrary radial gauge. Fixing $g_{rr} = 1$ enforces $H_{rr} = \sqrt{f}$, and the task of constructing an explicit non-extremal supersymmetric black string reduces to solving the three coupled first-order ordinary differential equations \eqref{eq:susy_diff} for $\qty{f,H_{\zeta\zeta}, \varpi_\zeta}$. In terms of the original parametrization \eqref{eq:ansatz_5d} (with $g_{rr} = 1$), the functions $\qty{g_{tt}, g_{\varphi\varphi}, g_\Sigma, w, a_t, a_\varphi}$ are then read off from \cref{eq:met_fn_non-susy_2_susy,eq:vec_fn_non-susy_2_susy}. 

\subsection{Imposing supersymmetry in the UV and the IR}\label{sec:susy_in_the_UV_and_IR}
In this section we impose supersymmetry on the asymptotic (UV) solution of \cref{sec:asymptotics} and on the near-horizon (IR) solution of \cref{sec:near-horizon}. This is a delicate, multi-step procedure. For clarity we introduce the notation
\begin{align}
  \eval{\texttt{expression}}_{\order{\eu^{-\frac{n r}{\ell_5}}}} \,, \quad \eval{\texttt{expression}}_{\order{r^n}} \,, 
\end{align}
to mean that \texttt{expression} is evaluated to \textit{and including} $n^{\text{th}}$ order in the UV and $n^{\text{th}}$ order in the IR, respectively.

\paragraph{Algebraic constraints}
First, we impose the constraint \eqref{eq:susy_constraint_special} at leading order in the UV and IR:
\begin{align}\label{eq:invertability_UV_and_IR}
  \begin{aligned}
    &\eval{\eqref{eq:susy_constraint_special}}_{\order{\eu^{\frac{4 r}{\ell_5}}}} & &\implies & c_0 ={}& \mathfrak{s}_{c_0} \frac{c_t}{c_\varphi \ell_5} + \Omega_\infty \,, \\
    &\eval{\eqref{eq:susy_constraint_special}}_{\order{r^0}} & &\implies & \mu_t ={}& j \Omega_H + \mathfrak{s}_{\mu_t} \frac{2\pi\iu \, C^2}{\beta \qty(\mathfrak{s}_{c_0} \frac{c_t}{c_\varphi \ell_5} - \Omega)} \,, \\
  \end{aligned}
\end{align}
where $\mathfrak{s}_{c_0}, \mathfrak{s}_{\mu_t} = \pm 1$ are independent sign choices. Note the explicit appearance of $\iu$ --- already at this stage there
 is a clue of complexification. From the IR regularity conditions \eqref{eq:IR_iom} we obtain $\mu_s + \mu_t \Omega_H = \frac{j}{2} \qty(1 + \Omega_H^2)$ by eliminating $L_\Sigma^2 L_\varphi$. Then
\begin{align}
  \begin{aligned}
  \mu_s ={}& \frac{j}{2}(1 - \Omega_H^2) - \mathfrak{s}_{\mu_t} \frac{2\pi\iu \, C^2 \, \Omega_H}{\beta \qty(\mathfrak{s}_{c_0} \frac{c_t}{c_\varphi \ell_5} - \Omega)} \,. 
  \end{aligned}
\end{align}
Thus, we must keep track of the four branches
\begin{align}
  \qty(\mathfrak{s}_{c_0}, \mathfrak{s}_{\mu_t}) = \qty{\qty(++), \qty(+-), \qty(-+), \qty(--)} \,. 
\end{align}
In each branch, we solve the algebraic supersymmetry constraints \eqref{eq:susy_algebraic} for $\qty{q,j}$ in terms of $\qty{a_\zeta, a_y}$, arriving at lengthy --- and at this stage not yet illuminating --- expressions for the IOMs $\qty{q, \widetilde{q}, j, \mu_s, \mu_t}$ in terms of 
\begin{align}\label{eq:after_algebraic_is_imposed}
    (C; \beta, \Omega_H, \Omega_\infty; \, c_t, c_\varphi; \, a_\zeta, a_y; \, \mathfrak{s}_{c_0}, \mathfrak{s}_{\mu_t}) \,. 
\end{align}

\paragraph{Supersymmetry in the UV} Next, we use the relations \eqref{eq:met_fn_non-susy_2_susy} to determine the UV expansion of the supersymmetric functions
\begin{align}
  \begin{aligned}
    f ={}& f^{(2,0)} \eu^{-\frac{2r}{\ell_5}} + f^{(4,0)} \eu^{-\frac{4r}{\ell_5}} + f^{(6,0)} \eu^{-\frac{6r}{\ell_5}} + \dots \,, \\
    H_{rr} ={}& H_{rr}^{(1,0)} \eu^{-\frac{r}{\ell_5}} + H_{rr}^{(3,0)} \eu^{-\frac{3r}{\ell_5}} + H_{rr}^{(5,0)} \eu^{-\frac{5r}{\ell_5}} + \dots \,, \\
    H_{\zeta\zeta} ={}& H_{\zeta\zeta}^{(-3,0)} \eu^{\frac{3r}{\ell_5}} + H_{\zeta\zeta}^{(-1,0)} \eu^{\frac{r}{\ell_5}} + H_{\zeta\zeta}^{(1,0)} \eu^{-\frac{r}{\ell_5}} + H_{\zeta\zeta}^{(3,0)} \eu^{-\frac{3r}{\ell_5}} + H_{\zeta\zeta}^{(5,0)} \eu^{-\frac{5r}{\ell_5}} + \dots \,, \\
    \varpi_\zeta ={}& \varpi_\zeta^{(-6,0)} \eu^{\frac{6r}{\ell_5}} + \varpi_\zeta^{(-4,0)} \eu^{\frac{4r}{\ell_5}} + \varpi_\zeta^{(-2,0)} \eu^{\frac{2r}{\ell_5}} + \varpi_\zeta^{(0,0)} + \varpi_\zeta^{(2,0)} \eu^{- \frac{2r}{\ell_5}} + \dots \,, 
  \end{aligned}
\end{align}
and relate the coefficients appearing above to the coefficients $g_{\bullet}^{(n,0)}$ evaluated at $\mathfrak{p} = \ell_5/3$. For example, the explicit leading-order coefficients read
\begin{align}\label{eq:UV_susy_coeff_leading}
  \begin{aligned}
  f^{(2,0)} ={}& \frac{C^2}{c_\Sigma^2 \ell_5^2} \,, & H_{rr}^{(1,0)} ={}& \frac{C}{c_\Sigma \ell_5} \,, \\
  H_{\zeta\zeta}^{(-3,0)} ={}& \mathfrak{s}_{c_0} \frac{c_t c_\varphi c_\Sigma \ell_5^2 }{C} \,, & \varpi_\zeta^{(-6,0)} ={}& - \mathfrak{s}_{c_0} \frac{c_t c_\varphi c_\Sigma^4 \ell_5^5}{C^4} \,. 
  \end{aligned}
\end{align}
We then place the differential supersymmetry relations \eqref{eq:susy_diff} and the Maxwell supersymmetry relations \eqref{eq:vec_fn_non-susy_2_susy} over a common denominator (to facilitate easier order-by-order matching) and substitute the expansions for $\qty{f, H_{rr}, H_{\zeta\zeta}, \varpi_\zeta}$ together with the IOM expressions \eqref{eq:after_algebraic_is_imposed}. Validity of \cref{eq:susy_diff,eq:vec_fn_non-susy_2_susy} to $\order{\eu^{- \frac{r}{\ell_5}}}$ fixes
\begin{align}\label{eq:as_UV}
  a_\zeta ={}& - \Psi_\infty \,, \quad a_y = - \Phi - \Psi_\infty \qty(\mathfrak{s}_{c_0} \frac{c_t}{c_\varphi \ell_5} - \Omega) \,. 
\end{align}
With the above relations imposed, the constraint \eqref{eq:susy_constraint_special}, to $\order{\eu^{- \frac{2r}{\ell_5}}}$, is satisfied only for the $\qty{(++),(+-)}$ branches. The $\qty{(-+),(--)}$ branches are therefore eliminated. Without loss of generality, we focus from now on on the $(++)$ branch, since the $(+-)$ branch is physically equivalent, as we shall see below. The final step is to observe that all remaining relations --- the maps \cref{eq:met_fn_non-susy_2_susy,eq:vec_fn_non-susy_2_susy}, the constraint \eqref{eq:susy_constraint_special}, and the differential supersymmetry constraints \eqref{eq:susy_diff} --- hold to $\order{\eu^{- \frac{4r}{\ell_5}}}$ if and only if
\begin{align}
  g_\Sigma^{(4,0)} ={}& 0 \,. 
\end{align}
The above implies that the Casimir contribution $\mathcal{E}$, identified in \eqref{eq:casimir}, vanishes. Thus, supersymmetry dynamically sets the v.e.v. of the holographic stress energy tensor along the $\Sigma_{\mathfrak{g}}$ direction to zero: $\langle \mathcal{T}\indices{^\theta_\theta} \rangle = \langle \mathcal{T}\indices{^\phi_\phi} \rangle = 0$, see \eqref{eq:stress-energy_explicit}. Now the IOMs $\qty{q, \widetilde{q}, j, \mu_s, \mu_t}$ are UV fixed in terms of
\begin{align}\label{eq:after_susy_UV_is_imposed}
    (C; \beta, \Omega_H, \Omega_\infty, \Phi, \Psi_\infty; \, c_t, c_\varphi) \,. 
\end{align}

\paragraph{Supersymmetry in the IR} 
Next, we examine the supersymmetry constraints in the IR. The IR expansions of the supersymmetric functions take the form
\begin{align}
  \begin{aligned}
    f ={}& f^{(0)} + f^{(2)} r^2 + \dots \,, \\
    H_{rr} ={}& H_{rr}^{(0)} + H_{rr}^{(2)} r^2 + \dots \,, \\
    H_{\zeta\zeta} ={}& H_{\zeta\zeta}^{(1)} r + H_{\zeta\zeta}^{(3)} r^3 + \dots \,, \\
    \varpi_{\zeta} ={}& \varpi_\zeta^{(0)} + \varpi_\zeta^{(2)} r^2 + \dots \,. 
  \end{aligned}
\end{align}
At $\order{r^0}$ the relations \eqref{eq:met_fn_non-susy_2_susy} determine
\begin{align}\label{eq:IR_susy_coeff_leading}
  f^{(0)} ={}& \frac{C^2}{L_\Sigma^2} \,, \quad H_{rr}^{(0)} = \frac{C}{L_\Sigma} \,, \quad \varpi_\zeta^{(0)} = \frac{1}{\frac{c_t}{c_\varphi \ell_5} - \Omega} \,. 
\end{align}
At $\order{r^0}$, the constraint \eqref{eq:susy_constraint_special} has already been imposed in \eqref{eq:invertability_UV_and_IR}. The Maxwell supersymmetry relations \eqref{eq:vec_fn_non-susy_2_susy} and the differential supersymmetry relations \eqref{eq:susy_diff} dictate
\begin{align}\label{eq:as_IR}
  \begin{aligned}
  a_\zeta ={}& - \frac{\frac{C^2}{L_\Sigma^2} + \frac{2\pi\iu \ell_5}{3\beta}}{\frac{c_t}{c_\varphi \ell_5} - \Omega} - \Psi_H \,, \\
  a_y ={}& - \frac{C^2}{L_\Sigma^2} - \Psi_H \qty(\frac{c_t}{c_\varphi \ell_5} - \Omega_H) \,, \\ 
  H_{\zeta\zeta}^{(1)} ={}& \frac{\frac{2\pi\iu C}{L_\Sigma \beta}}{\frac{c_t}{c_\varphi \ell_5} - \Omega_H} \,. 
  \end{aligned}
\end{align}
Proceeding to $\order{r^2}$, the relations \cref{eq:susy_constraint_special,eq:vec_fn_non-susy_2_susy,eq:susy_diff} fix $\qty{f^{(2)}, H_{rr}^{(2)}, H_{\zeta \zeta}^{(3)}, \varpi_\zeta^{(2)}}$ generating no new constraints between the boundary data parameters. Analogously, we can proceed to any desired order in the IR expansion without generating new constraints among the boundary data parameters; practically we have tested up to 10th order.

\paragraph{Linking the UV and the IR} The final, and arguably most important, step is to implement a UV-IR matching condition stemming from the fact that the gauge shift constants $\qty{a_\zeta, a_y}$, first introduced in \eqref{eq:refined_susy_ansatz}, are constant throughout the flow 
\begin{align}
  a_\zeta = a_{\zeta}^{\text{UV}} = a_{\zeta}^{\text{IR}} \,, \quad a_y = a_{y}^{\text{UV}} = a_{y}^{\text{IR}} \,.
\end{align}
Equating the expressions for $\qty{a_\zeta, a_y}$ found in \eqref{eq:as_UV} and \eqref{eq:as_IR} yields
\begin{align}\label{eq:matching_condition}
  \Psi \equiv \Psi_H - \Psi_\infty = \frac{\Phi - \frac{C^2}{L_\Sigma^2}}{\frac{c_t}{c_\varphi \ell_5} - \Omega} \,, \quad \beta \Phi = - \mathfrak{s}_{\mu_t} 2\pi\iu \frac{\ell_5}{3} \,,
\end{align}
where we have temporarily reinstated the sign $\mathfrak{s}_{\mu_t}$ to make manifest that the $(++)$ and $(+-)$ branches describe the same physics; otherwise we proceed in the $(++)$ branch\footnote{Note that the $\qty{(++),(+-)}$ branches correspond to the $\qty{-,+}$ signs in \cref{sec:convenient_vars}, respectively.}. The first condition is a new UV-IR relation stemming from supersymmetry, while the second condition is the \textit{supersymmetric linear constraint} for our black string\footnote{This is the analogue of $\beta\qty(\frac{1}{\ell_5} + \Omega_1 + \Omega_2 - \frac{3\Phi}{\ell_5}) = \pm 2\pi\iu$, obtained for the CCLP black hole \cite{Chong:2005hr} --- the gravity dual of the superconformal index on $S^1 \times S^3$ \cite{Cabo-Bizet:2018ehj}.}. 
We are not done fixing all the freedom since the equations \eqref{eq:IR_iom} have not yet been imposed. The first one is trivially satisfied, and the last two yield the same link between the constant $C^2$ and the IR boundary data
\begin{align}\label{eq:Crelation}
  C^2 ={}& -\iu L_\Sigma^2 L_\varphi \qty(\frac{c_t}{c_\varphi \ell_5} - \Omega) \,. 
\end{align}
We find it convenient to convert to the variables of \cref{sec:convenient_vars}. After eliminating $C^2$ through~\eqref{eq:Crelation}, the two supersymmetric conditions become
\begin{align}\label{eq:two_susy_conditions}
  \ell_\varphi ={}& - \frac{2\pi}{3\omega} - \iu \psi \,, \quad \Delta = - \frac{2\pi\iu}{3} \,,
\end{align}
where, as a reminder,
\begin{align}
  \psi ={}& \psi_H - \psi_\infty = \frac{1}{\ell_5} (\Psi_H - \Psi_\infty) \,, 
\end{align}
is the gauge invariant difference between the IR and UV holonomies along the $S^1_\varphi$. As a consistency check we note that the geometric quantity $\ell_\varphi$ indeed comes out gauge invariant.

\subsection{Thermodynamics of the supersymmetric black string}\label{sec:susy_thermo}
Recall that the non-supersymmetric thermodynamic quantities are given by
\begin{align}
  \begin{aligned}
    \vb{I} ={}&
    - \frac{9}{8} \frac{\widehat{\mathfrak{p}} \Delta^2}{\omega + \widetilde{\omega}} \mathfrak{c} 
    + \frac{\omega \widetilde{\omega}}{\omega + \widetilde{\omega}} J \,, \\
    \vb{L} ={}& 
    - \frac{9}{8} \frac{\widehat{\mathfrak{p}} \Delta^2}{\omega^2 - \widetilde{\omega}^2} \mathfrak{c} 
    - \frac{\widetilde{\omega}^2}{\omega^2 - \widetilde{\omega}^2} J \,, \\
    \widetilde{\vb{L}} ={}& 
    - \frac{9}{8} \frac{\widehat{\mathfrak{p}} \Delta^2}{\omega^2 - \widetilde{\omega}^2} \mathfrak{c} 
    - \frac{\omega^2}{\omega^2 - \widetilde{\omega}^2} J \,, \\
    \vb{Q} ={}& \frac{1}{\Delta}\qty(\frac{3\pi}{2} \ell_\Sigma^2 \ell_\varphi - \frac{9}{4} \frac{\widehat{\mathfrak{p}} \Delta^2}{\omega + \widetilde{\omega}}) \mathfrak{c} + \frac{2}{\Delta} \frac{\omega \widetilde{\omega}}{\omega + \widetilde{\omega}} J \,, \\
  \end{aligned}
\end{align}
where the right hand sides are gauge invariant and Casimir-subtracted. For the supersymmetric case the Casimir subtraction is immaterial since $\mathcal{E} = 0$ is dynamically imposed. Imposing the topological twist $\widehat{\mathfrak{p}} = 1/3$, the supersymmetric expressions for the IOMs in the UV \eqref{eq:after_susy_UV_is_imposed}, the relation for $C^2$ \eqref{eq:Crelation} and the two supersymmetric relations \eqref{eq:two_susy_conditions}, the expression for the angular momentum collapses to
\begin{align}
  J ={}& \frac{\pi \eta_\Sigma}{4G_5} \qty(j + 3(q - \mathfrak{p} \Psi_\infty) \Psi_\infty) \overset{\text{susy}}{=} \frac{\pi^2}{6 \omega^2} \mathfrak{c} \,, 
\end{align}
and the thermodynamic quantities simplify to
\begin{align}\label{eq:susy_thermo}
  \vb{I} ={}& \frac{\pi^2}{6\omega} \mathfrak{c} \,, \quad \vb{L} = \frac{\pi^2}{6 \omega^2} \mathfrak{c} \,, \quad \widetilde{\vb{L}} = 0 \,, \quad \vb{Q} = \frac{\iu}{4\omega} \qty[2\pi - 9 \ell_\Sigma^2 \qty(\frac{2\pi}{3} + \iu \psi \omega)] \mathfrak{c} \,. 
\end{align}
We see that indeed supersymmetry forces $\widetilde{\vb{L}} = 0$ and makes the thermodynamic quantities purely holomorphic. The entropy is obtained from the quantum statistical relation as
\begin{align}
  \mathcal{S} ={}& 2\pi \sqrt{\frac{\mathfrak{c}}{6} \vb{L}} - \frac{2\pi\iu}{3} \vb{Q} = - \frac{\pi^2}{3\omega} \mathfrak{c} + \frac{\pi}{6\omega} \qty[2\pi - 9 \ell_\Sigma^2 \qty(\frac{2\pi}{3} + \iu\psi\omega)] \mathfrak{c} \,.
\end{align}
This form of the thermodynamic expressions was anticipated in \cref{sec:convenient_vars}, except that now we have more detailed knowledge of the Maxwell charge $\vb{Q}$. Previously, it was abstractly given in terms of $\qty{\ell_\Sigma, \ell_\varphi, \omega}$ and now $\ell_\varphi = -\frac{2\pi}{3\omega} - \iu \psi$ has been related to $\qty{\omega, \psi}$. Note that in the above expression, in general all the parameters are complex and we work in conventions where
\begin{align}
  \Re(\omega) < 0 \quad \implies \quad \tau \in \text{upper half plane} \,. 
\end{align}
Sending $\vb{Q}$ to zero and enforcing $\Im(\omega) = 0$ is one way to guarantee real entropy, but it is not the only one. For example, any purely imaginary $\psi$ makes both $\ell_\varphi$ and the entropy real when $\Im(\omega) = 0$. As we shall see below, distinctively from the superconformal index, real entropy \textit{does not} imply extremality.

In terms of the modular parameter, our gauge invariant on-shell action
\begin{align}
  \vb I=-\frac{\pi\iu\mathfrak c}{12\tau} 
\end{align}
is in perfect agreement with $-\log Z_{T^2\times\Sigma_{\mathfrak g}}$ when
\begin{align}
  \log Z_{T^2 \times \Sigma_{\mathfrak{g}}} ={}& \frac{\iu\pi}{12\tau} c_l(\Delta, \widehat{\mathfrak{p}}) \,, 
\end{align}
evaluated for the universal twist $\Delta = - \frac{2\pi\iu}{3}$ and $\widehat{\mathfrak{p}} = \frac{1}{3}$. Indeed, for the universal twist
$c_l=\frac{32}{3}(\mathfrak g-1)a$, see \cref{sec:TTI_5d}, and using the standard holographic
dictionary $a=\frac{\pi\ell_5^3}{8G_5}$, valid to leading order at large $N$ for any 4d
$\mathcal N=1$ SCFT dual to minimal gauged supergravity, one finds \cite{Benini:2013cda,Benini:2012cz}
\begin{align}
  c_l=\frac{2\pi\eta_\Sigma}{3}\frac{\ell_5^3}{G_5}=\mathfrak c \,, 
\end{align}
as anticipated in \cref{sec:convenient_vars}. We emphasize again that no extremization is needed: the unique R-symmetry is
selected by default in minimal supergravity.

\subsection{Numerics}\label{sec:susy_numerics}
In this section we provide numerical evidence for the existence of the supersymmetric solutions, discussed above. First note that, in the $g_{rr} = 1$ radial gauge, our supersymmetric solutions have the following UV expansion 
\begin{align}\label{eq:susy_UV_exp}
\begin{split}
g_{tt} ={}& c_t^2\,\eu^{\frac{2r}{\ell_5}} - \frac{c_t^2}{6c_\Sigma^2}
 + \frac{c_t^2}{24 c_\Sigma^4}\qty(1-\frac{8c_\Sigma^2\pi^2}{3c_\varphi^2\omega^2})\eu^{-\frac{2r}{\ell_5}}\\
&- \frac{5c_t^2}{432 c_\Sigma^6}\Bigg[\,1 + \frac{8c_\Sigma^2\pi^2}{5c_\varphi^2\omega^2}
 - \qty(\frac{192 c_\Sigma^2\ell_\Sigma^2\pi}{5c_\varphi^2\omega^2} + \frac{32 c_\Sigma^2\ell_\Sigma^2\pi}{5 c_t c_\varphi\omega}\frac{\ell_5}{\beta})\qty(\tfrac{2\pi}{3}+\iu\psi\omega)\\
&\hphantom{-\frac{5c_t^2}{432 c_\Sigma^6}\Bigg[\,}
 + \qty(\frac{432 c_\Sigma^2\ell_\Sigma^4}{5c_\varphi^2\omega^2} + \frac{144 c_\Sigma^2\ell_\Sigma^4}{5 c_t c_\varphi\omega}\frac{\ell_5}{\beta} + \frac{72 c_\Sigma^2\ell_\Sigma^4}{5 c_t^2}\frac{\ell_5^2}{\beta^2})\qty(\tfrac{2\pi}{3}+\iu\psi\omega)^2\Bigg]\eu^{-\frac{4r}{\ell_5}} + \dots \,, \\
\frac{g_{\varphi\varphi}}{\ell_5^2} ={}& c_\varphi^2\,\eu^{\frac{2r}{\ell_5}} - \frac{c_\varphi^2}{6c_\Sigma^2}
 + \frac{c_\varphi^2}{24 c_\Sigma^4}\qty(1+\frac{8c_\Sigma^2\pi^2}{3c_\varphi^2\omega^2})\eu^{-\frac{2r}{\ell_5}}\\
&- \frac{5c_\varphi^2}{432 c_\Sigma^6}\Bigg[\,1 - \frac{8c_\Sigma^2\pi^2}{5c_\varphi^2\omega^2}
 + \qty(\frac{192 c_\Sigma^2\ell_\Sigma^2\pi}{5c_\varphi^2\omega^2} + \frac{32 c_\Sigma^2\ell_\Sigma^2\pi}{c_t c_\varphi\omega}\frac{\ell_5}{\beta})\qty(\tfrac{2\pi}{3}+\iu\psi\omega)\\
&\hphantom{-\frac{5c_\varphi^2}{432 c_\Sigma^6}\Bigg[\,}
 - \qty(\frac{432 c_\Sigma^2\ell_\Sigma^4}{5c_\varphi^2\omega^2} + \frac{144 c_\Sigma^2\ell_\Sigma^4}{c_t c_\varphi\omega}\frac{\ell_5}{\beta} + \frac{72 c_\Sigma^2\ell_\Sigma^4}{c_t^2}\frac{\ell_5^2}{\beta^2})\qty(\tfrac{2\pi}{3}+\iu\psi\omega)^2\Bigg]\eu^{-\frac{4r}{\ell_5}} + \dots \,, \\
\frac{g_{\Sigma}}{\ell_5^2} ={}& c_\Sigma^2\,\eu^{\frac{2r}{\ell_5}} + \frac13
 + \Bigg[\frac{2\pi\ell_\Sigma^2}{27 c_\Sigma^2 c_t c_\varphi\omega}\frac{\ell_5}{\beta}\qty(\tfrac{2\pi}{3}+\iu\psi\omega)\\
&\hphantom{c_\Sigma^2\,\eu^{\frac{2r}{\ell_5}} + \frac13 + \Bigg[}
 - \qty(\frac{\ell_\Sigma^4}{3 c_\Sigma^2 c_t c_\varphi\omega}\frac{\ell_5}{\beta} + \frac{\ell_\Sigma^4}{6 c_t^2 c_\Sigma^2}\frac{\ell_5^2}{\beta^2})\qty(\tfrac{2\pi}{3}+\iu\psi\omega)^2\Bigg]\eu^{-\frac{4r}{\ell_5}} + \dots \,, \\
w ={}& -\Omega_\infty - \frac{c_t}{c_\varphi\ell_5}\,\frac{\pi^2}{9 c_\Sigma^2 c_\varphi^2\omega^2}\,\eu^{-\frac{4r}{\ell_5}} + \dots \,, \\
\frac{a_t}{\ell_5} ={}& \frac{2\iu\pi}{3\beta} + \qty(\frac{c_t}{c_\varphi\ell_5}+\frac{\omega}{\beta})\psi_\infty
 - \iu\,\frac{c_t}{c_\varphi\ell_5}\qty(\frac{2\pi}{9 c_\Sigma^2\omega} - \frac{\ell_\Sigma^2}{c_\Sigma^2\omega}\qty(\tfrac{2\pi}{3}+\iu\psi\omega))\eu^{-\frac{2r}{\ell_5}}\\
&+ \iu\,\frac{c_t}{c_\varphi\ell_5}\qty(\iu\psi_\infty\,\frac{\pi^2}{9 c_\Sigma^2 c_\varphi^2\omega^2} + \frac{c_\varphi\ell_\Sigma^2}{6 c_t c_\Sigma^4}\qty(\tfrac{2\pi}{3}+\iu\psi\omega)\frac{\ell_5}{\beta})\eu^{-\frac{4r}{\ell_5}} + \dots \,, \\
\frac{a_\varphi}{\ell_5} ={}& -\psi_\infty
 + \iu\qty(\frac{2\pi}{9 c_\Sigma^2\omega} - \qty(\frac{\ell_\Sigma^2}{c_\Sigma^2\omega}+\frac{c_\varphi\ell_\Sigma^2}{c_t c_\Sigma^2}\frac{\ell_5}{\beta})\qty(\tfrac{2\pi}{3}+\iu\psi\omega))\eu^{-\frac{2r}{\ell_5}}\\
&+ \iu\,\frac{c_\varphi\ell_\Sigma^2}{6 c_t c_\Sigma^4}\qty(\tfrac{2\pi}{3}+\iu\psi\omega)\frac{\ell_5}{\beta}\,\eu^{-\frac{4r}{\ell_5}} + \dots \,, 
\end{split}
\end{align}
where the coefficients are carefully tuned in terms of
\begin{align}
  \qty{c_t, c_\varphi, c_\Sigma, \omega, \beta, \ell_\Sigma, \psi, \psi_\infty, \Omega_\infty} \,, 
\end{align}
such that all the supersymmetry constraints are satisfied, as explained in \cref{sec:susy_in_the_UV_and_IR}. The appearance of $\qty{\Omega_\infty, \psi_\infty}$ is due to $w$ not being frame invariant and $\qty{a_t, a_\varphi}$ not being gauge invariant. In the above formulation we use a mix of the old and new variables, see \cref{sec:convenient_vars}, where instead of $\qty{\beta, \Omega}$ or $\qty{\omega, \widetilde{\omega}}$ we use $\qty{\omega, \beta}$; the reason for that will become clear momentarily. Finally, the supersymmetry constraints \eqref{eq:two_susy_conditions} 
have been used to eliminate $\qty{\ell_\varphi, \Delta}$. Note that in practice, we have gone to 10th order.

Similarly, our supersymmetric solutions have the IR expansion
\begin{align}\label{eq:susy_IR_exp}
\begin{split}
g_{tt} ={}& \frac{4\pi^2}{\beta^2}\,r^2 + \dots \,, \\
\frac{g_{\varphi\varphi}}{\ell_5^2} ={}& \frac{\qty(\tfrac{2\pi}{3}+\iu\psi\omega)^2}{\omega^2}
 - \Bigg[\,\frac{4\pi^2}{\omega^2}
 - \frac{2\pi\qty(6-\tfrac{1}{\ell_\Sigma^2})\qty(\tfrac{2\pi}{3}+\iu\psi\omega)}{\omega^2}
 + \frac{\qty(54-\tfrac{21}{\ell_\Sigma^2}+\tfrac{1}{\ell_\Sigma^4})\qty(\tfrac{2\pi}{3}+\iu\psi\omega)^2}{9\,\omega^2}\,\Bigg]\frac{r^2}{\ell_5^2} + \dots \,, \\
\frac{g_\Sigma}{\ell_5^2} ={}& \ell_\Sigma^2 + \frac{\ell_\Sigma^2}{18}\qty(54-\frac{21}{\ell_\Sigma^2}+\frac{1}{\ell_\Sigma^4})\frac{r^2}{\ell_5^2} + \dots \,, \\
w ={}& -\Omega_H + \Bigg[\,\frac{\qty(6-\tfrac{1}{\ell_\Sigma^2})\pi\omega}{\tfrac{2\pi}{3}+\iu\psi\omega}
 - \frac{4\pi^2\omega}{\qty(\tfrac{2\pi}{3}+\iu\psi\omega)^2}\,\Bigg]\frac{1}{\beta}\,\frac{r^2}{\ell_5^2} + \dots \,, \\
\frac{a_t}{\ell_5} ={}& \Bigg[\,\frac{2\pi\iu\qty(3-\tfrac{1}{\ell_\Sigma^2})}{3\beta}
 + \psi_H\qty(\frac{\qty(6-\tfrac{1}{\ell_\Sigma^2})\pi\omega}{\tfrac{2\pi}{3}+\iu\psi\omega} - \frac{4\pi^2\omega}{\qty(\tfrac{2\pi}{3}+\iu\psi\omega)^2})\frac{1}{\beta}\,\Bigg]\frac{r^2}{\ell_5^2} + \dots \,, \\
\frac{a_\varphi}{\ell_5} ={}& -\psi_H + \Bigg[\,\frac{2\pi\iu\qty(3-\tfrac{1}{\ell_\Sigma^2})}{3\omega}
 - \frac{\iu\qty(54-\tfrac{21}{\ell_\Sigma^2}+\tfrac{1}{\ell_\Sigma^4})}{18\,\omega}\qty(\tfrac{2\pi}{3}+\iu\psi\omega)\,\Bigg]\frac{r^2}{\ell_5^2} + \dots \,, 
\end{split}
\end{align}
with the coefficients, tuned to their supersymmetric values, depending on
\begin{align}
  \qty{\omega, \beta, \ell_\Sigma, \psi, \psi_H, \Omega_H} \,. 
\end{align}
Once again, the dependence on $\qty{\psi_H, \Omega_H}$ is due to the non-frame invariance of $w$ and the non-gauge invariance of $\qty{a_t, a_\varphi}$, and, once again, in practice we have gone to 10th order.

The supersymmetric solution is generically complex, so how can we build a numerical shooter? The answer lies in the realization that the supersymmetric linear constraint
\begin{align}
  \Delta ={}& \frac{\beta \Phi}{\ell_5} = - \frac{2\pi\iu}{3} \,, 
\end{align}
does not mix the R-symmetric fugacity $\Delta$ with the $T^2$ fugacities $\qty{\omega, \widetilde{\omega}}$. Thus, there exists a special \textit{real supersymmetric sub-locus}, where we take
\begin{align}
  c_t, c_\varphi, c_\Sigma, \beta, \omega, \ell_\Sigma, \Omega_H, \Omega_\infty \in \mathbb{R} \,, \quad \psi_H, \psi_\infty \in \iu \mathbb{R} \,, 
\end{align}
ensuring that $\ell_\varphi = - \frac{2\pi}{3\omega} - \iu \psi \in \mathbb{R}$. Then the metric is purely real and the gauge field components orthogonal to the $\Sigma_{\mathfrak{g}}$, parametrized by $\qty{a_t, a_\varphi}$, are purely imaginary\footnote{Note that this is happening on the level of the Lorentzian supersymmetric solution
\begin{align*}
    \dd[]{s}^2 ={}& \underbrace{- g_{tt} \dd[]{t}^2 + g_{rr} \dd[]{r}^2 + g_{\varphi\varphi} \qty(\dd[]{\varphi} + w \dd[]{t})^2 + g_{\Sigma} \dd[]{s}^2_\Sigma}_{\mathbb{R}} \,, \quad \mathcal{A} = \underbrace{a_t \dd[]{t} + a_\varphi \qty(\dd[]{\varphi} + w \dd[]{t})}_{\iu \mathbb{R}} - \underbrace{\mathfrak{p} \, \omega_\Sigma}_{\mathbb{R}} \,. 
\end{align*}
Upon Wick rotation $t = -\iu t_E$ 
\begin{align*}
    \dd[]{s}^2 ={}& \underbrace{g_{tt} \dd[]{t_E}^2 + g_{rr} \dd[]{r}^2}_{\mathbb{R}} + \underbrace{g_{\varphi\varphi} \qty(\dd[]{\varphi} - \iu w \dd[]{t_E})^2}_{\mathbb{C}} + \underbrace{g_{\Sigma} \dd[]{s}^2_\Sigma}_{\mathbb{R}} \,, \quad \mathcal{A} = \underbrace{- \iu a_t \dd[]{t_E}}_{\mathbb{R}} + \underbrace{a_\varphi \qty(\dd[]{\varphi} - \iu w \dd[]{t_E})}_{\mathbb{C}} - \underbrace{\mathfrak{p} \, \omega_\Sigma}_{\mathbb{R}} \,,
\end{align*}
thus the Euclidean field configuration is complex unless we do further continuations to bring it on a real Euclidean section.}.
Said differently, on the level of the EOMs \cref{eq:evolution,eq:constraint} and the expansions \cref{eq:susy_UV_exp,eq:susy_IR_exp} we can analytically continue
\begin{align}
  \psi_H ={}& \iu \, \gamma_H \,, \quad \psi_\infty = \iu \, \gamma_\infty \,, \quad \psi = \iu \gamma \,, \quad a_t = -\iu \, \widetilde{a}_t \,, \quad a_\varphi = -\iu \, \widetilde{a}_\varphi \,, 
\end{align}
and then both the expansions and the EOMs are written in a manifestly real fashion.
Supersymmetry is ensured by the careful tuning of the coefficients, as described in detail in \cref{sec:susy_in_the_UV_and_IR}.

\begin{figure}[htbp]
  \centering
  \includegraphics[width=0.98\textwidth]{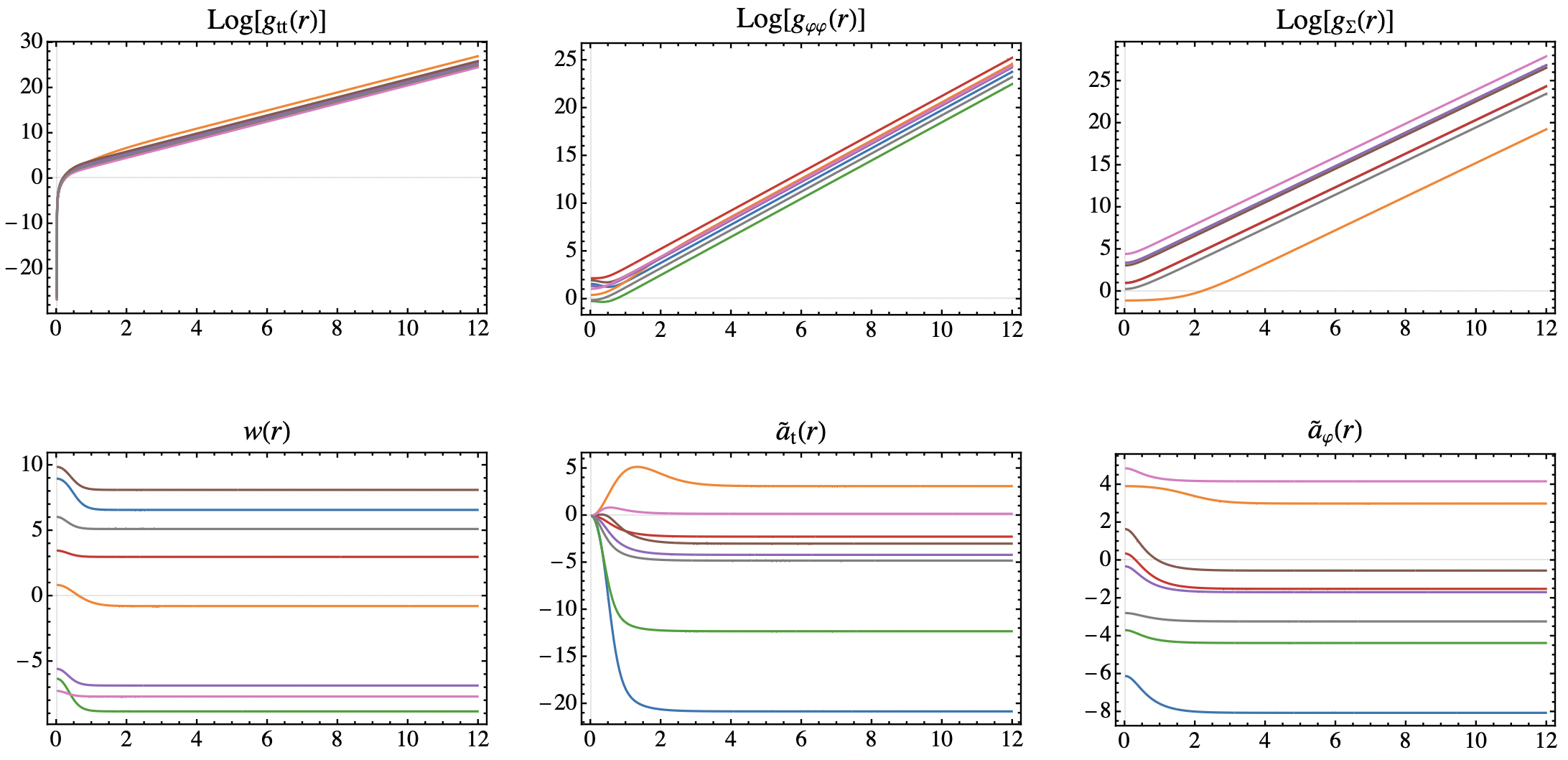}
  \caption{Radial profiles for a set of representative supersymmetric black strings.}
  \label{fig:susy_plots}
\end{figure}

With this preparation in mind we construct a supersymmetric shooter on the real supersymmetric sub-locus by supplying the initial conditions $\qty{\widetilde{\mathcal{G}}(\delta), \widetilde{\mathcal{G}}'(\delta)}$, where now $\widetilde{\mathcal{G}}(r) = \qty{g_{tt}(r), g_{\varphi\varphi}(r), g_{\Sigma}(r), w(r), \widetilde{a}_t(r), \widetilde{a}_\varphi(r)}$. As already alluded to, these initial conditions depend on the, now real, parameters
\begin{align}
  \qty{\omega, \beta, \ell_\Sigma, \gamma, \gamma_H, \Omega_H} \,,
\end{align}
as dictated by the IR expansion \eqref{eq:susy_IR_exp}.
Representative supersymmetric solutions are presented in \cref{fig:susy_plots}.

\noindent As before, the UV parameters are read off from the shooter as:
\begin{align}
  \begin{aligned}
  \log g_{tt} \sim{}& \frac{2r}{\ell_5} + b_{tt}^\infty \,, & \log g_{\varphi\varphi} \sim{}& \frac{2r}{\ell_5} + b_{\varphi\varphi}^\infty \,, & \log g_{\Sigma} \sim{}& \frac{2r}{\ell_5} + b_\Sigma^\infty \,, \\
  w \sim {}& w^\infty \,, & \widetilde{a}_t \sim{}& \widetilde{a}_t^\infty \,, & \widetilde{a}_\varphi \sim {}& \widetilde{a}_\varphi^\infty \,,
  \end{aligned}
\end{align}
where the constants $\qty{b_{tt}^\infty, b_{\varphi\varphi}^\infty, b_{\Sigma}^{\infty}}$ are determined by numerical linear fits near the boundary, and the constants $\qty{w^\infty, \widetilde{a}_{t}^\infty, \widetilde{a}_{\varphi}^{\infty}}$ are simply numerically read off near the boundary. From there, the UV boundary data is obtained as (see \cref{tab:susy_bb})
\begin{align}
  \begin{aligned}
  c_t = \eu^{\frac{b_{tt}^\infty}{2}} \,, \quad c_\varphi = \eu^{\frac{b_{\varphi\varphi}^\infty}{2}} \,, \quad c_\Sigma = \eu^{\frac{b_\Sigma^\infty}{2}} \,, \quad \Omega_\infty = - w^{\infty} \,, \quad \gamma_\infty = \frac{\widetilde{a}_\varphi^\infty}{\ell_5} \,. 
  \end{aligned}
\end{align}
\begin{table}
\definecolor{solA}{RGB}{31,119,180}% blue
\definecolor{solB}{RGB}{255,127,14}% orange
\definecolor{solC}{RGB}{44,160,44}% green
\definecolor{solD}{RGB}{214,39,40}% red
\definecolor{solE}{RGB}{148,103,189}% purple
\definecolor{solF}{RGB}{140,86,75}% brown
\definecolor{solG}{RGB}{227,119,194}% pink
\definecolor{solH}{RGB}{127,127,127}% gray
\newcommand{\sw}[1]{\textcolor{#1}{\rule{2.6ex}{1.4ex}}}%
\centering
\resizebox{\textwidth}{!}{%
\begin{tabular}{|c||c|c|c|c|c|c|c||c|c|c|c|c|c|}
\hline
\textbf{Color} & $\ell_5$ & $\beta$ & $\omega$ & $\Omega_H$ & $\gamma$ & $\gamma_H$ & $\ell_\Sigma$ & $c_t$ & $c_\varphi$ & $c_\Sigma$ & $\Omega_\infty$ & $\gamma_\infty$ & $\tilde{a}_t^{\infty}$ \\ \hline\hline
\sw{solA} & $1$ & $1.314$ & $-6.821$ & $-8.995$ & $1.945$ & $-6.092$ & $1.673$ & $2.605$ & $0.9297$ & $1.242$ & $-6.607$ & $-8.037$ & $-20.78$ \\ \hline
\sw{solB} & $1$ & $1.215$ & $-6.085$ & $-0.8698$ & $0.9192$ & $3.934$ & $0.5912$ & $4.747$ & $1.4$ & $0.09704$ & $0.7458$ & $3.015$ & $3.148$ \\ \hline
\sw{solC} & $1$ & $1.491$ & $-8.776$ & $6.3$ & $0.6766$ & $-3.672$ & $5.517$ & $1.662$ & $0.4904$ & $4.353$ & $8.798$ & $-4.349$ & $-12.27$ \\ \hline
\sw{solD} & $1$ & $1.391$ & $-1.8$ & $-3.493$ & $1.865$ & $0.3755$ & $1.69$ & $1.585$ & $1.939$ & $1.256$ & $-3.016$ & $-1.49$ & $-2.217$ \\ \hline
\sw{solE} & $1$ & $1.032$ & $-3.257$ & $5.547$ & $1.364$ & $-0.3017$ & $5.583$ & $2.203$ & $1.171$ & $4.406$ & $6.821$ & $-1.665$ & $-4.151$ \\ \hline
\sw{solF} & $1$ & $1.029$ & $-4.058$ & $-9.899$ & $2.191$ & $1.667$ & $4.76$ & $2.762$ & $1.266$ & $3.748$ & $-8.138$ & $-0.5246$ & $-2.958$ \\ \hline
\sw{solG} & $1$ & $1.314$ & $-1.999$ & $7.232$ & $0.6879$ & $4.878$ & $9.389$ & $1.38$ & $1.265$ & $7.437$ & $7.663$ & $4.19$ & $0.2108$ \\ \hline
\sw{solH} & $1$ & $1.16$ & $-4.026$ & $-6.071$ & $0.442$ & $-2.767$ & $1.172$ & $1.799$ & $0.7064$ & $0.8036$ & $-5.149$ & $-3.21$ & $-4.765$ \\ \hline
\end{tabular}%
}
\caption{Shooting data for eight representative supersymmetric black strings: the IR inputs $\{\ell_5,\beta,\omega,\Omega_H,\gamma,\gamma_H,\ell_\Sigma\}$ (left of the double rule) and the UV data $\{c_t,c_\varphi,c_\Sigma,\Omega_\infty,\gamma_\infty,\tilde{a}_t^{\infty}\}$ read off from the numerical flow (right), with $\gamma_\infty=\tilde{a}_\varphi^\infty/\ell_5$.}
\label{tab:susy_bb}
\end{table}
Thus, we have shot from the IR aiming along the direction of the real supersymmetric sub-locus. If the shooter remains on the real supersymmetric sub-locus, then the following consistency conditions must be satisfied
\begin{align}\label{eq:consistency_conditions}
  \omega ={}& \beta \qty(\Omega_H - \Omega_\infty - \frac{c_t}{c_\varphi\ell_5}) \,, \quad -\qty(\frac{\widetilde{a}_t^{\infty}}{\ell_5} + \frac{2\pi}{3\beta}) = \qty(\frac{\omega}{\beta} + \frac{c_t}{c_\varphi\ell_5}) \gamma_\infty \,. 
\end{align}
\begin{figure}[htbp]
  \centering
  \includegraphics[width=0.98\textwidth]{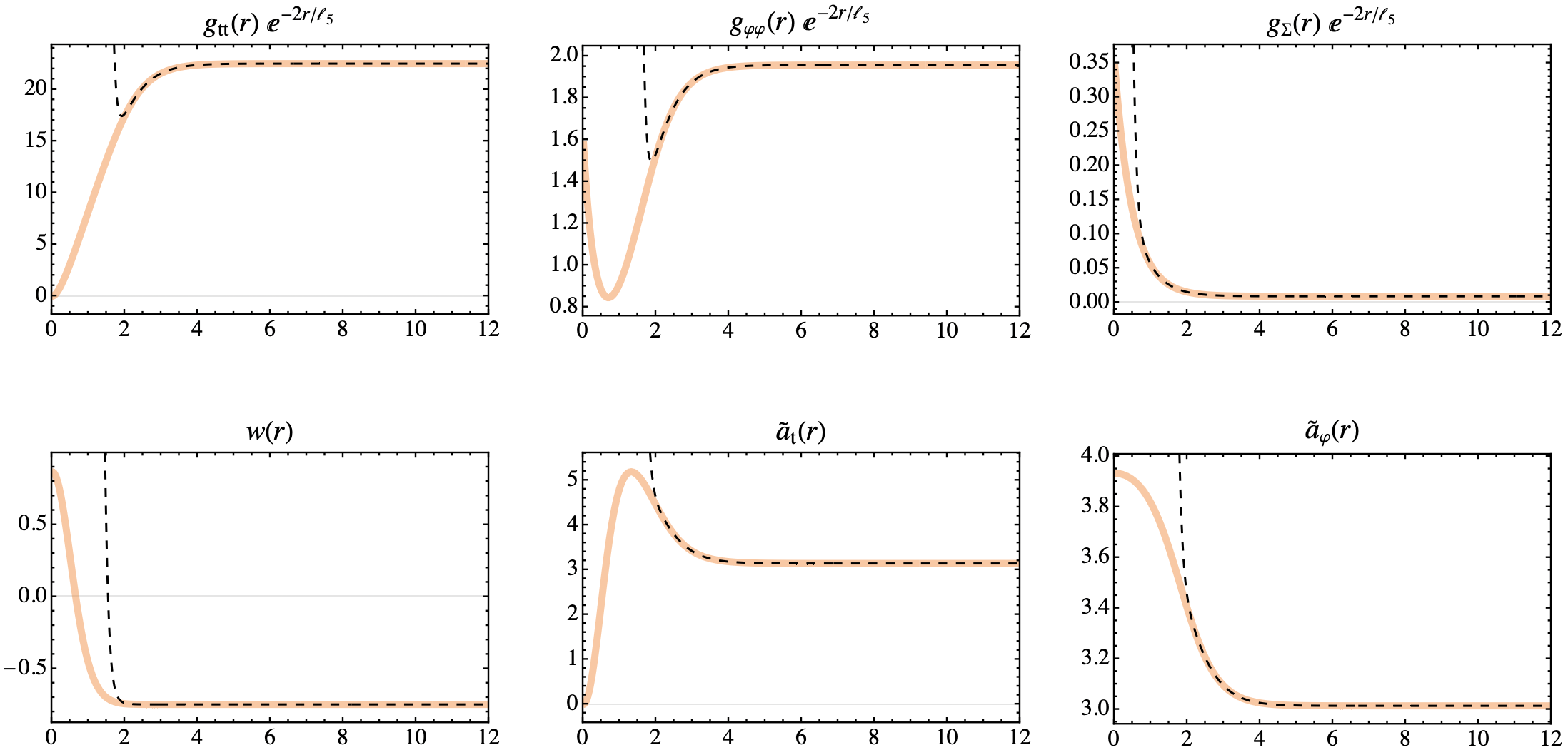}
  \caption{Comparison between radial profiles for one representative supersymmetric solution: exact UV asymptotics (dashed) vs numerical solution (solid colored).}
  \label{fig:uv_vs_numeric}
\end{figure}
The first condition is simply the definition of $\omega$ as in \cref{sec:convenient_vars}, and the second condition is read off from the leading term of $a_t$ in \eqref{eq:susy_UV_exp}. The consistency conditions \eqref{eq:consistency_conditions} are indeed satisfied $\qty(\text{to} \sim 10^{-8})$ when \texttt{WorkingPrecision} is set to 20. What is more, in \cref{fig:uv_vs_numeric} we overlay the exact supersymmetric UV expansion \eqref{eq:susy_UV_exp} on top of one of these solutions and observe close agreement $\qty(\text{to} \sim 10^{-8})$ near the boundary.

Finally, note that detailed analysis of a table similar to \cref{tab:susy_bb}, but for a vastly larger number of representative solutions, reveals that
\begin{align}\label{eq:cSigma_to_lSigma}
  c_\Sigma^2 ={}& \mathcal{C}(\ell_\Sigma) \qty(\ell_\Sigma^2 - \frac{1}{3}) \,,
\end{align}
where $\mathcal{C}(\ell_\Sigma)$ is a function that we could not determine in closed form reliably. Thus, the UV breathing mode of the $\Sigma_{\mathfrak{g}}$ is solely determined by the IR size of the $\Sigma_{\mathfrak{g}}$, and something special occurs when $\ell_\Sigma = \frac{1}{\sqrt{3}}$. That special case will be discussed in the next section.

\subsection{Analytical limits of our supersymmetric solution}\label{sec:analytical_limits}
In certain limits our non-extremal supersymmetric solution simplifies and in fact can be written in closed form, analytically. The first special case is the limit to extremality; it is related to (4.17--4.18) of \cite{Bernamonti:2007bu}. The second special case is the limit $\ell_\Sigma \rightarrow \frac{1}{\sqrt{3}}$; it is related to a certain (complex) finite $\beta$ supersymmetric $\text{BTZ} \times \Sigma_{\mathfrak{g}}$ with Wilson lines along $\mathbb{R}_t \times S^1_\varphi$.

\paragraph{Extremal limit}
The first limit we study is that of extremality. It is obtained as
\begin{align}
  \beta \rightarrow {}& \infty \,, \quad \Omega \rightarrow \frac{c_t}{c_\varphi \ell_5} \,, 
\end{align}
at the same rate. Or equivalently
\begin{align}
  \omega ={}& \beta\qty(\Omega - \frac{c_t}{c_\varphi \ell_5}) = \text{fixed} \,, \quad \widetilde{\omega} = \beta\qty(\Omega + \frac{c_t}{c_\varphi \ell_5}) \rightarrow \infty \,. 
\end{align}
In that limit the IR expansion \eqref{eq:susy_IR_exp} is nonsensical --- the capped-off region simply does not exist. However, the UV expansion \eqref{eq:susy_UV_exp} is well behaved. Indeed in the $\qty{\beta,\omega}$ variables, we simply send $\beta \rightarrow \infty$ at fixed $\omega$. Remarkably, the expansions for $\qty{g_\Sigma, a_\varphi}$ truncate and the remaining functions re-sum as:
\begin{align}\label{eq:resummed_extremal}
  \begin{aligned}
    g_\Sigma(r) ={}& c_\Sigma^2 \ell_5^2 \eu^{\frac{2r}{\ell_5}} + \frac{\ell_5^2}{3} \,, \\
    g_{tt}(r) ={}& \frac{3 c_t^2 c_\varphi^2 \ell_5^2 \omega^2 \, \eu^{\frac{4r}{\ell_5}}}{g_\Sigma(r) \, \texttt{poly}(r) + \ell_5 c_\Sigma \sqrt{g_\Sigma(r)} \qty(8\pi^2 + \frac{3c_\varphi^2 \omega^2}{c_\Sigma^2} - \frac{64}{35} a_0^2) \eu^{\frac{r}{\ell_5}} } \,, \\
    g_{\varphi\varphi}(r) ={}& \frac{c_t^2 c_\varphi^2 c_\Sigma^2 \ell_5^4 \, \eu^{\frac{6r}{\ell_5}}}{g_{tt}(r) g_\Sigma(r)} \,, \\
    w(r) ={}& - \Omega_\infty - \frac{c_t}{c_\varphi \ell_5} + \sqrt{\frac{g_{tt}(r)}{g_{\varphi\varphi}(r)}}\,, \\
    a_\varphi(r) ={}& - \psi_\infty \ell_5 + \iu \frac{a_0 \ell_5}{9 c_\Sigma^2 \omega} \eu^{- \frac{2r}{\ell_5}} \,, \\
    a_t(r) ={}& - a_\varphi(r) \sqrt{\frac{g_{tt}(r)}{g_{\varphi\varphi}(r)}} \,, 
  \end{aligned}
\end{align}
where
\begin{align}
  \begin{aligned}
  a_0 ={}& 2 \pi - 9 \ell_\Sigma^2 \qty(\frac{2\pi}{3} + \iu \psi \omega) \,, \\
  \texttt{poly}(r) ={}& 4\pi^2 \qty(\frac{\eu^{- \frac{2r}{\ell_5}}}{3c_\Sigma^2} - 2) + \qty(\frac{\eu^{- \frac{6r}{\ell_5}}}{63 c_\Sigma^6} + \frac{8 \eu^{- \frac{4r}{\ell_5}}}{105 c_\Sigma^4} - \frac{32 \eu^{- \frac{2r}{\ell_5}}}{105 c_\Sigma^2} + \frac{64}{35}) a_0^2 \,. 
  \end{aligned}
\end{align}
We directly verify that the above functions solve the EOMs \cref{eq:evolution,eq:constraint}, and we rely on the fact that we have re-summed a supersymmetric solution in the UV to claim that this configuration is supersymmetric. The above supersymmetric and extremal rotating dyonic solution can in fact be precisely mapped to (4.17--4.18) of \cite{Bernamonti:2007bu}. To see this first note that
\begin{align}
  g^{\text{there}} ={}& \frac{1}{\ell_5} \,, \quad A^{\text{there}} = \frac{\sqrt{3}}{2} \mathcal{A} \,. 
\end{align}
Then we perform the coordinate transformation
\begin{align}
  \begin{aligned}
  u ={}& \frac{c_t}{c_\Sigma} \qty(\frac{c_\varphi \ell_5}{c_t} \qty(\varphi - \Omega_\infty t) - t) \,, \\
  v ={}& \frac{c_t}{c_\Sigma} \qty(\frac{c_\varphi \ell_5}{c_t} \qty(\varphi - \Omega_\infty t) + t) \,, \\
  \rho ={}& \frac{\ell_5^2}{\sqrt{g_\Sigma}} \quad \implies \quad h = 1 - \frac{\rho^2}{3 \ell_5^2} \,,
  \end{aligned}
\end{align}
where $\qty{u,v,\rho,h}$ are the same as in \cite{Bernamonti:2007bu}. Then the field configuration is directly coming out as (4.17--4.18) for the $k^{\text{there}} = -1, \alpha^{\text{there}} = 0$ case, with
\begin{align}
  \beta^{\text{there}} ={}& \frac{\iu \sqrt{3} a_0 c_\Sigma \ell_5}{72 c_\varphi \omega} \,. 
\end{align}
Comparing to \eqref{eq:susy_thermo} we see that
\begin{align}
  \vb{Q} ={}& \frac{\iu a_0}{4\omega} \mathfrak{c} \,,
\end{align}
thus, the constant $a_0$ is proportional to the Maxwell charge. When we turn off the electric charge as 
\begin{align}
  \vb{Q} \propto{}& a_0 = 0 \,, \quad \psi_\infty = 0 \,, 
\end{align}
we recover a further specialized supersymmetric and extremal solution discussed in detail in \cite{Hong:2021bzg}, see (3.6) there. The explicit parameter mapping to \cite{Hong:2021bzg} is
\begin{align}
  c^{\text{there}} ={}& 4 c_\Sigma^2 \,, \quad q_0^{\text{there}} = - \frac{\pi^2}{9\omega^2} \,, \quad h^{\text{there}} = c_\varphi^2 c_\Sigma + \frac{8\pi^2 c_\Sigma^3}{3 \omega^2} \,,
\end{align}
where \cite{Hong:2021bzg} also sets $\ell_5 = 1$. Three remarks are in order. First, the asymptotic
value of \cite{Hong:2021bzg}'s structure function is $H(\infty)=h+3q_0c^{3/2}=c_\varphi^2c_\Sigma$, so the
regularity conditions $c>0$, $q_0<0$, $H(\infty)>0$ of \cite{Hong:2021bzg} are \emph{automatically}
satisfied for arbitrary real boundary data --- our UV parametrization covers exactly the regular
patch of the solution. Second, the parameter $q_0$, which sets the near-horizon extremal-BTZ
radius through $\rho_+^2=-9q_0$, is determined purely by the modular parameter:
$\rho_+=-\pi/\omega=\iu/(2\tau)$. This identifies our electric-off extremal solution as the
$(c,d)=(1,0)$ member of the $SL(2,\mathbb Z)$ family of \cite{Hong:2021bzg}, with the remaining
members corresponding to modular images of the boundary torus. Third, at $\vb{Q} \propto a_0\neq0$ one can view the solution of \cite{Bernamonti:2007bu} as an ``electrified'' version of \cite{Hong:2021bzg}.

The attentive reader will notice that we constructed the supersymmetric UV expansion \eqref{eq:susy_UV_exp} within the timelike class. On the other hand \cite{Bernamonti:2007bu} work exclusively in the null class. Thus we observe an intriguing class-switching phenomenon when $\beta \rightarrow \infty$. To see how this works first note that \cite{Gauntlett:2003fk} show:
\begin{align}
  \overline{\epsilon^c} \cdot \epsilon \propto f \,, 
\end{align}
where $f$ is the function appearing in \cref{sec:susy_generalities}. Thus, in the timelike class $f > 0$, and in the null class $f = 0$. Of course, in the null class the parametrization in \cref{sec:susy_generalities} breaks down. However, the parametrization in terms of $\qty{g_\bullet, a_\bullet}$ is agnostic about $f$ --- both classes can be written as a generic cohomogeneity one ansatz. In particular, within the UV expansion \eqref{eq:susy_UV_exp} one can certainly imagine a limit that sends $f \rightarrow 0$ and, in effect, switch class. We verify, up to 10th order in the UV, that every coefficient of $f$ scales as $1/\beta$ or higher. Thus, the extremal limit
$\beta \rightarrow \infty$ indeed sends $f \rightarrow 0$ and implements class-switching. The phenomenon observed here --- a family of supersymmetric solutions in the timelike class whose
extremal limit lands in the null class --- has appeared before. 
Already in the classifications of
\cite{Gauntlett:2003fk} the null case must be analyzed separately, and the norm of
the supersymmetric Killing vector can degenerate on loci or in limits of timelike-class solutions;
explicit examples of supersymmetric geometries of minimal 5d supergravity in which the bilinear
norm vanishes to all orders at a point were constructed in \cite{Pasini:2015zlx}, and the analogous
timelike degeneration in 4d minimal gauged supergravity was discussed in
\cite{Caldarelli:2003pb}. See also \cite{Boruch:2025qdq,Larsen:2026sav,Nanda:2026mbp} where, in the context of ungauged supergravity, supersymmetric non-extremal
saddles connect continuously to extremal BPS backgrounds of a qualitatively different character.

Finally, note that the thermodynamic relations \eqref{eq:susy_thermo} remain unmodified in the extremal limit $\widetilde{\omega} \rightarrow \infty$ at $\omega$ fixed, as they simply do not depend on $\widetilde{\omega}$. Observe also that for the TTI the notion of demanding real entropy (either by going on the real supersymmetric sub-locus as described above or by simply setting $\vb{Q} = 0$) is \textit{distinct} from the notion of 
extremality. In contrast, for the SCI once the \textit{non-linear constraint} between the charges is implemented, the entropy becomes real and the solution automatically becomes extremal and free of CTCs. It is an open problem whether our new supersymmetric but non-extremal solutions have any good Lorentzian interpretation away from extremality.

\paragraph{Non-extremal ``electrified'' $\text{BTZ} \times \Sigma_{\mathfrak{g}}$ limit}
We already alluded in \eqref{eq:cSigma_to_lSigma} that something special occurs when $\ell_\Sigma \rightarrow 1/\sqrt{3}$ at finite $\beta$. To see what we apply that limit to the IR expansion \eqref{eq:susy_IR_exp}. We observe that $\qty{g_\Sigma, a_\varphi}$ truncate to constants and the remaining functions re-sum as:
\begin{align}\label{eq:resummed_btz_like}
  \begin{aligned}
  g_\Sigma ={}& \frac{\ell_5^2}{3} \,, \\
  g_{tt}(r) ={}& \frac{4\pi^2 \ell_5^2}{\beta^2} \frac{\qty(\frac{2\pi}{3} + \iu\psi\omega)^2 \sinh^{2}{\frac{3r}{\ell_5}}}{9 \qty(\frac{2\pi}{3} + \iu\psi\omega)^2 + 4\pi \widehat{a}_0 \qty(1 - \cosh^{}{\frac{3r}{\ell_5}})} \,, \\
  g_{\varphi\varphi}(r) ={}& \frac{4\pi^2 \ell_5^6}{27 \beta^2 \omega^2} \frac{\qty(\frac{2\pi}{3} + \iu \psi \omega)^2 \sinh^{2}{\frac{3r}{\ell_5}}}{g_{tt}(r) g_\Sigma} \,, \\
  w(r) ={}& - \Omega_H - \frac{4\pi \ell_5^2}{3\beta\omega} \frac{\qty(\frac{2\pi}{3} - \iu\psi\omega) \sinh^{2}{\frac{3r}{2\ell_5}}}{g_{\varphi\varphi}(r)} \,, \\
  a_\varphi ={}& - \psi_H \ell_5 \,, \\
  a_t(r) ={}& - a_\varphi \qty(w(r) + \Omega_H) \,,
  \end{aligned}
\end{align}
where 
\begin{align}
  \widehat{a}_0 ={}& \eval{a_0}_{\ell_\Sigma = \frac{1}{\sqrt{3}}} = - 3 \iu \psi\omega \,. 
\end{align}
We directly verify that the above functions solve the EOMs \cref{eq:evolution,eq:constraint}, and we rely on the fact that we have re-summed a supersymmetric solution in the IR to claim that this configuration is supersymmetric. This field configuration is clearly not 
asymptotically locally $\text{AdS}_5$ as $g_{\Sigma}$ never grows. Instead, it is a certain ``electrified'' version of $\text{BTZ} \times \Sigma_{\mathfrak{g}}$ with Wilson lines turned on along $\mathbb{R}_t \times S^1_\varphi$. To see that first note that the Wilson lines are pure gauge along $\dd[]{t}$ and $\dd[]{\varphi}$, i.e.\ $\mathcal{F} = - \mathfrak{p} \Omega_\Sigma$. Then the 3d block $\qty(t,r,\varphi)$ obeys
\begin{align}
  R_{\mu\nu}^{(3)} ={}& - \qty(\frac{4}{\ell_5^2} + \frac{\mathfrak{p}^2}{2 g_\Sigma^2}) g_{\mu\nu}^{(3)} \equiv - \frac{2}{\ell_3^2} g_{\mu\nu}^{(3)} \quad \implies \quad \ell_3 = \frac{2\ell_5}{3} \,, 
\end{align}
thus the 3d part of the metric is constant curvature with the $\text{AdS}_3$ scale $\ell_3$ related to the $\text{AdS}_5$ scale $\ell_5$ as above. Even more explicitly, under the coordinate transformation
\begin{align}
  T ={}& \frac{\omega^2 R_+}{\beta \widehat{a}_0} t \,, \quad \Phi = \varphi + \qty(\frac{\omega^2 R_-}{\ell_3 \beta \widehat{a}_0} - \Omega_H) t \,, 
\end{align}
with
\begin{align}
  R_\pm = \frac{\ell_5}{\omega} \qty(\frac{2\pi}{3} \pm \iu \psi \omega)\,,
\end{align}
the field configuration turns to
\begin{align}
  \begin{aligned}
    \dd[]{s}^2 ={}& - N^2 \dd[]{T}^2 + \dd[]{r}^2 + R^2 \qty(\dd[]{\Phi} + N^\Phi \dd[]{T})^2 + \frac{\ell_5^2}{3} \dd[]{s}^2_{\Sigma}\,, \\
  \mathcal{A} ={}& - \psi_H \ell_5 \dd[]{\Phi} + \frac{3 \psi_H R_-}{2 R_+} \dd[]{T} - \mathfrak{p} \, \omega_\Sigma \,, 
  \end{aligned}
\end{align}
where
\begin{align}
  \begin{aligned}
    N^2 ={}& \frac{\qty(R^2 - R_+^2) \qty(R^2 - R^2_-)}{\ell_3^2 R^2} \,, & N^\Phi ={}& - \frac{R_+R_-}{\ell_3 R^2} \,, \\
    R^2 ={}& \frac{R_+^2 + R_-^2}{2} + \frac{R_+^2 - R_-^2}{2} \cosh^{}{\frac{2r}{\ell_3}} \,. 
  \end{aligned}
\end{align}
The above is recognized as $\text{BTZ} \times \Sigma_{\mathfrak{g}}$ written in standard coordinates. The specific form of the Wilson lines is what allows for supersymmetry to be preserved even away from extremality. Note that when we turn off the electric charge
\begin{align}
  \eval{\vb{Q}}_{\ell_\Sigma = \frac{1}{\sqrt{3}}} \propto {}& \widehat{a}_0 \propto \psi = 0 \,, \quad \psi_H = 0 \,,
\end{align}
the two horizons $R_+ = \frac{\ell_3 \pi}{\omega} = R_-$ coalesce and the Wilson lines vanish. Then the solution becomes purely
magnetically charged $\text{extremal-BTZ} \times \Sigma_{\mathfrak{g}}$. Note that the class-switching phenomenon discussed above is once again realized in that further extremal limit.

Away from extremality, a similar (complex) non-extremal $\text{BTZ} \times S^2$ was written down in \cite{Boruch:2025qdq} in the context of ungauged supergravity, see also \cite{Larsen:2026sav} for a recent discussion on (complex) non-extremal $\text{BTZ}$ directly in 3d. To our knowledge, above we provide the first example of (complex) non-extremal $\text{BTZ} \times \Sigma_{\mathfrak{g}}$ embedded into 5d gauged supergravity.
  
%%%%%%%
\section{Conclusions}\label{sec:conclusion}
%%%%%%%

In this paper we have constructed a family of asymptotically locally AdS$_5$ solutions of five-dimensional minimal gauged supergravity with $\mathbb{R}_t \times S^1_\varphi \times \Sigma_{\mathfrak{g}}$ boundary, showcasing a number of novel features. 
The symmetry of the problem allowed us to employ a cohomogeneity one ansatz, thus reducing the equations to a set of ODEs. Despite not being able to find generic analytic solutions to this complicated set of equations, we employed perturbative expansions together with numerical shooting, in order to establish the existence of these solutions. Similar methods had been used in \cite{Cassani:2014zwa,Cassani:2018mlh} to study cohomogeneity one ans\"atze, for solutions with $S^1\times S^3$ boundary topology, that are somewhat simpler than those presented here. In particular, the underlying topology of the solutions we considered, involving a compact Riemann surface $\Sigma_{\mathfrak{g}}$, implies that 
the gauge field must be defined locally in two patches covering the space-time manifold, with important consequences for the definition and the computation of conserved charges as well as of the on-shell action. Such subtleties, related to the presence of Chern-Simons terms, were previously discussed in 
\cite{Hanaki:2007mb}, motivated by the explicit black ring solutions of minimal five-dimensional ungauged supergravity \cite{Emparan:2001wn,Elvang:2004rt}. The patch-wise integration of the Chern-Simons term in the 
on-shell action plays a crucial role in the localization calculations in \cite{Colombo:2025yqy}. However, the results of this paper are not applicable to our backgrounds, because they are not toric and the boundary metric is not conformally flat. Indeed, it would be very interesting to generalize the analysis of \cite{Colombo:2025yqy}, in order to include solutions such as those presented in this work, and generalizations. 
 
It is worth emphasizing that our analysis relies on numerical studies only as a means of establishing existence of the solutions, a step that is entirely missing 
in the context of recent localization approaches for the calculation of the supergravity on-shell action, see~\cite{Sparks:2026taw} for a review. Indeed, precisely as in those approaches, we are able to compute the on-shell action entirely analytically, and express it as a function of the UV boundary data and the physical charges through the use of UV-IR relations. 
In this respect, our study shares some similarities with that presented in~\cite{Cassani:2014zwa,Cassani:2018mlh,Ntokos:2021duk}, concerning cohomogeneity one AlAdS$_5$ solutions. However, the different global topologies of the solutions result in important technical as well as conceptual differences. 
Firstly, in the supersymmetric case we consider families of non-extremal solutions, that are therefore complex valued, for which we can calculate 
 the thermodynamics. We also identify a sub-family of these solutions that are ``almost real'' in the sense that they have real entropy and angular momenta, and still manage to preserve supersymmetry thanks to the presence of an imaginary Wilson line for the gauge field\footnote{This feature has been recently 
 observed in related contexts in \cite{Larsen:2026sav} and \cite{Nanda:2026mbp}.}. For this sub-family we employed numerical methods to establish existence of the solutions. Another novel aspect of our solutions is that the electric charge that naturally appears in the thermodynamic relations is the Maxwell charge, instead of the Page charge or the holographic charge, as advocated in \cite{Cassani:2018mlh,Ntokos:2021duk}. This is clearly related to the different space-time topologies, but it would be nice to understand this feature at a more fundamental level.

 Perhaps unexpectedly, we have shown (numerically and perturbatively) that the non-extremal supersymmetric solutions all lie in the \emph{timelike} class of 
 \cite{Gauntlett:2003fk} and switch to the null class in the extremal limit. We conjecture that this is a generic feature of supersymmetric solutions of theories that possess both classes of BPS solutions, such as gauged or ungauged five-dimensional supergravity. This conjecture is supported by the explicit analytic extremal and non-extremal solutions presented in \cref{sec:analytical_limits} which are indeed in the null and timelike class, respectively. If true, this conjecture may provide a way forward to study extremal solutions in the null class, using localization techniques. 
  
The solutions we constructed have a constant curvature metric on the Riemann surface. Based on the analysis in~\cite{Anderson:2011cz,Bobev:2020jlb}, as well as the fact that the dual 4d SCFT is topologically twisted on $\Sigma_{\mathfrak{g}}$, it should be possible to generalize these backgrounds by allowing an arbitrary non-constant curvature metric on $\Sigma_{\mathfrak{g}}$. Constructing such solutions explicitly will be hard since they will not be cohomogeneity one and thus one should solve PDEs in order to find them. Nevertheless, it is natural to expect that at least in the supersymmetric setup the on-shell action of these more general solutions is independent of the metric on $\Sigma_{\mathfrak{g}}$. It will be very interesting to establish this, perhaps using supergravity localization techniques. The constant curvature metric on $\Sigma_{\mathfrak{g}}$ with $\mathfrak{g}>1$ has $3\mathfrak{g}-3$ complex structure moduli. The TTI is manifestly independent of these parameters\footnote{From the perspective of the IR 2d $\mathcal{N}=(0,2)$ SCFT on $T^2$ these parameters correspond to exactly marginal deformations.} which is mirrored by the fact that the supergravity solutions we constructed and their on-shell action also do not depend on them. Notably, this is true even in the absence of supersymmetry. It will be interesting to understand this fact further from the perspective of the SCFT non-supersymmetric path integral on $T^2\times \Sigma_{\mathfrak{g}}$.

Based on our findings, there are a number of directions that can be explored further. The TTI can be defined also on $T^2\times S^2$ and in this case it can be (equivariantly) refined by the addition of a fugacity for azimuthal rotations of $S^2$. Extremal solutions of this type have been studied in \cite{Hosseini:2019lkt,Hosseini:2020vgl} and they should admit an appropriate complex non-extremal generalization. We therefore expect that there should exist supergravity solutions with $\mathbb{R}^{1,1} \times S^1 \times S^2$ topology and possessing 
$\mathbb{R}\times U(1)^2$ isometry. These solutions fall outside our ansatz and it would be interesting to try to generalize this, to include squashing and tilting of the $S^2$. However, generically these new solutions will be cohomogeneity two and with current technologies finding such backgrounds appears prohibitively
difficult. 
 
A generalization that should be more accessible, because it retains a cohomogeneity one ansatz, is to extend our construction to solutions with the same topology and symmetry, in five-dimensional supergravity coupled to vector multiplets. In particular, the 5d $\mathcal{N}=2$ STU model can be uplifted to
 type IIB supergravity with a neat holographic interpretation in terms of the 4d $\mathcal{N}=4$ SYM and some of its orbifolds. 
 This new class of solutions would be a generalization of the static magnetically charged Lorentzian solutions in~\cite{Maldacena:2000mw,Benini:2013cda} and would provide
 the bulk dual of the TTI studied in~\cite{Hosseini:2016cyf,Hong:2018viz,Hosseini:2020vgl,Hong:2021bzg}, with flavor fugacities and magnetic fluxes. Indeed, some of the features of our solutions, such as the BPS constraints, should become more transparent in this more general context. 

Our solutions should also admit a generalization to 7d gauged supergravity where they would describe the partition function of the dual 6d SCFT on $T^2 \times \mathcal{M}_4$ with an appropriate topological twist on the compact manifold $\mathcal{M}_4$. Static, extremal and supersymmetric solutions of this type have been studied in the literature, see for example \cite{Gauntlett:2000ng,Benini:2013cda,Bobev:2017uzs} and it will be interesting to generalize them along the lines of our work. For the particular case when $\mathcal{M}_4$ is a product of two Riemann surfaces these solutions will describe the TTI for the 4d $\mathcal{N}=1$ theories of class $\mathcal{S}$ that we briefly discussed in \cref{sec:TTI_5d}.

The simple form of the QFT result for the TTI in~\eqref{eq:TTItauintro} is derived using the Cardy limit. In the supergravity derivation of this result that we presented here we did not use the Cardy limit. This is similar to the situation for the SCI for 4d $\mathcal{N}=1$ SCFTs and its holographic description, see~\cite{Cabo-Bizet:2018ehj,Cassani:2021fyv}. It will be most interesting to understand the reason behind this extended regime of validity of the supergravity on-shell action and the interplay with the large $N$ limit. Relatedly, it will be important to investigate the interplay between the Cardy and the large $N$ limits in the calculation of the TTI and to understand whether there are interesting phase transitions in the TTI as one varies continuous parameters like flavor or rotational fugacities. More generally, it will be important to investigate the analog of the family of ``Farey tail'' solutions discussed in~\cite{Dijkgraaf:2000fq} in the context of the 2d elliptic genus and AdS$_3$/CFT$_2$. As stressed in~\cite{Hong:2021bzg} there should be a similar family of gravitational solutions for the TTI and it will be important to combine the analysis in~\cite{Hong:2021bzg} with our results here to understand this better.

Finally, it should be possible to use the results and methods in~\cite{Bobev:2021qxx,Bobev:2022bjm,Cassani:2022lrk,Cassani:2023vsa} to find the leading four-derivative correction to the supergravity on-shell action. Note that to do this we do not need to correct the solutions constructed in this work and we need to modify only the calculation of the on-shell action as in~\cite{Bobev:2022bjm,Cassani:2022lrk}. It will be interesting to perform this calculation and confront it with the first subleading correction in the large $N$ limit of the TTI in the dual SCFT.

\acknowledgments
We would like to thank Edoardo Colombo, Junho Hong, Sameer Murthy, Ioannis Papadimitriou, Jaeha Park, Antonio Pittelli, and Alberto Zaffaroni for illuminating discussions. NB is supported by the FWO projects G003523N, G094523N, and G0E2723N, and the KU Leuven C1 project C16/25/01.

\appendix
\section{Patchwork}\label{sec:patchwork}

\subsection{Magnetic monopoles over a Riemann surface}
An explicit construction of magnetic monopoles on Riemann surfaces of arbitrary genus is presented in \cite{Martin:1996ch}. However, for our purposes it suffices to take into account some basic features of the construction, as we recall below. We start by defining two open sets: $\qty{U_N, U_S}$ covering $\Sigma_{\mathfrak{g} > 1}$\footnote{To really cover $\Sigma_{\mathfrak{g} > 1}$ with a good atlas one generically needs more than two patches, however, for the purposes of discussing the topology of $\text{U}\qty(1)$ connections on $\Sigma_{\mathfrak{g} > 1}$ two are enough.}. To construct them we consider a point $p \in \Sigma_{\mathfrak{g}}$ and two small open disks $D_- \subset D_+$ around $p \in D_-$. Then the open sets and 
their intersection are defined as
\begin{align}
  U_S \equiv {}& D_+ \,, \quad U_N \equiv \Sigma_{\mathfrak{g}} \, \backslash \, \bar{D}_- \,, \quad U_N \cap U_S \equiv D_+ \, \backslash \, \bar{D}_- \cong S^1 \times (0,1) \,,
\end{align}
where the bar on $D_-$ denotes closure. For the local metric on $\Sigma_{\mathfrak{g} > 1}$ we take
\begin{align}
  \dd[]{s}^2 ={}& \dd[]{\theta}^2 + \sinh^{2}{\theta} \dd[]{\phi}^2 \,. 
\end{align}
Take a smooth connection one-form on $U_S$
\begin{align}
  \omega_S ={}& \qty(\cosh^{}{\theta} - 1) \dd[]{\phi} \,, \quad \dd[]{\omega_S} = \Omega_\Sigma \,,
\end{align}
where $\Omega_\Sigma$ is the volume form on $\Sigma_{\mathfrak{g}}$. The $(- 1)$ above is included such that $\omega_S \sim \frac{1}{2} \theta^2 \dd[]{\phi}$ vanishes smoothly at the point $p$ near $\theta = 0$. The effect of this is that the Dirac string is moved out of $U_S$ from the perspective of $\omega_S$. To define how the connection one-form looks on $U_N$ we make use of the scalar Green's function $G_p$ at the point $p$ on $\Sigma_{\mathfrak{g}}$, which has the property
\begin{align}
  \dd[] \star_\Sigma \dd[] G_p ={}& \delta_p - \frac{\Omega_\Sigma}{2\pi \eta_\Sigma} \,, \quad \int_\Sigma G_p \Omega_\Sigma = 0 \,, 
\end{align}
where $\delta_p$ is the unit mass Dirac two-form at $p$. Then
\begin{align}
  \omega_N ={}& - 2\pi \eta_\Sigma \star_\Sigma \dd[]{G_p} \,, \quad \dd[]{\omega_N} = \Omega_\Sigma \,. 
\end{align}
From the perspective of $\omega_N$ the Dirac string is now placed at $p$ which is outside of $U_N$. Now consider a curve $C \cong S^1$ inside of $U_N \cap U_S$ at $\theta = \theta_0$. We use Stokes' theorem to evaluate
\begin{align}
  \begin{aligned}
  \oint_C \omega_S ={}& \int_{D_{\theta_0}} \Omega_\Sigma = 2\pi \qty(\cosh^{}{\theta_0} - 1) \,, \\
  \oint_{-C} \omega_N ={}& \int_{\Sigma \, \backslash \, D_{\theta_0}} \Omega_\Sigma = 2 \pi \eta_\Sigma - 2\pi \qty(\cosh^{}{\theta_0} - 1) \,,
  \end{aligned}
\end{align}
and arrive at the \textit{transition identity}
\begin{align}\label{eq:transition_identity}
  \oint_C \qty(\omega_N - \omega_S) ={}& - 2\pi \eta_\Sigma = - \int_\Sigma \Omega_\Sigma \,. 
\end{align}
Said differently, the difference
\begin{align}
  \omega_N - \omega_S ={}& \dd[]{\lambda_{NS}} \,, \quad \lambda_{NS} = - \eta_\Sigma \phi + \text{(smooth single valued function)} \,, 
\end{align}
is closed, $\lambda_{NS}$ is the transition function, and we have $\oint_C \dd[]{\lambda_{NS}} = - \eta_\Sigma \oint_C \dd[]{\phi} = - 2\pi \eta_\Sigma$. Thus, indeed $\qty{\omega_S, \omega_N}$ are local representatives of a connection one-form $\omega_\Sigma$ of a $\text{U}\qty(1)$ bundle over $\Sigma_{\mathfrak{g}}$ whose curvature is the globally well-defined $\Omega_\Sigma$, and the first Chern number of the connection is quantized
\begin{align}
  c_1(\omega_\Sigma) ={}& \frac{1}{2\pi} \int_\Sigma \Omega_\Sigma = \eta_\Sigma = 2(\mathfrak{g} - 1) \in \mathbb{Z} \,. 
\end{align}

\subsection{Odd dimensional integrals on manifolds involving $\Sigma_{\mathfrak{g}}$}
In the main text, the gauge field and its field strength are defined as
\begin{align}
  \begin{aligned}
  \mathcal{A} ={}& \mathcal{B} - \mathfrak{p} \, \omega_\Sigma \,, \\
  \mathcal{B} ={}& a_t \dd[]{t} + a_\varphi \qty(\dd[]{\varphi} + w \dd[]{t}) = \mathcal{A}_t \dd[]{t} + \mathcal{A}_\varphi \dd[]{\varphi} \,, \\
  \mathcal{F} ={}& \dd[]{\mathcal{B}} - \mathfrak{p} \, \Omega_\Sigma \,. 
  \end{aligned}
\end{align}
We encounter integrals of the type
\begin{align}
  I_{\text{3d}}^{\text{CS}} = \int_{\mathcal{N}} \mathcal{A} \wedge \mathcal{F} \,, \quad I^{\text{CS}}_{\text{5d}} = \int_{\mathcal{M}} \mathcal{A} \wedge \mathcal{F} \wedge \mathcal{F} \,,
\end{align}
where the 3d Chern-Simons integrals are evaluated on $\mathcal{N} = S^1_\varphi \times \Sigma$ or $\mathcal{N} = \mathbb{R}_t \times \Sigma$ and $\mathcal{M} \cong \mathbb{R}_t \times [0, \infty] \times S^1_\varphi \times \Sigma$\footnote{In Euclidean signature we will of course be taking $\mathbb{R}_t \rightarrow S^1_t$.}. Formally we evaluate these integrals by extending
\begin{align}
   \mathcal{N} \rightarrow {}& \overline{\mathcal{N}} \,, \quad \mathcal{M} \rightarrow \overline{\mathcal{M}} \,, 
\end{align}
where $\partial \overline{\mathcal{N}} = \mathcal{N}$ and $\partial \overline{\mathcal{M}} = \mathcal{M}$ as
\begin{align}
  I_{\text{3d}}^{\text{CS}} = \int_{\overline{\mathcal{N}}} \overline{\mathcal{F}} \wedge \overline{\mathcal{F}} \,, \quad I^{\text{CS}}_{\text{5d}} = \int_{\overline{\mathcal{M}}} \overline{\mathcal{F}} \wedge \overline{\mathcal{F}} \wedge \overline{\mathcal{F}} \,,
\end{align}
where we have also extended the fundamental fields. We split the extended manifolds into patches
\begin{align}
  \overline{\mathcal{N}} \rightarrow {}& \qty{\overline{\mathcal{N}}_S, \overline{\mathcal{N}}_N} \,, \quad \overline{\mathcal{M}} \rightarrow \qty{\overline{\mathcal{M}}_S, \overline{\mathcal{M}}_N} \,, 
\end{align}
where
\begin{align}
  \begin{aligned}
  \partial \overline{\mathcal{N}}_S ={}& \mathcal{N}_S + \overline{\Gamma}_2 \,, & \partial \overline{\mathcal{N}}_N ={}& \mathcal{N}_N - \overline{\Gamma}_2 \,, \\
  \partial \overline{\mathcal{M}}_S ={}& \mathcal{M}_S + \overline{\Gamma}_4 \,, & \partial \overline{\mathcal{M}}_N ={}& \mathcal{M}_N - \overline{\Gamma}_4 \,, \\
  \end{aligned}
\end{align}
and $\qty{\overline{\Gamma}_2,\overline{\Gamma}_4}$ are, respectively, the 3d and 5d extensions of
\begin{align}
  \Gamma_2 ={}& \begin{cases}
  S^1_\varphi \times S^1_\phi \,, \\
  \mathbb{R}_t \times S^1_\phi \,,
  \end{cases} \quad \Gamma_4 = \mathbb{R}_t \times [0, \infty] \times S^1_\varphi \times S^1_\phi \,,
\end{align}
with $\partial \overline{\Gamma}_2 = \Gamma_2$, $\partial \overline{\Gamma}_4 = \Gamma_4$. Then using Stokes repeatedly we have
\begin{align}\label{eq:expand_I3d}
  I_{\text{3d}}^{\text{CS}} ={}& \int_{\partial \overline{\mathcal{N}}_S} \overline{\mathcal{A}}_S \wedge \overline{\mathcal{F}} + \int_{\partial \overline{\mathcal{N}}_N} \overline{\mathcal{A}}_N \wedge \overline{\mathcal{F}} \notag \\
   ={}& \int_{\mathcal{N}_S} \mathcal{A}_S \wedge \mathcal{F} + \int_{\mathcal{N}_N} \mathcal{A}_N \wedge \mathcal{F} + \int_{\overline{\Gamma}_2} \qty(\overline{\mathcal{A}}_S - \overline{\mathcal{A}}_N) \wedge \overline{\mathcal{F}} \notag \\
   ={}& \int_{\mathcal{N}_S} \mathcal{A}_S \wedge \mathcal{F} + \int_{\mathcal{N}_N} \mathcal{A}_N \wedge \mathcal{F} - \int_{\partial \overline{\Gamma}_2} \qty(\overline{\mathcal{A}}_S - \overline{\mathcal{A}}_N) \wedge \overline{\mathcal{A}}_S \notag \\
   ={}& \int_{\mathcal{N}_S} \mathcal{A}_S \wedge \mathcal{F} + \int_{\mathcal{N}_N} \mathcal{A}_N \wedge \mathcal{F} + \int_{\Gamma_2} \mathcal{A}_N \wedge \mathcal{A}_S \,. 
\end{align}
We now formally define the na\"{\i}ve 3d Chern-Simons integral
\begin{align}
\fint_{\mathcal{N}} \mathcal{A} \wedge \mathcal{F} \equiv {}& \int_{\mathcal{N}_S} \mathcal{A}_S \wedge \mathcal{F} + \int_{\mathcal{N}_N} \mathcal{A}_N \wedge \mathcal{F} \,, 
\end{align}
as the straightforward sum of the two patches. In our specific background, these can be evaluated as
\begin{align}\label{eq:naive_3d_integral_calculated}
  \fint_{\mathcal{N}} \mathcal{A} \wedge \mathcal{F} ={}& \sum_{a = \qty{S,N}} \int_{\mathcal{N}_a} \qty(\mathcal{B} - \mathfrak{p} \, \omega_a) \wedge \qty(\dd[]{\mathcal{B}} - \mathfrak{p} \, \Omega_\Sigma) \notag \\
   ={}& \sum_{a = \qty{S,N}} \int_{\mathcal{N}_a} \qty(\qty(\mathcal{B} - \mathfrak{p} \, \omega_a) \wedge \dd[]{\mathcal{B}} - \mathfrak{p} \, \mathcal{B} \wedge \Omega_\Sigma) \notag \\
    ={}& - \mathfrak{p} \int_{\mathcal{N}_S \cup \mathcal{N}_N} \mathcal{B} \wedge \Omega_\Sigma \,, 
\end{align}
where in the second equality we have used that $\dd[]{\mathcal{B}} = \dd[]{r} \wedge \qty(\mathcal{A}_t' \dd[]{t} + \mathcal{A}_\varphi' \dd[]{\varphi})$ and its pullback on a surface of constant $r$ vanishes. The last term in \eqref{eq:expand_I3d} is the interface integral which can be rewritten as
\begin{align}\label{papero}
  \int_{\Gamma_2} \mathcal{A}_N \wedge \mathcal{A}_S ={}& \int_{\Gamma_2} \qty(\mathcal{B} - \mathfrak{p} \, \omega_N) \wedge \qty(\mathcal{B} - \mathfrak{p} \, \omega_S) \notag \\
  ={}& \mathfrak{p} \int_{\Gamma_2} \mathcal{B} \wedge (\omega_N - \omega_S) \notag \\
   ={}& - \mathfrak{p} \int_{\mathcal{N}} \mathcal{B} \wedge \Omega_\Sigma \notag \\
  \equiv {}& \fint_{\mathcal{N}} \mathcal{A} \wedge \mathcal{F} \,, 
\end{align}
where we have used the transition identity \eqref{eq:transition_identity} and we have noticed that the interface integral is formally equal to the na\"{\i}ve integral in \eqref{eq:naive_3d_integral_calculated}.
Overall, the 3d Chern-Simons integral on a constant $r$ slice is
\begin{align}
  I^{\text{CS}}_{\text{3d}} = \int_{\mathcal{N}} \mathcal{A} \wedge \mathcal{F} ={}& 2 \fint_{\mathcal{N}} \mathcal{A} \wedge \mathcal{F} \,,
\end{align}
for any $\mathcal{N} = \mathbb{R}_t \times \Sigma$ or $\mathcal{N} = S^1_\varphi \times \Sigma$.

We proceed analogously with the 5d Chern-Simons integral
\begin{align}\label{eq:expand_I5d}
  I_{\text{5d}}^{\text{CS}} ={}& \int_{\partial\overline{\mathcal{M}}_S} \overline{\mathcal{A}}_S \wedge \overline{\mathcal{F}} \wedge \overline{\mathcal{F}} + \int_{\partial\overline{\mathcal{M}}_N} \overline{\mathcal{A}}_N \wedge \overline{\mathcal{F}} \wedge \overline{\mathcal{F}} \notag \\
   ={}& \int_{\mathcal{M}_S} \mathcal{A}_S \wedge \mathcal{F} \wedge \mathcal{F} + \int_{\mathcal{M}_N} \mathcal{A}_N \wedge \mathcal{F} \wedge \mathcal{F} + \int_{\overline{\Gamma}_4} \qty(\overline{\mathcal{A}}_S - \overline{\mathcal{A}}_N) \wedge \overline{\mathcal{F}} \wedge \overline{\mathcal{F}} \notag \\
   ={}& \int_{\mathcal{M}_S} \mathcal{A}_S \wedge \mathcal{F} \wedge \mathcal{F} + \int_{\mathcal{M}_N} \mathcal{A}_N \wedge \mathcal{F} \wedge \mathcal{F} - \int_{\partial \overline{\Gamma}_4} \qty(\overline{\mathcal{A}}_S - \overline{\mathcal{A}}_N) \wedge \overline{\mathcal{A}}_S \wedge \overline{\mathcal{F}} \notag \\
   ={}& \int_{\mathcal{M}_S} \mathcal{A}_S \wedge \mathcal{F} \wedge \mathcal{F} + \int_{\mathcal{M}_N} \mathcal{A}_N \wedge \mathcal{F} \wedge \mathcal{F} + \int_{\Gamma_4} \mathcal{A}_N \wedge \mathcal{A}_S \wedge \mathcal{F} \,. 
\end{align}
Again we define the na\"{\i}ve integral as
\begin{align}
\fint_{\mathcal{M}} \mathcal{A} \wedge \mathcal{F} \wedge \mathcal{F} \equiv {}& \int_{\mathcal{M}_S} \mathcal{A}_S \wedge \mathcal{F} \wedge \mathcal{F} + \int_{\mathcal{M}_N} \mathcal{A}_N \wedge \mathcal{F} \wedge \mathcal{F} \,. 
\end{align}
On our background, it is evaluated as
\begin{align}
  \fint_{\mathcal{M}} \mathcal{A} \wedge \mathcal{F} \wedge \mathcal{F} ={}& \sum_{a = \qty{S,N}} \int_{\mathcal{M}_a} \qty(\mathcal{B} - \mathfrak{p} \, \omega_a) \wedge \qty(\dd[]{\mathcal{B}} - \mathfrak{p} \, \Omega_\Sigma) \wedge \qty(\dd[]{\mathcal{B}} - \mathfrak{p} \, \Omega_\Sigma) \notag \\
   ={}& - 2\mathfrak{p} \int_{\mathcal{M}_S \cup \mathcal{M}_N} \mathcal{B} \wedge \dd[]{\mathcal{B}} \wedge \Omega_\Sigma \,. 
\end{align}
The last term in \eqref{eq:expand_I5d} is the interface integral which evaluates to
\begin{align}
  \int_{\Gamma_4} \mathcal{A}_N \wedge \mathcal{A}_S \wedge \mathcal{F} ={}& \int_{\Gamma_4} \qty(\mathcal{B} - \mathfrak{p} \, \omega_N) \wedge \qty(\mathcal{B} - \mathfrak{p} \, \omega_S) \wedge \qty(\dd[]{\mathcal{B}} - \mathfrak{p} \, \Omega_\Sigma) \notag \\
  ={}& \mathfrak{p} \int_{\Gamma_4} \mathcal{B} \wedge \dd[]{\mathcal{B}} \wedge \qty(\omega_N - \omega_S) \notag \\
   ={}& - \mathfrak{p} \int_{\mathcal{M}} \mathcal{B} \wedge \dd[]{\mathcal{B}} \wedge \Omega_\Sigma \notag \\
    \equiv {}& \frac{1}{2} \fint_{\mathcal{M}} \mathcal{A} \wedge \mathcal{F} \wedge \mathcal{F} \,,
\end{align}
where we have again used the transition identity \eqref{eq:transition_identity} and have noticed that the interface integral is $\qty(1/2) \times$ the na\"{\i}ve integral.
Overall, the full 5d Chern-Simons integral is
\begin{align}
  I^{\text{CS}}_{\text{5d}} ={}& \int_{\mathcal{M}} \mathcal{A} \wedge \mathcal{F} \wedge \mathcal{F} = \frac{3}{2} \fint_{\mathcal{M}} \mathcal{A} \wedge \mathcal{F} \wedge \mathcal{F} \,. 
\end{align}

\subsection{Radially conserved charges: non-closed forms without interface integrals}
It is convenient to consider the following one-parameter three-forms
\begin{align}\label{eq:source_three_forms_onepara}
  \begin{aligned}
  \mathcal{Q}^{c_1} ={}& \star_5 \mathcal{F} + c_1 \mathcal{A} \wedge \mathcal{F} \,, \\
   \mathcal{J}^{c_2} ={}& \star_5 \dd[]{\qty((\partial_\varphi)^{\flat}) } + 3 \qty(\iota_{\partial_\varphi} \mathcal{A}) \qty(\star_5 \mathcal{F} + c_2 \mathcal{A} \wedge \mathcal{F}) \,,
  \end{aligned}
\end{align}
where $c_1$ and $c_2$ are two arbitrary constants. 
Recalling that $\dd \mathcal{Q}^{c_1=1}=\dd \mathcal{J}^{c_2=2/3} =0$ we have that
\begin{align}\label{nonclosedthreeforms}
  \begin{aligned}
 \dd \mathcal{Q}^{c_1} ={}& (c_1-1) \mathcal{F} \wedge \mathcal{F} \,, \\
 \dd \mathcal{J}^{c_2} ={}& \frac{3c_2-2}{2} \left[ (3\iota_{\partial_\varphi} \mathcal{A}) \mathcal{F} \wedge \mathcal{F} - \iota_{\partial_\varphi} ( \mathcal{A} \wedge \mathcal{F} \wedge \mathcal{F}) \right]\, .
  \end{aligned}
\end{align}

Let us start from $\mathcal{Q}^{c_1}$. In each patch we take 
\begin{align}
  \begin{aligned}\label{Apatch}
   \mathcal{A}_a ={}& a_t \dd[]{t} + a_\varphi \qty(\dd[]{\varphi} + w \dd[]{t}) - \mathfrak{p} \, \omega_a \,,
  \end{aligned}
\end{align}
where $\omega_a$ are the well-defined connections on $U_a$ with $a\in \{N,S\}$ and $\dd \omega_a= \Omega_\Sigma$. On each patch then 
 \begin{align}
 \begin{aligned}
  \mathcal{Q}_a^{c_1} ={}& \star_5 \mathcal{F} + c_1 \mathcal{A}_a \wedge \mathcal{F} \,, \\
\end{aligned}
\end{align}
are regular three-forms and we can use Stokes' theorem legitimately. At fixed $t$, consider the four-dimensional slab ${\cal S} = [r_1,r_2]\times S^1_\varphi \times \Sigma_{\mathfrak{g}}$, 
with $\partial {\cal S} = S^1_\varphi \times \Sigma_{r_1} - S^1_\varphi \times \Sigma_{r_2}$. We cover this slab with the two patches 
${\cal S}_a = [r_1,r_2]\times S^1_\varphi \times U_a$. 
 The boundary of each patch is given by
  \begin{align}
 \begin{aligned}
\partial {\cal S}_S = S^1_\varphi \times U_S |_{r_1} - S^1_\varphi \times U_S |_{r_2} + \Gamma_3 \, , \qquad \partial {\cal S}_N = S^1_\varphi \times U_N |_{r_1} - S^1_\varphi \times U_N |_{r_2} - \Gamma_3 \, ,
  \end{aligned}
\end{align}
where $\Gamma_3 =[r_1,r_2] \times S^1_\varphi \times S^1_\phi$ is the 3d interface of the 4d slab $\mathcal{S}$. Then we have
  \begin{align}
 \begin{aligned}
   (c_1-1)\int_{\cal S} \mathcal{F} \wedge \mathcal{F} ={}& \int_{{\cal S}_S}\dd \mathcal{Q}_S^{c_1} + \int_{{\cal S}_N}\dd \mathcal{Q}_N^{c_1} \\
   ={}& \int_{\partial {\cal S}_S} \mathcal{Q}_S^{c_1} + \int_{\partial {\cal S}_N} \mathcal{Q}_N^{c_1} \\
   ={}& \int_{S^1_\varphi \times U_S |_{r_1}} \mathcal{Q}_S^{c_1} - \int_{S^1_\varphi \times U_S |_{r_2}} \mathcal{Q}_S^{c_1} + \int_{\Gamma_3} \mathcal{Q}_S^{c_1} \\
&+ \int_{S^1_\varphi \times U_N |_{r_1}} \mathcal{Q}_N^{c_1} - \int_{S^1_\varphi \times U_N |_{r_2}} \mathcal{Q}_N^{c_1} - \int_{\Gamma_3} \mathcal{Q}_N^{c_1} \, . 
  \end{aligned}
\end{align}
The above identity can be rewritten as 
\begin{align} 
 \begin{aligned}
  \fint_{S^1_\varphi \times \Sigma_{r_2}} \mathcal{Q}^{c_1} - \fint_{S^1_\varphi \times \Sigma_{r_1}} \mathcal{Q}^{c_1} = \int_{\Gamma_3} \left( \mathcal{Q}_S^{c_1} - \mathcal{Q}_N^{c_1} \right) -(c_1-1)\int_{\cal S} \mathcal{F} \wedge \mathcal{F} \,.
  \end{aligned}
\end{align}
It is then clear that for $c_1=1$, the na\"{\i}ve Page charge, defined without the interface integral, is \emph{not} radially conserved. 
The second term on the right hand side evaluates to 
\begin{align}
 \begin{aligned}
- (c_1-1)\int_{{\cal S}} \mathcal{F}\wedge \mathcal{F} = (c_1-1)4\pi \mathfrak{p} (a_\varphi (r_1)- a_\varphi (r_2) ) 2\pi \eta_\Sigma\, .
  \end{aligned}
\end{align}
To evaluate the $\Gamma_3$ interface integral we use
 \begin{align}
 \begin{aligned}
 \mathcal{A}_a \wedge \mathcal{F} |_{\Gamma_3} = -\mathfrak{p} a_\varphi' \omega_a \wedge \dd r \wedge \dd \varphi \,. 
  \end{aligned}
\end{align}
Then
 \begin{align}
 \begin{aligned}
 \int_{\Gamma_3} \left( \mathcal{Q}_S^{c_1} - \mathcal{Q}_N^{c_1} \right) &= c_1\int_{\Gamma_3} \left( \mathcal{A}_S - \mathcal{A}_N \right) \wedge \mathcal{F} \\
 &= - c_1 2\pi\mathfrak{p} (a_\varphi (r_2) - a_\varphi(r_1)) \oint_{S^1_\phi} (\omega_S-\omega_N) \\
  &= -c_1\int_{\cal S} \mathcal{F} \wedge \mathcal{F}\, . 
  \end{aligned}
\end{align}
Summing the two contributions gives 
\begin{align}
 \label{patchyintegrals}
 \begin{aligned}
  \fint_{S^1_\varphi \times \Sigma_{r_2}} \mathcal{Q}^{c_1} - \fint_{S^1_\varphi \times \Sigma_{r_1}} \mathcal{Q}^{c_1} = (2-c_1)\int_{\cal S} \mathcal{F} \wedge \mathcal{F} \, , \end{aligned}
  \end{align}
and therefore the charge $q$, corresponding to $c_1=2$, is radially conserved. An analogous calculation can be performed by replacing $S^1_\varphi$ with $\mathbb{R}_t$ to show the conservation of $\widetilde{q}$.

Let us now discuss more briefly the generalized Noether-Wald charge $\mathcal{J}^{c_2}$. Now, integrating $\dd \mathcal{J}^{c_2}$ over the slab ${\cal S}$ and noting that 
$ \iota_{\partial_\varphi} ( \mathcal{A} \wedge \mathcal{F} \wedge \mathcal{F}) |_{\cal S}=0$, we find
\begin{align} 
 \begin{aligned}
  \fint_{S^1_\varphi \times \Sigma_{r_2}} \mathcal{J}^{c_2} - \fint_{S^1_\varphi \times \Sigma_{r_1}} \mathcal{J}^{c_2} = \int_{\Gamma_3} \left( \mathcal{J}_S^{c_2} - \mathcal{J}_N^{c_2} \right) -3\frac{3c_2-2}{2}\int_{\cal S} (\iota_{\partial_\varphi} \mathcal{A}) \mathcal{F} \wedge \mathcal{F} \,.
  \end{aligned}
\end{align}
It is then clear that for $c_2=2/3$, the na\"{\i}ve Noether-Wald charge is not radially conserved. 
The second term on the right hand side evaluates to 
\begin{align}
 \begin{aligned}
-3\frac{3c_2-2}{2}\int_{\cal S} (\iota_{\partial_\varphi} \mathcal{A}) \mathcal{F} \wedge \mathcal{F} = 3(3c_2-2) \pi\mathfrak{p} (a_\varphi^2 (r_2) - a^2_\varphi(r_1)) 2 \pi \eta_\Sigma \,.
  \end{aligned}
\end{align}
To evaluate the $\Gamma_3$ interface integral we use
 \begin{align}
 \begin{aligned}
( \iota_{\partial_\varphi} \mathcal{A}_a) \mathcal{A}_a \wedge \mathcal{F} |_{\Gamma_3} = -\mathfrak{p} a_\varphi a_\varphi' \omega_a \wedge \dd r \wedge \dd \varphi \,. 
  \end{aligned}
\end{align}
Then 
 \begin{align}
 \begin{aligned}
 \int_{\Gamma_3} \left( \mathcal{J}_S^{c_2} - \mathcal{J}_N^{c_2} \right) &= 3c_2\int_{\Gamma_3} ( \iota_{\partial_\varphi} \mathcal{A}) \left( \mathcal{A}_S - \mathcal{A}_N \right) \wedge \mathcal{F} \\
 &=\frac{3}{2}c_2 \int_{\cal S} (\iota_{\partial_\varphi} \mathcal{A}) \mathcal{F} \wedge \mathcal{F} \,.
  \end{aligned}
\end{align}
Summing the two contributions gives 
\begin{align}
 \label{patchyintegrals2}
 \begin{aligned}
  \fint_{S^1_\varphi \times \Sigma_{r_2}} \mathcal{J}^{c_2} - \fint_{S^1_\varphi \times \Sigma_{r_1}} \mathcal{J}^{c_2} = 3(1-c_2) \int_{\cal S} (\iota_{\partial_\varphi} \mathcal{A}) \mathcal{F} \wedge \mathcal{F} \, ,
  \end{aligned}
\end{align}
and therefore the charge $j$, corresponding to $c_2=1$, is radially conserved.

\subsection{Radially conserved charges: closed forms with interface integrals}
\label{app:patchwise-angular-momentum}

We report below the computations that show that the charges defined with the three-forms 
\begin{align}
  \begin{aligned}
    \mathcal{Q}_{\text{Page}} \equiv {}& \star_5 \mathcal{F} + \mathcal{A} \wedge \mathcal{F} \,, \\
    \mathcal{J}_{\text{Wald}} \equiv {}& \star_5 \dd[]{\qty((\partial_\varphi)^{\flat}) } + 3 \qty(\iota_{\partial_\varphi} \mathcal{A}) \qty(\star_5 \mathcal{F} + \frac{2}{3}\mathcal{A} \wedge \mathcal{F}) \,, 
  \end{aligned}
\end{align}
are radially conserved if defined by supplementing the three-dimensional integrals with the correct interface integrals. These computations were originally presented in \cite{Hanaki:2007mb}. Below we rewrite them in our present conventions and then show that for our ansatz the corresponding charges reduce to those discussed in the previous subsection, using non-closed three-forms, without supplementing the three dimensional integrals with two-dimensional interfaces.

In order to discuss $ \mathcal{Q}_{\text{Page}}$, we can consider a family of local three-forms as before, and denote these with tildes, 
\begin{align}
    \mathcal{\tilde Q}^{\tilde c_1} ={}& \star_5 \mathcal{F} + \tilde c_1 \mathcal{A} \wedge \mathcal{F} \,, 
\end{align}
because we will deal with the integral differently from the previous section. In particular, 
on a three-dimensional hypersurface
of constant $r$ and $t$,
we define 
\begin{align}
4\pi^2 \eta_\Sigma\, \tilde q_{\tilde c_1}\equiv \int_{S^1_\varphi \times U_S } \mathcal{\tilde Q}_S^{\tilde c_1} + \int_{S^1_\varphi \times U_N } \mathcal{\tilde Q}_N^{\tilde c_1} 
+ \tilde c_1 \int_{S^1_\varphi \times S^1_\phi } \mathcal{A}_N \wedge \mathcal{A}_S \, .
\end{align}
We can evaluate this on our background and compare with the previous definition. 
In particular, using 
\begin{align}
  \begin{aligned}
 \fint_{S^1_\varphi \times \Sigma } \mathcal{A}\wedge \mathcal{F} = \int_{S^1_\varphi \times S^1_\phi } \mathcal{A}_N \wedge \mathcal{A}_S 
  \end{aligned}
\end{align}
we have the equality 
\begin{align}
\tilde q_{\tilde c_1} = q_{2\tilde c_1}\, , 
\end{align}
which means the definition with the interface term is equal ``numerically'' to the na\"{\i}ve definition, with $c_1=2\tilde c_1$. 
In particular, 
\begin{align}
 \int_{S^1_\varphi \times \Sigma}
\mathcal{Q}_{\text{Page}} = \fint_{S^1_\varphi \times \Sigma} \mathcal{\hat Q}_{\text{Page}} \, .
 \end{align}
 Indeed, with this definition one can repeat the slab calculation of the previous section to show that
the Page charge is radially conserved \cite{Hanaki:2007mb}. Take $\tilde c_1=1$, then
 \begin{align}
 \begin{aligned}
0 &= \int_{{\cal S}_S}\dd \mathcal{Q}_S^{1} + \int_{{\cal S}_N}\dd \mathcal{Q}_N^{1} = \int_{\partial {\cal S}_S} \mathcal{Q}_S^{1} + \int_{\partial {\cal S}_N} \mathcal{Q}_N^{1} \\
&= \int_{S^1_\varphi \times U_S |_{r_1}} \mathcal{Q}_S^{1} - \int_{S^1_\varphi \times U_S |_{r_2}} \mathcal{Q}_S^{1} + \int_{\Gamma_3} \mathcal{Q}_S^{1} \\
&+ \int_{S^1_\varphi \times U_N |_{r_1}} \mathcal{Q}_N^{1} - \int_{S^1_\varphi \times U_N |_{r_2}} \mathcal{Q}_N^{1} - \int_{\Gamma_3} \mathcal{Q}_N^{1} \, . 
  \end{aligned}
\end{align}
On the other hand, we have
 \begin{align}
 \begin{aligned}
 \int_{\Gamma_3} \left( \mathcal{Q}_S^{1} - \mathcal{Q}_N^{1} \right) &= \int_{\Gamma_3} \left( \mathcal{A}_S - \mathcal{A}_N \right) \wedge \mathcal{F} \\
 & = - \int_{S^1_\varphi \times S^1_\phi |_{r_2}} \mathcal{A}_N \wedge \mathcal{A}_S +\int_{S^1_\varphi \times S^1_\phi |_{r_1}} \mathcal{A}_N \wedge \mathcal{A}_S \,, 
  \end{aligned}
\end{align}
so that 
 \begin{align}
 \begin{aligned}
0 &= \int_{S^1_\varphi \times U_S |_{r_1}} \mathcal{Q}_S^{1} - \int_{S^1_\varphi \times U_S |_{r_2}} \mathcal{Q}_S^{1} \\
&+ \int_{S^1_\varphi \times U_N |_{r_1}} \mathcal{Q}_N^{1} - \int_{S^1_\varphi \times U_N |_{r_2}} \mathcal{Q}_N^{1} \\
& +\int_{S^1_\varphi \times S^1_\phi |_{r_1}} \mathcal{A}_N \wedge \mathcal{A}_S - \int_{S^1_\varphi \times S^1_\phi |_{r_2}} \mathcal{A}_N \wedge \mathcal{A}_S \\
& = \tilde q_1 |_{r_1} - \tilde q_1 |_{r_2} \, .
  \end{aligned}
\end{align}
So the Page charge, supplemented with interface terms, is conserved \cite{Hanaki:2007mb}.

Let us now move to $\mathcal{J}_{\text{Wald}}$. We will begin considering a general background and a generic Killing vector $\xi$, and later we will restrict to our background, where the interface contribution simplifies. We will work in gauges such that
\begin{equation}
 \mathcal{L}_\xi \mathcal{A}_N=0=\mathcal{L}_\xi \mathcal{A}_S\, ,
\label{eq:A4-invariance}
\end{equation}
and define $ b_\xi^{(a)}\equiv\iota_\xi \mathcal{A}_a$. Cartan's identity then implies that 
\begin{equation}
\dd b_\xi^{(a)} = - \iota_\xi \mathcal{F}\, .
 \label{eq:A4-dphi}
\end{equation}

It is useful to distinguish literal equality of differential forms from
equality after pullback to the invariant hypersurfaces relevant for the charge. We
write
\begin{equation}
 \alpha\doteq\gamma
 \label{eq:A4-doteq}
\end{equation}
when $\alpha-\gamma$ is of the form
$\iota_\xi\Omega$, with $\Omega$ an invariant top-degree form. The
pullback of such a term to a $\xi$-invariant chain vanishes. 
Following \cite{Hanaki:2007mb} we define the local three-form
\begin{equation}
 X_\xi[\mathcal{A}_a]
 \equiv
 b_\xi^{(a)}
 \left( \star_5\mathcal{F}+\frac{2}{3}\mathcal{A}_a\wedge \mathcal{F}
 \right).
 \label{eq:A4-X-def}
\end{equation}
Using the Maxwell equation, a straightforward computation shows that 
\begin{align}
 \dd X_\xi[ \mathcal{A}]
 &=
 -(\iota_\xi \mathcal{F})\wedge\star_5 \mathcal{F}
 -\frac{1}{3}\iota_\xi(\mathcal{A}\wedge \mathcal{F}\wedge \mathcal{F})
 \doteq
 -(\iota_\xi \mathcal{F})\wedge\star_5\mathcal{F}.
 \label{eq:A4-dX}
\end{align}
For a Killing vector, the Killing identity gives
\begin{equation}
 \dd\star_5 \dd\xi^\flat
 =
 2\star_5 (R_5)_{\mu\nu}\xi^\mu \dd x^\nu
 \label{eq:A4-Killing-identity}
\end{equation}
and the Einstein equations then imply 
\begin{equation}
 \dd\star_5\dd\xi^\flat
 \doteq
 3(\iota_\xi \mathcal{F})\wedge\star_5\mathcal{F}.
 \label{eq:A4-Komar-divergence}
\end{equation}
Then \eqref{eq:A4-dX} and \eqref{eq:A4-Komar-divergence} show that
\begin{equation}
\dd \mathcal{J}_{\text{Wald}} = \dd\left(\star_5\dd\xi^\flat+3X_\xi[\mathcal{A}]\right)\doteq 0.
 \end{equation}
Strictly speaking, the adjective ``closed'' refers to the
equivalence relation $\doteq$, which is precisely what is required for
conservation of its integral on a $\xi$-invariant cycle.
The integral is finally defined as 
\begin{align}
 \int_{S^1_\varphi \times \Sigma}
 \mathcal{J}_{\rm Wald} [\xi]
 \equiv{}&
 \int_{S^1_\varphi \times U_S}
 \mathcal{J}_{{\rm Wald},S} [\xi]
 +\int_{S^1_\varphi \times U_N} \mathcal{J}_{{\rm Wald},N} [\xi]
 +\int_{\Gamma_2}Y_\xi[\mathcal{A}_N,\mathcal{A}_S]\, ,
 \label{eq:A4-modified-integral-general}
\end{align}
where the interface two-form $Y_\xi[\mathcal{A}_N,\mathcal{A}_S]$ must obey
\begin{equation}
 \dd Y_\xi \doteq \mathcal{J}_{{\rm Wald},S} - \mathcal{J}_{{\rm Wald},N} \,. 
 \label{eq:A4-Y-condition}
\end{equation}
To construct a symmetric representative, it is useful to introduce
\begin{equation}
\mathcal{\overline A} \equiv\frac{1}{2}(\mathcal{A}_S+\mathcal{A}_N)\, , \qquad
 \overline b_\xi\equiv\iota_\xi \mathcal{\overline A}\, ,
 \qquad \beta=\mathcal{A}_S-\mathcal{A}_N\, , \qquad b_\xi\equiv\iota_\xi\beta \, ,
 \label{eq:A4-midpoint}
\end{equation}
with
\begin{equation}
 \dd b_\xi =\mathcal{L}_\xi\beta-\iota_\xi \dd\beta = 0\, .
 \label{eq:A4-db}
\end{equation}
Denoting by $\overline Z$ a two-form on the overlap satisfying
\begin{equation}
\dd \overline Z=\star_5 \mathcal{F}+ \mathcal{\overline A}\wedge \mathcal{F}\, ,
 \label{eq:A4-Zbar}
\end{equation}
by direct computation one can show that a representative of the interface two-form is given by
\begin{equation}
 Y_\xi [ \mathcal{A}_N, \mathcal{A}_S] =
 3b_\xi\,\overline Z + \overline b_\xi\, \mathcal{\overline A}\wedge \beta = 3b_\xi\,\overline Z +\overline b_\xi\, \mathcal{A}_N\wedge \mathcal{A}_S \, .
 \label{eq:A4-Y-general}
\end{equation}
In summary, the final result is 
\begin{align}
 \int_{S^1_\varphi \times \Sigma}
 \mathcal{J}_{\rm Wald} [\xi] ={}&
 \fint_{S^1_\varphi \times \Sigma}
 \mathcal{J}_{\rm Wald}[\xi]
 +\int_{\Gamma_2}\left( 3b_\xi\,\overline Z +\overline b_\xi\, \mathcal{A}_N\wedge \mathcal{A}_S\right) \, . 
\end{align}

Let us now specify this to our particular background, taking $\xi=\partial_\varphi$. On the overlap we have
\begin{equation}
 \beta\equiv \mathfrak{p} (\omega_N-\omega_S) \qquad \Rightarrow \qquad b_{\partial_\varphi} = 0 \, 
 \label{eq:A4-beta}
\end{equation}
and 
\begin{equation}
 \iota_{\partial_\varphi} \mathcal{A}_N=\iota_{\partial_\varphi}\mathcal{A}_S=\overline b_{\partial_\varphi} = a_\varphi \, ,
 \label{eq:A4-phi-aphi}
\end{equation}
so that the term containing the implicit primitive $\overline Z$
vanishes identically, and 
the interface term assumes the simple form
\begin{equation}
 Y_{\partial_\varphi} [\mathcal{A}_N,\mathcal{A}_S] = a_\varphi \,\mathcal{A}_N\wedge \mathcal{A}_S\, .
 \label{eq:A4-Y-final}
\end{equation}

Let us now show that \eqref{eq:A4-modified-integral-general} is radially conserved. At fixed $t$, consider the four-dimensional slab
$ \mathcal{S} = [r_1,r_2]\times S_\varphi^1\times\Sigma_{\mathfrak{g}}$, 
covered by the two patches
\begin{equation}
 \mathcal{S}_a
 =
 [r_1,r_2]\times S_\varphi^1\times U_a,
 \qquad
 a\in\{N,S\}.
 \end{equation}
The common three-dimensional interface is $ \Gamma_3 = [r_1,r_2]\times S_\varphi^1\times S_\phi^1$. Applying Stokes' theorem separately on $\mathcal{S}_S$ and
$\mathcal{S}_N$, the interface contributions combine into
\begin{equation}
 \int_{\Gamma_3}
 \left(
 \mathcal{J}_{{\rm Wald},S} 
 -\mathcal{J}_{{\rm Wald},N} 
 \right) = \int_{\Gamma_3}\dd Y_{\partial_\varphi} =\int_{S^1_\varphi \times \Sigma |_{r_2}} Y_{\partial_\varphi} - \int_{S^1_\varphi \times \Sigma |_{r_1}} Y_{\partial_\varphi}\, .
\end{equation}
It follows that
\begin{equation}
 \int_{S^1_\varphi \times \Sigma |_{r_1}}
 \mathcal{J}_{{\rm Wald}} - \int_{S^1_\varphi \times \Sigma |_{r_2}}
 \mathcal{J}_{{\rm Wald}} =0\, .
  \end{equation}
Thus the globally defined radially conserved quantity is
\begin{align}
 j= \frac{1}{4\pi^2\eta_\Sigma}
 \left[ \fint_{S_\varphi^1\times \Sigma}
\mathcal{J}_{{\rm Wald}} 
 +\int_{\Gamma_2}a_\varphi \mathcal{A}_N\wedge \mathcal{A}_S
 \right].
 \label{eq:A4-j-patchwise}
\end{align}

The same argument does not establish conservation of the integral over
$\mathbb{R}_t\times\Sigma_{\mathfrak{g}}$. The reason is that
$\xi=\partial_\varphi$ is not tangent to that three-cycle, and therefore
the pullback of the terms represented by $\doteq$ is not forced to
vanish.

It is straightforward to check the equivalence of the present definition with that of the previous section. Indeed we have
\begin{align}
 \int_{S^1_\varphi \times \Sigma}
 \mathcal{J}_{{\rm Wald}} =& \fint_{S^1_\varphi \times \Sigma}
  \left[ \star_5 \dd(\partial_\varphi)^\flat
 +3a_\varphi\star_5\mathcal{F}
 +2a_\varphi \mathcal{A}\wedge \mathcal{F}
 \right] +a_\varphi\int_{\Gamma_2}\mathcal{A}_N\wedge \mathcal{A}_S \nonumber\\
 =& \fint_{S^1_\varphi \times \Sigma}
  \left[ \star_5 \dd(\partial_\varphi)^\flat
 +3a_\varphi\star_5 \mathcal{F}
 +3a_\varphi \mathcal{A}\wedge \mathcal{F}
 \right] = \fint_{S^1_\varphi \times \Sigma} \mathcal{\hat J}_{{\rm Wald}} \, ,
 \end{align}
where we used the fact that on a constant-$r$ slice, $a_\varphi$ is constant and employed the identity \eqref{papero}.
 
% Bibliography 

%% [A] Recommended: using JHEP.bst file
\newpage
\bibliographystyle{JHEP}
\bibliography{biblio.bib}

%% or
%% [B] Manual formatting (see below)
%% (i) We suggest to always provide author, title and journal data or doi:
%% in short all the informations that clearly identify a document.
%% (ii) please avoid comments such as "For a review'', "For some examples",
%% "and references therein" or move them in the text. In general, please leave only references in the bibliography and move all
%% accessory text in footnotes.
%% (iii) Also, please have only one work for each \bibitem.

%\begin{thebibliography}{99}

%\bibitem{a}
%Author,
%\emph{Title},
%\emph{J. Abbrev.} {\bf vol} (year) pg.

%\bibitem{b}
%Author,
%\emph{Title},
%arxiv:1234.5678.

%\bibitem{c}
%Author,
%\emph{Title},
%Publisher (year).

%\end{thebibliography}
\end{document}